\documentclass[amssymb,amsmath,prd,twocolumn,superscriptaddress,nofootinbib]{revtex4-2}

\usepackage{graphicx}
\usepackage{bm}
\usepackage{amssymb,amsmath}
\usepackage{mathrsfs}
\usepackage{latexsym}
\usepackage{color}
\usepackage[normalem]{ulem}
\usepackage{dcolumn}
\usepackage[colorlinks=true,citecolor=blue,urlcolor=blue]{hyperref}
\usepackage[usenames,dvipsnames]{xcolor}
\usepackage{xspace}
\usepackage{graphicx}

\allowdisplaybreaks

\newcommand{\CCA}{\affiliation{Center for Computational Astrophysics, Flatiron Institute, 162 5th Ave, New York, NY 10010, USA}}
\newcommand{\XGI}{\affiliation{eXtreme Gravity Institute, Department of Physics, Montana State University, Bozeman, Montana 59717, USA}}
\newcommand{\GT}{\affiliation{Center for Relativistic Astrophysics and School of Physics, Georgia Institute of Technology, Atlanta, GA 30332, USA}}
\newcommand{\CIT}{\affiliation{Department of Physics, California Institute of Technology, Pasadena, California 91125, USA}}
\newcommand{\CITLab}{\affiliation{LIGO Laboratory, California Institute of Technology, Pasadena, CA 91125, USA}}
\newcommand{\US}{\affiliation{Mathematical Sciences and STAG Research Centre, University of Southampton, SO17 1BJ, Southampton, UK}}
\newcommand{\GSI}{\affiliation{GSI Helmholtzzentrum f\"ur Schwerionenforschung,
	Planckstra{\ss}e 1, 64291 Darmstadt, Germany}}
\newcommand{\HFHF}{\affiliation{Helmholtz Research Academy Hesse for FAIR (HFHF), GSI
	Helmholtz Center for Heavy Ion Research, Campus Darmstadt,
	Planckstra{\ss}e 1, 64291 Darmstadt, Germany}}

\definecolor{kcmagenta}{rgb}{0.54, 0.17, 0.88}
\definecolor{chorange}{rgb}{0.851, 0.372, 0.007}
\definecolor{tlteal}{rgb}{0,.55,.55}
\definecolor{jcpink}{rgb}{1.0, 0.0, 0.5}
\definecolor{mmgreen}{rgb}{0.0, 0.8, 0.6}
\definecolor{bbsalmon}{rgb}{1.0, 0.47, 0.42}
\definecolor{mworange}{rgb}{1.0, 0.60, 0.33}

\newcommand{\code}{\texttt}

\newcommand{\BayesWave}{{\tt BayesWave}\xspace}

\newcommand{\ttrig}{t_{\textrm{trig}}}

\newcommand{\Mc}{{\cal{M}}}
\newcommand{\hcbc}{h^{\mathrm{cbc}}}
\newcommand{\hw}{h^{\mathrm{w}}}
\newcommand{\cbcw}{{\mathrm{cbc/w}}}

\newcommand{\fpeak}{f_{\mathrm{peak}}}
\newcommand{\Rtyp}{R_{1.4}}

\newcommand{\cbc}[1]{#1^\mathrm{cbc}}
\begin{document}

\title{Probing neutron stars with the full premerger and postmerger gravitational wave signal from binary coalescences}

\author{Marcella Wijngaarden}  \CCA\US
\author{Katerina Chatziioannou} \CIT \CITLab 
\author{Andreas Bauswein}\GSI\HFHF
\author{James A. Clark}\CITLab\GT
\author{Neil J. Cornish}\XGI

\date{\today}

\begin{abstract}
The gravitational wave signal emitted during the coalescence of two neutron stars carries information about the stars' internal structure. 
During the long inspiral phase the main matter observable is the tidal interaction between the binary components, an effect that can be parametrically modeled
with compact-binary solutions to general relativity.  
After the binary merger the main observable is frequency modes of the remnant, most commonly giving rise to a short-duration signal
accessible only through numerical simulations. The complicated morphology and the decreasing detector sensitivity 
 in the relevant frequencies currently hinder detection of the postmerger signal and motivate separate analyses for the premerger and postmerger data. 
However, planned and ongoing detector improvements could soon put the postmerger signal within reach.
In this study we target the whole premerger and postmerger signal without an artificial separation at the binary merger.
We construct a hybrid analysis that models the inspiral with templates based on analytical calculations and calibrated to numerical relativity and the postmerger signal with a flexible morphology-independent analysis. Applying this 
analysis to GW170817 we find, as expected, that the postmerger signal remains undetected. We further study simulated signals and find that we can reconstruct the full signal and simultaneously estimate both the premerger tidal deformation and the postmerger signal frequency content. Our analysis allows us to study neutron star physics using all the data available and directly test the premerger and postmerger signal for consistency thus probing effects such as the onset of the hadron-quark phase transition.
\end{abstract}

\maketitle

\section{Introduction}
\label{sec:intro}

Understanding of properties of supranuclear matter in the core of neutron stars (NSs), commonly encoded through their Equation of State (EoS) is an ongoing challenge~\cite{Lattimer:2015nhk,Ozel:2016oaf,Oertel:2016bki,Baym:2017whm}. Electromagnetic and gravitational wave (GW) observations of the coalescence of two NSs provide an additional tool to study the ultra-dense interiors of NSs, as they affect the emitted signals. Two such GW events, GW170817~\cite{TheLIGOScientific:2017qsa} and GW190425~\cite{Abbott:2020uma}, have already been detected by LIGO~\cite{TheLIGOScientific:2014jea} and Virgo~\cite{TheVirgo:2014hva} and planned, improved GW detectors promise more and louder detections in the future~\cite{Aasi:2013wya}. These data will add to the interdisciplinary effort to study the EoS, including observations of macroscopic NS properties such as heavy pulsar observations~\cite{Antoniadis:2013pzd,Cromartie:2019kug,Fonseca:2021wxt}, X-ray pulse-profiling with NICER~\cite{Miller:2019cac,Riley:2019yda,Miller:2021qha,Riley:2021pdl}, and nuclear theory and experiment, e.g.~\cite{Raaijmakers:2019dks,Dietrich:2020efo,Landry:2020vaw,Al-Mamun:2020vzu,Reed:2021nqk,Essick:2020flb,Raaijmakers:2021uju,Biswas:2021yge,Pang:2021jta,Legred:2021hdx,Essick:2021kjb,Essick:2021ezp}.

The GW signal from a binary NS (BNS) coalescence consists of two parts: a premerger and a postmerger. In the premerger phase the binary components inspiral toward each other as they lose orbital energy to GWs and eventually collide~\cite{Blanchet:2013haa}. In the case of GW170817, the detectable premerger signal increased in frequency from  $\sim23$Hz to many hundreds of Hz as the orbital separation between the two NSs rapidly decayed over $\sim2$ minutes~\cite{Abbott:2018wiz} before finally merging at a frequency of $\sim1500$Hz~\cite{Torres-Rivas:2018svp}. During the late stages of the inspiral, tidal interactions accelerate the binary evolution and leave an imprint on the signal that depends on the NS matter properties and can thus be used to infer the EoS~\cite{Chatziioannou:2020pqz}. The premerger GW signal is typically modeled with compact binary coalescence (CBC) templates that are based on approximate solutions to the general relativity field equations and numerical relativity (NR) simulations, see~\cite{Dietrich:2020eud} for a recent review.

After the inevitable merger, the remnant star evolves in a way that depends on the system mass and the EoS, see~\cite{2019JPhG...46k3002B,2019PrPNP.10903714B,2021GReGr..53...59S,Bernuzzi:2020tgt} for reviews. For most EoSs and NS masses, a hypermassive NS supported by differential rotation and thermal effects is expected to be formed and sustained for $\sim10-100$ms~\cite{Baumgarte:1999cq}. During this time, the NS remnant emits a short-duration postmerger GW signal with a characteristic peak at  $1500-4000$Hz~\cite{Xing:1994ak,Ruffert:2001gf,Shibata:2005ss, 2005PhRvL..94t1101S,shibata:06bns, Oechslin:2007gn, 2011PhRvD..83l4008H, 2012PhRvL.108a1101B, Bauswein:2012ya, hotokezaka:13, 2014PhRvL.113i1104T, 2014arXiv1412.3240T, 2014arXiv1411.7975K, 2015arXiv150401764B, bauswein:15, Foucart2016, Lehner:2016lxy, 2016CQGra..33x4004E, Dietrich2017}
depending on the mass and EoS, a promising frequency range for upcoming GW detectors\footnote{Depending on the NS mass and EoS, other possibilities for the postmerger remnant include direct collapse to a black hole (BH) whose ringdown signal is at $\sim6$kHz, and a long-lived NS that emits a signal that can last for minutes or more.}. In what follows this is the case we consider. While the exact physics governing the postmerger phase is not fully understood and NR simulations are complicated by factors such as thermal effects, turbulence, magnetohydrodynamical instabilites, neutrinos, and possible phase transitions, simulations make a robust empirical prediction for the frequency of the peak of the postmerger spectrum given the NS mass and EoS. 

Information extracted from a postmerger signal about NS properties, is complementary to the premerger. Analysis of a postmerger signal would probe different density~\cite{2020PhRvD.102l3023B} and temperature~\cite{Raithel:2021hye,Hammond:2021vtv} regimes of the EoS, as the more massive, hot remnant will have a higher maximum density than the premerger NSs~\cite{Ruffert:2001gf}. This complementarity can lead to insights about high-density phenomena such as phase-transitions between the densities probed by the premerger and postmerger signals~\cite{Radice:2016rys,Most:2018eaw,Bauswein:2018bma,Bauswein:2019skm,Weih:2019xvw,Bauswein:2020ggy,Liebling:2020dhf,Prakash:2021wpz,2021PhRvD.104d3011L}. Finally, detection of a postmerger signal would allow us to determine the nature of the merger remnant, namely NS or BH, which has implications for a potential electromagnetic counterpart and its interpretation~\cite{Margalit:2019dpi} as well as the determination of the threshold mass for prompt collapse~\cite{Bauswein:2020aag}.

In the case of GW1708017 and GW190425 only piecewise analyses of the GW signal have been performed focusing either on the premerger or on the postmerger part of the signal~\cite{Abbott:2018wiz,Abbott:2017dke,LIGOScientific:2018urg,Abbott:2020uma}. This reduces the computational cost significantly, as the postmerger signal requires a large sampling rate (typically $8194$Hz) while the premerger analysis involves long duration data segments (typically $128$s). As no postmerger signal is detectable yet~\cite{Abbott:2018wiz,Abbott:2017dke}, excluding it from the premerger analysis did not bias the results for the binary parameters~\cite{Dudi:2018jzn}. However, looking forward and in the case of a postmerger signal detection, separate analyses will not be able to exploit phase coherence through merger or parameter relations, such as those between the tidal deformability from the premerger signal and the frequency peak of the postmerger spectrum~\cite{2015arXiv150401764B,2020PhRvD.102l3023B,2020PhRvD.101h4039V}.

Another reason for separate premerger and postmerger analyses is that the morphology and details of the postmerger signal are not well understood, making it difficult to construct a first-principles physical model such as the premerger CBC templates. While approximate analytic models for the postmerger signal have been proposed with the aid of NR simulations~\cite{Hotokezaka:2013iia,Bauswein:2015vxa,Bose:2017jvk,Tsang:2019esi,Breschi:2019srl,Easter:2020ifj,Soultanis:2021oia}, these are still limited in accuracy
as well as by uncertainties of the simulations they are based on~\cite{Breschi:2019srl}. An alternative are morphology-independent analyses that are not limited
to a specific signal type~\cite{Clark:2014wua,Clark:2015zxa,Chatziioannou:2017ixj}. Specifically, a model-agnostic approach with the \BayesWave algorithm~\cite{Cornish:2014kda,Cornish:2020dwh,bayeswave} has been shown to accurately determine the main features of the postmerger signal~\cite{Chatziioannou:2017ixj,Torres-Rivas:2018svp}, and in some cases do so more accurately than tailored models~\cite{Easter:2020ifj}. 

\BayesWave models a GW signal with a sum of sine-Gaussian wavelets. Both the number of wavelets and their parameters are marginalized over, resulting in a flexible analysis
that has been applied to a variety of signals~\cite{Kanner:2016,Littenberg:2016,becsy:2017,Tsang:2018uie,Tsang:2019zra,Pankow:2018qpo,Millhouse:2018dgi,Ghonge:2020suv,Dalya:2020gra,Chatziioannou:2021mij}.
 \BayesWave's sine-Gaussian wavelets are particularly adept for postmerger signals that are dominated by distinct frequency components, as each component can be approximately modeled by a wavelet. 
In this context, \BayesWave has been used to search for a short-duration, high-frequency signal after both GW170817~\cite{Abbott:2018wiz} and GW190425~\cite{Abbott:2020uma}, returning
null results and upper limits on the energy content~\cite{Abbott:2018wiz}.  A detailed study of \BayesWave's sensitivity to postmerger signals is presented in~\cite{Chatziioannou:2017ixj}.

In this paper we construct a hybrid analysis of the full BNS GW signal that targets both the premerger and the postmerger data. We simultaneously analyze the full signal using 1) a CBC template to describe the well-modeled premerger phase and 2) sine-Gaussian wavelets to capture the less well understood postmerger. The parameters of the CBC template and the wavelets are simultaneously sampled over to obtain the combined multidimensional posterior for all components of the model. Those parameters include the premerger tidal deformability that quantifies the inspiral tidal deformation and the peak frequency of the postmerger spectrum. Our analysis extracts both simultaneously and allows for direct consistency comparisons.

After validating the hybrid modeled/unmodeled analysis on a toy model based on GW150914 data, we apply it to GW170817 and find that the postmerger signal
remains undetected, as expected from the upper limit estimates of~\cite{Abbott:2018wiz}. We further analyze simulated signals of high signal-to-noise ratio (SNR) for which the postmerger signal
is detectable. We show that our analysis can reconstruct the premerger signal as well as the main components of the postmerger signal such as the dominant frequency mode. We simulate signals for which the premerger and postmerger parts are consistent (corresponding to a hadronic EoS) and inconsistent (corresponding to an EoS with a strong phase transition in the relevant density scales) and demonstrate that our analysis can extract either behavior.

The rest of the paper is organized as follows.
In Sec.~\ref{sec:code} we describe in detail our analysis and algorithm.
In Sec.~\ref{sec:150914} we present a proof-of-principle analysis on GW150914.
In Sec.~\ref{sec:170817} we reanalyze the GW170817 data.
In Sec.~\ref{sec:injections} we apply our analysis to simulated data.
In Sec.~\ref{sec:conclusions} we conclude.

\section{Methodology and Models}
\label{sec:code}

In order to analyze a GW signal that contains both modeled and unmodeled features we employ a hybrid analysis that makes use of both CBC templates and 
flexible models for the signal. We base our analysis on the morphology-agnostic data analysis algorithm \BayesWave~\cite{Cornish:2014kda,Cornish:2020dwh} 
by extending it to account for CBC templates similar to~\cite{Chatziioannou:2021ezd}. In its core functionality \BayesWave simultaneously models 
a GW \textit{signal} (referred to as the ``signal model" in \BayesWave literature), instrumental \textit{glitches} (the ``glitch model"), and the gaussian
detector \textit{noise} (the ``PSD model"). \BayesWave uses minimal assumptions, i.e., no physical model, to describe these components. Instead, the \textit{signal} and the \textit{glitch} are modeled as the sum of a variable number of sine-Gaussian wavelets, whereas the \textit{noise} model describes the power spectral density (PSD) of the Gaussian noise with a variable number of spline points and Lorentzians. For this work and~\cite{Chatziioannou:2021ezd} a fourth component has been added, which models
GW signals through CBC templates obtained by solving the two-body problem in general relativity; the \textit{CBC} model.

\subsection{Overview of models}

\BayesWave uses sampling methods to characterize the multi- and variable-dimension posterior distribution for independent models that target
different features of the GW data. 
The goal is to draw samples from the posterior distribution function $p(M|d)$, the probability that a model $M$ describes the data $d$ and a prior $p(M)$, defined as
\begin{align}
	p(M|d) = \frac{p(M) p(d|M)}{p(d)},
\end{align}
where $p(d)$ is the evidence, and $p(d|M)$ expresses the likelihood of the data for a given $M$.  
We model the data in each interferometric detector $d_I$ as a linear combination of multiple components
\begin{align}
	d_I =  g_I + \hcbc_I + \hw_I + n_I \equiv M_I + n_I,
\end{align}
where $g_I$ denotes any detector glitches\footnote{For the remainder of the paper we will work with data without instrumental glitches
and thus ignore the glitch model $g_I$. As we will not be making use of the glitch model, we will also refer to the wavelet signal model simply as the \textit{wavelet} model to avoid confusion with the CBC signal model.}, the full GW signal ($\hcbc_I + \hw_I$) consists of the CBC waveform model $\hcbc_{I}$ and any signal (such as the postmerger) that is not included in the CBC template and is captured by wavelets $\hw_I$, and $n_I$ is the Gaussian component of the noise. The combination of a signal, glitch, and noise defines the full model $M$. Given such a model, the likelihood function $p(d_I|M_I)$, in short $L_I$, can be computed by noting that the residual $r_I = d_I - M_I$ is consistent with Gaussian noise (see also Eq. (3) of~\cite{Cornish:2020dwh}):
\begin{align}
	\log{L_I} = -\frac{1}{2} (r_I | r_I) +C_I,\label{eq:logL}
\end{align}
where $C_I$ is a constant that depends on the PSD of the detector noise $S_{n,I}(f)$, and the noise weighted inner product is defined as
\begin{align}
	(a|b) \equiv 2 \int \frac{a^*(f) b(f) + b^*(f) a(f)}{S_{n,I}(f)} df, \label{eq:innerproduct}
\end{align}
where an asterisk denotes a complex conjugate.
The remaining ingredients of the analysis are a specific model for 
each data component, the corresponding prior, and the sampling procedure.

We express the GW signal at geocenter and then project it onto each detector in the network to compute the detector response in the frequency domain
\begin{equation}
	\begin{aligned}
	h^{\cbcw}_I(f) &= e^{2 \pi i \Delta t(\alpha, \delta)} \\ &\left[F^{+}(\alpha, \delta, \psi) h^{\cbcw}_{+} + F^{\times}(\alpha, \delta, \psi) h^{\cbcw}_{\times} \right] .\label{eq-proj}
	\end{aligned}
\end{equation}
The intrinsic part of the signal for each GW polarization mode is given by $h^{\cbcw}_{+}$ and $h^{\cbcw}_{\times}$ and it is different under the CBC or the wavelet model.
The projection then involves a time delay in the arrival of the signal in the different detectors $\Delta t(\alpha, \delta)$, the detector antenna patterns $F^+(\alpha, \delta, \psi)$, $F^{\times}(\alpha, \delta, \psi)$ for each GW polarization mode, parametrized through the sky location $\alpha$, $\delta$ corresponding to the right ascension and declination respectively, and the polarization angle $\psi$. The projection is independent of whether the signal
is expressed through CBC templates or wavelets so the functional form of Eq.~\eqref{eq-proj} and the parameters $(\alpha,\delta,\psi)$ are common. The ratio $h^{\cbcw}_{+}/h^{\cbcw}_{\times}$ is related to the 
binary inclination $\iota$ and is also shared between the CBC and wavelet models.

By construction \BayesWave breaks down the full parameter space defined by all models in different blocks and uses a blocked Gibbs sampler to explore the posterior. Each block of parameters is 
sampled over with different (reversible jump~\cite{10.1093/biomet/82.4.711}) Markov Chain Monte Carlo (RJMCMC) samplers. Parallel tempering is employed both for efficient posterior exploration and evidence calculation~\cite{10.1080/10635150500433722}, so all samplers share the same number and temperature of chains and exchanges between the chains
proceed for all models simultaneously. 
In order for this procedure to lead to efficient exploration of the posterior distribution, each block of parameters needs to be selected to be as independent from and uncorrelated to other blocks as possible.

\begin{figure}
	\centering
	\includegraphics[width=0.49\textwidth]{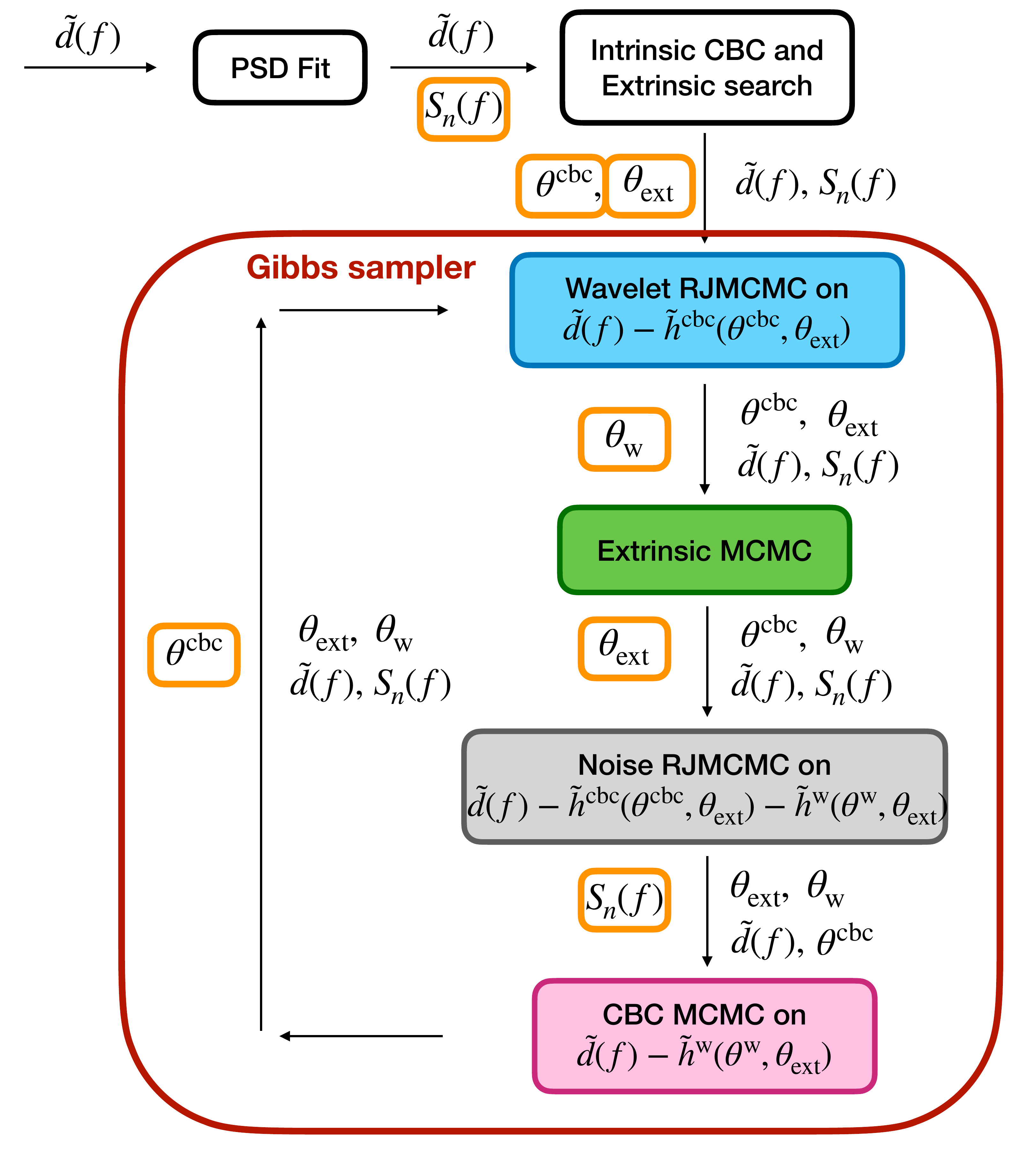}
	\caption{General code workflow. After some preconditioning, the data are used to obtain a quick fit to the noise PSD $S_n(f)$ and CBC $\theta^{\mathrm{cbc}}$ and extrinsic $\theta_{\mathrm{ext}}$ parameters. These are then used as a starting point for the 
	blocked Gibbs sampler (red box) which consists of $4$ independent samplers. Each iteration of the Gibbs sampler consists of a wavelet RJMCMC (blue box) that updates the wavelet parameters
	$\theta^{\mathrm{w}}$, an extrinsic MCMC (green box) that updates the extrinsic parameters $\theta_{\mathrm{ext}}$, a noise RJMCMC (gray box) that updates $S_n(f)$, and a CBC MCMC (pink box)
	that updates $\theta^{\mathrm{cbc}}$. Orange rectangles enclose the parameter that is being updated from the
	preceding box. Each block consists of ${\cal{O}}(10^2)$ iterations, while the entire red Gibbs box consists of ${\cal{O}}(10^4)$ iterations. }
	\label{fig:workflow}
\end{figure}

The sampler ensemble consists of (1) a CBC MCMC, Sec.~\ref{sec:cbcsampler}, (2) an RJMCMC sampler for the wavelet intrinsic parameters, Sec.~\ref{sec:intrinsicsampler}, (3) an MCMC sampler for the common extrinsic parameters for the CBC and wavelet models, Sec.~\ref{sec:extrinsicsampler}, and (4) an RJMCMC noise model sampler, Sec.~\ref{sec:psdsampler}. The structure is given in Fig.~\ref{fig:workflow} 
and it is similar to the code used in~\cite{Chatziioannou:2021ezd} but instead of 
nonoverlapping parameter blocks, we now allow for a set of common parameters to be updated in multiple blocks as detailed in the subsequent subsections. 
Sampling then proceeds iteratively both within each block (${\cal{O}}(10^2)$ iterations at a time) and between different blocks (${\cal{O}}(10^4)$ iterations).
In the rest of this section, we discuss each (RJ)MCMC, which parameters it updates, and how it relates to the other samplers. The discussion is structured 
to be as self-contained and detailed as possible; readers primarily interested on the results can find them starting in Sec.~\ref{sec:150914}.

%

\subsection{CBC-specific parameters: CBC sampler}
\label{sec:cbcsampler}

A nonprecessing quasicircular inspiral of two compact objects is characterized by a set of $6$ intrinsic source parameters and a set of $7$ extrinsic parameters that describe the binary's location and orientation relative to the detectors. 
The CBC sampler updates the parameters that are specific to $\hcbc_{+}$ and $\hcbc_{\times}$, namely masses, spins, tidal parameters (for BNSs), distance, time, and phase of coalescence.
In Fig.~\ref{fig:workflow} the CBC sampler is depicted by the pink box ``CBC MCMC" and the relevant parameters are collectively denoted by $\theta^{\mathrm{cbc}}$.

\subsubsection{Waveform model and Parameters}

The CBC sampler is integrated with the \code{LALSimulation} \cite{lalsuite} waveforms suite and can make use of any non-precessing waveform model; we here focus on PhenomD~\cite{Husa:2015iqa,Khan:2015jqa} and PhenomDNRT~\cite{Dietrich:2017aum,Dietrich:2018uni,Dietrich:2019kaq}. The former includes the full inspiral-merger-ringdown (IMR) signal expected from a BBH coalescence. The latter targets the inspiral of BNS systems as it includes the effect of 
tidal interactions between the NSs, manifesting as a
change in the waveform phase. PhenomDNRT models the long inspiral phase and instead of terminating abruptly at merger, it decays rapidly through a window function after the merger frequency~\cite{Dietrich:2018uni,Dietrich:2019kaq}. PhenomDNRT is well suited for our purposes regarding BNS signals as the waveform remains smooth and continuous, naturally allowing the resulting CBC (inspiral) + wavelet (postmerger) model to be smooth without any forced stitching. 

A total of $9$ CBC parameters are updated by the CBC sampler $\theta^{\mathrm{cbc}}\equiv (m_1,m_2,\chi_1,\chi_2,\Lambda_1,\Lambda_2, D_L, t_c, \phi_c)$. 
\begin{itemize}
\item The masses of the binary components are $m_1$ and $m_2$ with the convention $m_1 \geq m_2$. We also parametrize the masses through the total mass $M = m_1 + m_2$, chirp mass $\Mc= \left(m_1 m_2\right)^{3/5}/\left(m_1 + m_2\right)^{1/5}$, mass ratio $q=m_2/m_1<1$, and symmetric mass ratio $\eta = (\Mc/M)^{5/3}<1/4$.
\item The dimensionless spin components aligned with the orbital angular momentum are $\chi_1$ and $\chi_2$. Current detector sensitivities make individual spin components hard to measure~\cite{Purrer:2015nkh,Chatziioannou:2018wqx} 
so a commonly used spin combination is the mass-weighted effective spin parameter,
\begin{align}
\chi_\text{eff} =  \frac{m_1 \chi_1 + m_2 \chi_2}{m_1 + m_2},
\end{align}
which is conserved under spin-precession to at least the second post-Newtonian order~\cite{Racine:2008qv}.
\item Tidal interactions in BNS signals are encoded by the NS dimensionless tidal deformabilities 
\begin{align}
\Lambda_{1,2} = \frac{2}{3} k_{1,2} \left( \frac{R_{1,2}}{m_{1,2}} \right)^5,
\end{align}
where $k_{1,2}$ is the tidal Love number, $R_{1,2}$ is the radius, and $m_{1,2}$ is the mass of each binary component.
To leading order, the tidal effects are imprinted in the waveform through the binary tidal deformability~\cite{Favata:2013rwa,Wade:2014vqa}
\begin{align}
\tilde{\Lambda} = \frac{16}{13} \frac{(12q + 1)\Lambda_1 + (12 + q)q^4 \Lambda_2}{(1 + q)^5},
\end{align}
with subleading terms introducing another tidal parameter that is generally not measurable, $\delta\tilde{\Lambda}$~\cite{Wade:2014vqa}.
For BBH systems, $\Lambda_1=\Lambda_2=0$~\cite{Binnington:2009bb,Poisson:2020vap}. 
\item The CBC sampler further handles $3$ extrinsic parameters either because they do not explicitly appear in the wavelet model (time of coalescence $t_c$, luminosity distance $D_L$) 
or because they are 
correlated with intrinsic parameters and need
to be simultaneously sampled for efficiency (phase of coalescence $\phi_c$). 
Despite the well-known correlation between the distance and the inclination~\cite{Cutler:1994ys}, we do \emph{not} update the latter here as the inclination affects the wavelet model as well.

\end{itemize}

\subsubsection{Priors}
The priors on the binary intrinsic parameters depend on whether we consider BNSs or BBHs.
We sample the detector frame chirp mass $\Mc\in\left[0.23, 2.6\right] M_{\odot}$ and total mass $M \in\left[1.0, 6.0\right] M_{\odot}$ for BNSs and $\Mc\in\left[0.23 , 174 \right] M_{\odot}$, $M \in\left[1.0, 400.0 \right] M_{\odot}$ for BBHs and 
apply a Jacobian correction that makes the prior flat for the individual component masses $m_1$ and $m_2$~\cite{Callister:2021gxf}. We use flat priors for the dimensionless spin components $\chi_1, \chi_2\in\left[-0.05, 0.05\right]$ for BNSs and $\chi_1, \chi_2\in\left[-1,1\right]$ for BBHs. In the BNS case we sample the effective dimensionless tidal parameters uniform $\tilde{\Lambda}\in\left[0, 1000\right]$ and $\delta\tilde{\Lambda}\in\left[-500, 500\right]$ with the constraint that the corresponding individual component tidal deformabilities cannot be negative. 
We can optionally directly sample the individual component tidal parameters uniformly with the default prior range $\Lambda_{1,2}$ $\in$ $\left[0, 1000\right]$ though all results presented here use the former tidal prior. In all cases, the individual component tidal deformabilities are assumed to be independent of each other and not linked under any assumption about the NS EoS~\cite{Chatziioannou:2018vzf,Zhao:2018nyf,Carson:2019rjx}. 

We sample the merger time relative to the GPS trigger time ($\ttrig$) with a default window of $t_c$ $\in$ $\left[\ttrig - 0.5\text{s} , \ttrig + 1.5 \text{s}\right]$. Alternatively, it is possible to use a symmetric window of arbitrary size around the trigger time. We also sample in the logarithmic luminosity distance with an adjustment to the Jacobian such that the prior is uniform in volume (i.e., uniform in $~\text{D}_L^3$) with $\text{D}_L$ $\in$ $\left[1, 10000\right]\text{Mpc}$. Finally, the coalescence phase prior is uniform in $\phi_c \in \left[0, 2\pi \right]$.

\subsubsection{Data}

Since the CBC and the wavelet models are active simultaneously, the overall likelihood has to take into account both CBCs and wavelets. The blocked nature of the code means that the
wavelet model is not updated in the CBC sampler, so the wavelet-subtracted data $\cbc{d}\equiv d-\hw$ are constant as the CBC parameters are being updated. We compute $\cbc{d}=d-\hw$
once at the beginning of the CBC sampler block given the current wavelet parameters. The wavelet-subtracted data $\cbc{d}$
then play the role of the data in the sense that $\cbc{d}-\hcbc$ is the residual during the likelihood computation.

\subsubsection{Proposals}

We employ the CBC sampler described in detail in~\cite{Cornish:2021wxy} within the \BayesWave blocked Gibbs sampler, see Fig.~\ref{fig:workflow}. The sampler performs by default 300 iterations
that update the CBC parameters 
described above using a mixture of proposals: \\
\textit{Fisher jumps}: Gaussian jumps along the eigenvectors of the Fisher information matrix. The Fisher information matrix is recomputed at the start of each CBC sampler block for each parallel chain and using each chain's current location. \\
\textit{Differential evolution~\cite{2006S&C....16..239T}}: Two samples from each chain's history propose a new location.  The initial history array is drawn from the prior and then updated every 20 iterations. As each CBC sampler block does 300 iterations by default, the history array contains information about previous blocks. \\
%
\textit{Jiggle}: Small Gaussian jumps in individual parameter directions. The size of the proposed jump is scaled with the chain temperature such that high temperature chains have larger jumps and can explore a larger area of the parameter space. The jiggle proposal allows for exploring areas outside those covered by Fisher matrix jumps and are especially useful early on. \\
\textit{Mixed}: Uniform proposal for the tidal parameters and differential evolution (see above) for the other parameters. This proposal is employed $5\%$ of the time and aids convergence for cases where the tidal parameters cannot be tightly constrained. The proposal facilitates taking larger steps for the tidal parameters whereas the other better constrained parameters evolve with smaller steps, thus leading to a higher acceptance rate than a fully uniform proposal.
 
\subsubsection{Heterodyned likelihood}
\label{sec:cbcheterodyne}

We integrate the heterodyned likelihood technique~\cite{2010arXiv1007.4820C,Cornish:2021wxy} in the \BayesWave CBC (and extrinsic, to be described later) sampler in order to speed up the likelihood evaluation. The technique is based on the fact that all CBC waveforms that contribute to the posterior (i.e. have a likelihood high enough that the MCMC proposal can be accepted) are 
similar. The likelihood of a model $\hcbc$ can then be computed based on a fixed and precomputed reference model $\hcbc_{\mathrm{ref}}$ that is similar to $\hcbc$, and given the noise PSD and the data. 

We can rewrite the residuals $r$ using the reference heterodyne, omitting the detector subscript $I$ for clarity, as:
\begin{align}
	r = \cbc{d} - \hcbc = \cbc{d} - (\hcbc_{\mathrm{ref}} - \Delta \hcbc) = r_{\mathrm{ref}} + \Delta \hcbc,
\end{align} 
where $\Delta \hcbc$ is the difference between the reference waveform and the exact waveform at a given set of CBC parameters, $\cbc{d}$ is the  wavelet-subtracted data as given to the CBC sampler, $\cbc{d} = d - \hw$, and $r_{\mathrm{ref}}\equiv\cbc{d}-\hcbc_{\mathrm{ref}}$. The likelihood from Eq. \eqref{eq:logL} then becomes
\begin{align}
	\log{L} \sim -\frac{1}{2} (r_{\mathrm{ref}} + \Delta \hcbc | r_{\mathrm{ref}} + \Delta \hcbc),
\end{align}
where we omit the second term of Eq.~\eqref{eq:logL} as the PSD is held fixed during the CBC updates and the term does not affect the relative likelihoods. This form of the likelihood enables several 
computational advantages as we split the inner product in its components
\begin{equation}
	\begin{aligned}
	\label{eq:loghetparts}
	-2 \log{L} \sim &(r_{\mathrm{ref}} |r_{\mathrm{ref}}) + (\Delta \hcbc|\Delta \hcbc)  \\
	&+ \left[(r_{\mathrm{ref}}|\Delta \hcbc) + (\Delta \hcbc|r_{\mathrm{ref}}) \right].
	\end{aligned}
\end{equation}

A detailed description of how each components of the heterodyned likelihood above is calculated is available in~\cite{2010arXiv1007.4820C,Cornish:2021wxy}; we here
briefly discuss how this is adapted and implemented within the \BayesWave blocked Gibbs sampler. The first term of the inner product $(r_{\mathrm{ref}} |r_{\mathrm{ref}})$ is the most straightforward as it
is computed once at the beginning
of each CBC sampler block using the current wavelet-subtracted data and CBC parameters for each chain. 

The key performance improvement of the heterodyne likelihood is achieved by recognizing that the latter two terms in Eq.~\eqref{eq:loghetparts} can be further split up into the product of two terms:
a slowly varying term that is computed at each iteration, and a rapidly varying term that is computed once. Both terms are evaluated through 
a sum over frequencies with bin width $\Delta f$ using a Legendre polynomial expansion~\cite{Cornish:2021wxy}. The rapidly-varying term requires a small $\Delta f$ but this is computed once. 
The slowly-varying term that needs to be calculated on each likelihood evaluation can instead make use of a coarse frequency grid of $\Delta f \leq$~4 Hz thus significantly reducing computational cost.

The term $(\Delta \hcbc|\Delta \hcbc)$ in Eq.~\eqref{eq:loghetparts} introduces a slowly varying phase difference as a function of frequency between the reference waveform and the proposed waveform. 
However, in the noise weighted inner product, Eq.~\eqref{eq:innerproduct} the denominator might not be slowly varying if the PSD includes sharp spectral lines. Following the approach discussed above, we split the PSD in a smooth broadband component that is slowly varying as a function of frequency, and line features which are only evaluated at discrete intervals. 
If the PSD is held fixed throughout the analysis, the smooth part is computed once at the beginning. When using a variable PSD model instead, 
we update the smooth PSD component at the beginning of the CBC sampler block. 

The final term of Eq.~\eqref{eq:loghetparts} combining $r_{\mathrm{ref}}$ and $\Delta \hcbc$ is split up into the product of a slowly and rapidly varying part by introducing the smooth component of the PSD in the numerator and denominator of, respectively, the rapidly and slowly varying part. Due to this, the heterodyned difference in the waveforms $\Delta \hcbc$ is effectively whitened using only the smooth PSD component on a coarse frequency grid. The rapidly varying part consisting of the heterodyned whitened residuals $r_{\mathrm{ref}}$ now includes an extra product with the smooth part of the PSD, but since this is only computed at the start of the CBC sampler block, it adds little computational cost.

The heterodyned likelihood speeds up the likelihood calculation by a factor of $\sim T_{\mathrm{obs}} \Delta f$, where $T_{\mathrm{obs}}$ is the duration of the analyzed data, thus making it especially useful for long duration signals such as BNSs. The heterodyne implementation
in \BayesWave is more complex than the one in~\cite{Cornish:2021wxy} as it has to account for a potentially changing PSD and the wavelet model. We thus expect more modest overall computational improvements, though the speed-up of each individual likelihood calculation is consistent with~\cite{Cornish:2021wxy}.

\subsection{Wavelet parameters: Wavelet sampler}
\label{sec:intrinsicsampler}

\BayesWave has a common RJMCMC sampler for its \textit{signal} and \textit{glitch} models where the non-Gaussian features (either a GW signal or a detector glitch) are modeled as a sum of a variable number of sine-Gaussian wavelets which can be expressed in the time-domain as
\begin{align}
	\label{eq:wavelets}
	\psi(t;A, f_0, \tau, t_0, \phi_0) = Ae^{-(t-t_0)^2/\tau^2} \cos \left( 2\pi f_0 (t-t_0) + \phi_0 \right),
\end{align}
where $\tau \equiv Q/(2\pi f_0)$ is the decay time, $Q$ is the quality factor, $A$ is the wavelet amplitude, $f_0$ is the wavelet central frequency, $t_0$ is the wavelet central time and $\phi_0$ is the phase offset.
The glitch model uses independent sums of wavelets for the different detectors, whereas the signal models use a set of wavelets at geocenter and projects them onto detectors through Eq.~\eqref{eq-proj} with
\begin{align}
	\hw_{+}(f) &= \sum_{i=0}^{N_s} \psi (f;A_i,f_{0,i},Q_i,t_{0,i},\phi_{0,i}) \\
	\hw_{\times}(f) &= \epsilon \hw_{+}(f) e^{i \pi /2},\label{eq:polarization}
\end{align}
where $\epsilon$ is the ellipticity parameter and $N_s$ is the number of wavelets in the signal model. Details about the wavelet sampler and priors/proposals for each wavelet parameter and the number of wavelets are
discussed in~\cite{Cornish:2014kda,Cornish:2020dwh} and remain mostly unaltered in our analysis. The prior on the number of wavelets has an upper limit of $100$.
We extend the standard settings by adding an option to limit the central frequency range of the wavelets 
to $f_0\in\left[f^{\mathrm{w}}_{\mathrm{min}},f^{\mathrm{w}}_{\mathrm{max}}\right]$  which defaults to the analysis bandwidth. 
Since the ellipticity $\epsilon$ is related to the remaining extrinsic parameters and to the binary inclination, it is not sampled by the wavelet sampler but the extrinsic sampler discussed next.

In Fig.~\ref{fig:workflow} the wavelet sampler is depicted by the blue box termed ``wavelet RJMCMC" where $\theta^{\mathrm{w}}$ corresponds to the relevant parameters: the number of wavelets and the amplitude,
quality factor frequency, time, and phase of each wavelet.
Similarly to the CBC sampler, the wavelet sampler makes use of the CBC-subtracted data $d^{\mathrm{w}}\equiv d-\hcbc$, which are computed once at the beginning of each wavelet sampler block using the
current CBC sample.

\subsection{Common parameters: Extrinsic sampler}
\label{sec:extrinsicsampler}

Since both the CBC and wavelet models target the same GW source, they share a number of extrinsic source parameters that govern the signal projection onto the detector network, Eq.~\eqref{eq-proj}. These common 
parameters $\theta_{\mathrm{ext}}$ are updated by the extrinsic sampler block, denoted as ``extrinsic MCMC'' in Fig.~\ref{fig:workflow}.

\subsubsection{Parameters and priors}
\label{sec:extrinsicparams}

The extrinsic parameters shared between the wavelet and CBC models are $\theta_{\mathrm{ext}}\equiv(\alpha,\delta,\psi,\epsilon,\varphi_0,A^\mathrm{cbc})$.

\begin{itemize}
\item The source sky location is given by the right ascension $\alpha \in \left[0, 2\pi \right]$ and the declination $\sin{\delta} \in \left[0, 1\right]$, and the corresponding prior is flat on the sphere.
\item The polarization angle $\psi$ is sampled with a uniform prior $\psi \in \left[0, 2\pi\right]$. 
\item The degree of elliptical polarization in the signal is encoded in $\epsilon \in \left[-1, 1 \right]$, Eq.~\eqref{eq:polarization}. For elliptically polarized signals, the ellipticity parameter 
can be related to the orbital inclination with respect to the line of sight to the binary $\iota$ as
\begin{align}
	\cos{\iota} = \frac{1 - \sqrt{1 - \epsilon^2}}{\epsilon}.
\end{align}
We use a uniform prior in $\cos{\iota} \in \left[-1, 1\right]$, which is a deviation from previous works where \BayesWave samples uniformly in $\epsilon$ instead. 
\item An overall phase shift $\varphi_0$ is sampled uniformly as  $\varphi_0 \in \left[0, 2\pi \right]$ and it is equivalent to a phase shift in all wavelets simultaneously (see Eq.~\ref{eq:wavelets}). Despite the degeneracy, introducing this parameter leads to more efficient convergence in the wavelet model.  
\item We also introduce an overall amplitude scaling parameter $A^\mathrm{cbc}$, which is sampled uniformly and is degenerate with the CBC luminosity distance $D_L$. We restrict $A^\mathrm{cbc}$ to correspond to values for $D_L$ within the $D_L$ prior range and add a Jacobian factor such that the prior remains uniform in volume, i.e., $(D_L/A^\mathrm{cbc})^3$.
\end{itemize}
 
The extrinsic parameters can be efficiently sampled independently from the CBC intrinsic parameters by noting that a change in the extrinsic parameters induces an amplitude, time, and phase shift in the signal 
seen in each detector, while leaving its phase evolution unaltered. We therefore compute the geocenter CBC waveform given the fixed intrinsic parameters once and then apply the shifts accordingly to evaluate the waveform at each new proposed set of extrinsic parameters. This has the advantage that we can sample the extrinsic parameters efficiently while not having to recompute the full, expensive CBC waveform phase at each iteration. 

In practice we write the CBC response at a given detector $I$ using the geocenter waveform $\hcbc_{+}$ and applying the shifts as
\begin{align}
	\hcbc_I(f) = \hcbc_{+}(f) F^\mathrm{cbc}_I e^{i\lambda_I} e^{2 \pi i f \Delta t_I},
\end{align}
where $\Delta t_I$ is the arrival time at the detector relative to the geocenter, the phase shift is
\begin{align}
	\lambda_I \equiv \arctan[\epsilon F^{\times}(\alpha,\delta,\psi)/F^+(\alpha,\delta,\psi)] + \phi_c,
\end{align}
 and $F^\mathrm{cbc}_I$ is the magnitude of the detector response 
\begin{align}
	F^\mathrm{cbc}_I \equiv A^\mathrm{cbc} \sqrt{\left[\frac{\cos{\iota}}{\epsilon} F^+(\alpha,\delta,\psi) \right]^2 + \left[\cos{\iota}\,F^{\times}(\alpha,\delta,\psi) \right]^2},\label{eq:fmagcbc}
\end{align}
Even though $A^\mathrm{cbc}$ is completely degenerate with $D_L$, we find that it increases sampling efficiency in the extrinsic sampler due to the distance-inclination degeneracy~\cite{Cutler:1994ys}.

At the end of the block of extrinsic updates, the scaling and shift parameters are reset and propagated to the corresponding CBC parameters before proceeding to the CBC sampler 
\begin{align}
	D_{L} &\rightarrow \frac{D_{L}}{A^\mathrm{cbc}} \\
	t_{c} &\rightarrow t_{c} + \Delta t_{a} \\
	\phi_{c} &\rightarrow \phi_{c} - \Delta \varphi_0
\end{align}
In these equations $\Delta t_{a}$ and $\Delta \varphi_0$ correspond to the overall time and phase shift induced on the waveform by the change in extrinsic parameters. The time of arrival of the merger at geocenter $t_c$ changes due to the new sky location and this shift is encoded in $\Delta t_{a}$.

\subsubsection{Proposals}
\label{sec:extrinsicproposals}

To update the parameters $\theta_{\mathrm{ext}}$ the extrinsic sampler uses a mixture of proposals, which are described in detail in \cite{Cornish:2014kda,Cornish:2020dwh}. In the following we focus on some 
CBC-specific updates made to the existing proposals.

The most commonly used proposal is the Fisher proposal, where jumps are proposed along the eigenvectors of the Fisher information matrix. The matrix elements, eigenvalues, and eigenvectors are computed once at the start of the extrinsic sampler block by finite differencing using the current chain position. We extend the existing \BayesWave Fisher calculation by adding the CBC model such that the final Fisher matrix captures 
information from the full CBC+wavelet signal. When the sky location is known and fixed (for example for sources with an electromagnetic counterpart), the Fisher matrix is reduced to the parameters that are varied.

The \BayesWave extrinsic sampler also employs a two-part proposal that exploits a degeneracy of the extrinsic parameters when multiple detectors are used (also proposed by \cite{Raymond:2014uha}). The first part consists of a sky-ring proposal that finds a new sky location ($\alpha$, $\sin{\delta}$) such that the time delays between the detectors are preserved \cite{Veitch:2014wba,Cornish:2014kda}. The second part of the proposal updates the remaining parameters ($\psi$, $\cos{\iota}$, $A^\mathrm{CBC}$ and $\varphi_0$) either uniformly in their prior range or deterministically such that the waveform is identical at the new sky location compared to the waveform at the previous sky location, see \cite{Cornish:2021wxy} for details. Since the waveform is conserved, so is
the likelihood, though this does not guarantee that the proposed sky location will be accepted due the necessary Jacobian factor~\cite{Cornish:2021wxy}.

Finally, the \BayesWave extrinsic sampler includes a uniform proposal that draws from the prior of each parameter. These jumps are particularly useful for higher temperature chains to explore the entire space available  
as well as early on in the sampling if a signal has not been found. 

\subsubsection{Heterodyned likelihood}

A change in the extrinsic parameters $\theta_{\mathrm{ext}}$ will affect both the CBC and the wavelet models, therefore both need to be taken into account when computing the
likelihood. However, we find that the CBC part of the model dominates the computational cost so we again use the heterodyned likelihood to speed up the calculation. 
In this case, there are some differences compared to the implementation for the CBC sampler as discussed in Sec.~\ref{sec:cbcheterodyne} because we can no longer 
assume that the wavelet-subtracted data are fixed. Indeed, a new set of extrinsic parameters changes the full signal projection and the full residual $r = d - \hw - \hcbc$ has to be recomputed on each iteration.  

Despite this, we can achieve a computational improvement by splitting up the heterodyne computation even further to avoid recomputing all components of the heterodyne when the residuals change, i.e., at every iteration. All subcomponents of the heterodyne that do not involve the residuals are computed and stored once at the start of each extrinsic MCMC sampler block, whereas the remaining components are computed at each iteration. If the signal wavelet model is not employed (i.e. an analysis with only the CBC model), 
the heterodyne likelihood implementation falls back to the same, more efficient, version as for the CBC sampler.

Since the likelihood calculation is more involved and parts of the heterodyne procedure need to be recomputed in each iteration, we expect more modest computational
improvements compared to those of Sec.~\ref{sec:cbcheterodyne}.

\subsection{Noise parameters: PSD sampler with \code{BayesLine}}
\label{sec:psdsampler}

The PSD of the Gaussian noise is modeled as a combination of splines and Lorenzians targeting the broadband behavior and spectral lines respectively 
and marginalized over through an RJMCMC in \code{BayesLine}~\cite{Littenberg:2014oda,Chatziioannou:2019,Cornish:2020dwh}.
The analysis presented here 
makes no updates on this noise marginalization procedure and
can optionally include \code{BayesLine} as one of the blocks, corresponding to the ``Noise RJMCMC'' gray box in Fig.~\ref{fig:workflow}. Alternatively, the 
analysis can skip the noise marginalization and use a predetermined fixed PSD.
If the PSD is modeled by \code{BayesLine}, the heterodyne procedure needs to be repeated at the end of the \code{BayesLine} sampling block and before moving to CBC sampling with the latest PSD sample.
However, this coincides with a recomputing of the heterodyne elements due to a change of the wavelet model that is subtracted from the data 
so no further computations are necessary. 
In the current implementation~\cite{Littenberg:2014oda}, the wavelet model is subtracted from the data before entering the \code{BayesLine} sampling block and we extend this subtraction to include the CBC model.

\section{Toy model: GW150914}
\label{sec:150914}

To illustrate the wavelet+CBC model, we begin with a toy model by analyzing the first detected BBH merger, GW150914 ~\cite{Abbott:2016blz}. In reality, the full BBH signal can be modeled with existing CBC templates, something
 confirmed in the case of GW150914 through comparisons between different models and physical effects~\cite{Abbott:2016wiq}. In our case, and in order to test our hybrid analysis on what is probably the best
studied signal, we pretend we lacked efficient modeling of a BBH merger and waveform templates terminated shortly before merger. 

We analyze 4s of data from LIGO Hanford and LIGO Livingston available through the Gravitational Wave Open Science Center (GWOSC)~\cite{GWOSC,Abbott:2019ebz} 
in the frequency band from 16Hz to 2048Hz. The noise PSD is
marginalized over as discussed in Sec.~\ref{sec:psdsampler}. For the CBC model, we use either the PhenomD waveform or the PhenomDNRT waveform. 
PhenomD is appropriate for BBH systems as it models the full IMR signal.
On the other hand, we do not expect PhenomDNRT to accurately model the BBH data as it lacks an accurate merger and ringdown portion due to tapering~\cite{Dietrich:2018uni}.
Regardless, we use it to demonstrate how any leftover signal that is not captured by the CBC waveform can be captured by the wavelets. 

\begin{table*}
\centering
\begin{tabular}{c|c|c|c|c|c} 
	Label & GPS time (s)  
	& Model & Waveform &\begin{tabular}[c]{@{}c@{}}Wavelet \\ window (s) \end{tabular}&\begin{tabular}[c]{@{}c@{}}Wavelet \\ bandwidth (Hz) \end{tabular}\\ 
	\hline
	\rule{0pt}{10pt}CBC$_{\mathrm{IMR}}$ &  1126259462.391   & CBC/noise & PhenomD & N/A & N/A\\
	\rule{0pt}{10pt}CBC$_{\mathrm{Insp}}$&  1126259462.391   & CBC/noise & PhenomDNRT & N/A & N/A\\
	\rule{0pt}{10pt}CBC$_{\mathrm{Insp}}$+wavelet& 1126259462.416  & CBC/wavelet/noise & PhenomDNRT & (-0.025, 0.025) & (150, 2048) \\
	\hline
\end{tabular}
\caption{
Settings for the runs of Sec.~\ref{sec:150914} on GW150914. All runs use a segment length of 4s, a low frequency cut off of 16Hz, and a sampling rate of 4096Hz. From left to right, columns correspond to the run label in Fig.~\ref{fig:GW150914wf_multi}, the GPS time ($\ttrig$), the models active, the CBC waveform, the wavelet window around the trigger time, and the wavelet bandwidth. 
}
\label{tab:150914settings}
\end{table*}

\begin{figure*}
	\centering
	\includegraphics[width=0.49\textwidth]{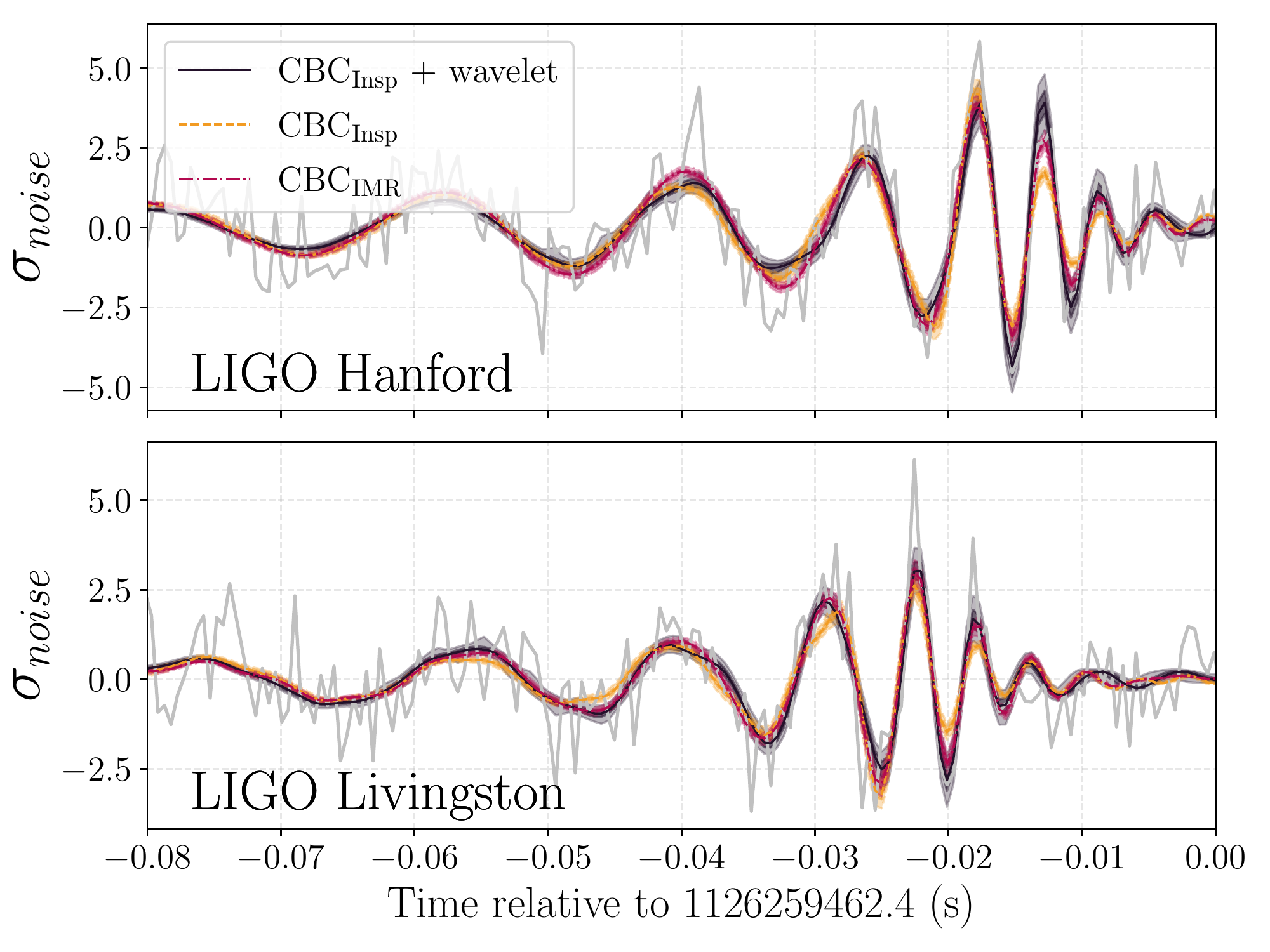}
	\includegraphics[width=0.49\textwidth]{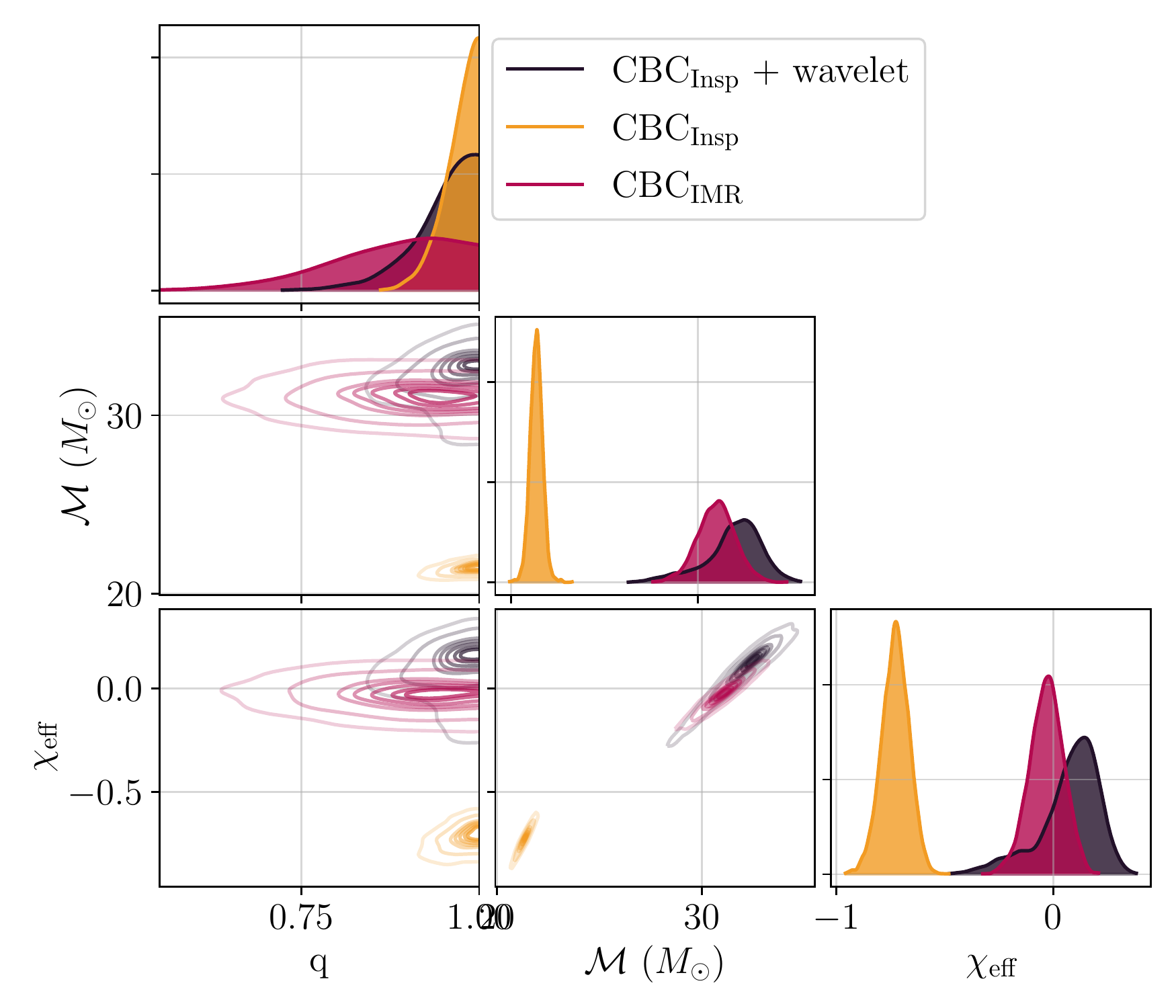}
	\caption{Results on the GW150914 data with different waveforms and analyses. The left panel presents 90\% credible intervals for the whitened signal reconstructions in each detector while the right panel shows one- and two-dimensional 
	marginalized posterior distributions for selected CBC parameters: mass ratio $q$, detector-frame chirp mass ${\cal{M}}$, and the effective spin $\chi_\text{eff}$.
	In both panels red corresponds to an analysis with the CBC model and PhenomD (CBC$_{\mathrm{IMR}}$), yellow corresponds again to a CBC model analysis but with PhenomDNRT (CBC$_{\mathrm{Insp}}$), and black corresponds to an analysis with the CBC+wavelet model with PhenomDNRT (CBC$_{\mathrm{Insp}}$+wavelet). 
	In the left panel the gray dashed line indicates the data whitened with a fair draw PSD from the noise model posterior.
	}
	\label{fig:GW150914wf_multi}
\end{figure*}

We perform three analyses with different models and CBC templates:
\begin{enumerate}
\item  A full ``CBC$_{\mathrm{IMR}}$" analysis where the data are described only by the CBC model with PhenomD. 
\item A ``CBC$_{\mathrm{Insp}}$" analysis where the data are again described only by the CBC model but with PhenomDNRT.
\item A ``CBC$_{\mathrm{Insp}}$+wavelets" analysis where the data are described by a combination of the CBC model with PhenomDNRT and the wavelet model. We restrict the 
wavelet model to times $t_0$ $\in$ $\left[\ttrig - 0.025\text{s} , \ttrig + 0.025 \text{s}\right]$ and frequencies above 150Hz, targeting the merger portion of the signal that is missed by PhenomDNRT. 
\end{enumerate}
The run labels and relevant settings are given in Table~\ref{tab:150914settings}.

In Fig.~\ref{fig:GW150914wf_multi} we show the signal reconstructions and parameter posteriors for these three analyses. The left panel shows the whitened data and the 90$\%$ credible intervals for each reconstruction in LIGO Hanford (top) and LIGO Livingston (bottom). All analyses are able to identify the GW signal and they further lead to 
consistent signal reconstructions of the early portion of the signal. However, as the signal approaches the merger phase, the CBC$_{\mathrm{Insp}}$ analysis deviates from the reference CBC$_{\mathrm{IMR}}$ analysis as well as the hybrid CBC$_{\mathrm{Insp}}$+wavelets analysis by underpredicting the strength of the signal. The hybrid CBC$_{\mathrm{Insp}}$+wavelets agrees well with the full CBC$_{\mathrm{IMR}}$, though the uncertainty of the former is larger in the merger phase where the signal is 
no longer modeled with CBC templates but with wavelets.

The recovered source parameters are given in the right panel of Fig.~\ref{fig:GW150914wf_multi}. The CBC$_{\mathrm{IMR}}$ analysis leads to consistent results with those reported in previous studies~\cite{LIGOScientific:2016vlm,LIGOScientific:2016dsl,LIGOScientific:2018mvr}, while the CBC$_{\mathrm{Insp}}$ analysis shows a significant bias away from the expected posteriors. The bias is most evident in the chirp mass: PhenomDNRT compensates for the lack of a merger and ringdown by decreasing the chirp mass, thus leading to a longer inspiral phase in an attempt to capture part of the missing merger cycles. Despite this bias, the CBC$_{\mathrm{Insp}}$ analysis is still not able to capture the full signal as seen on the left panel. 
The CBC$_{\mathrm{Insp}}$+wavelets proof-of-concept model results in recovered parameter posteriors that are consistent with the full CBC$_{\mathrm{IMR}}$ analysis. The posteriors are not 
expected to be identical as now the CBC model has access only to the inspiral portion of the signal and therefore lacks information available to the 
CBC$_{\mathrm{IMR}}$ analysis. However, the wavelet model can efficiently capture the missing portion of the signal, allowing the CBC model to recover unbiased system parameters from the inspiral
phase only.

\begin{figure}
	\centering
	\includegraphics[width=0.49\textwidth]{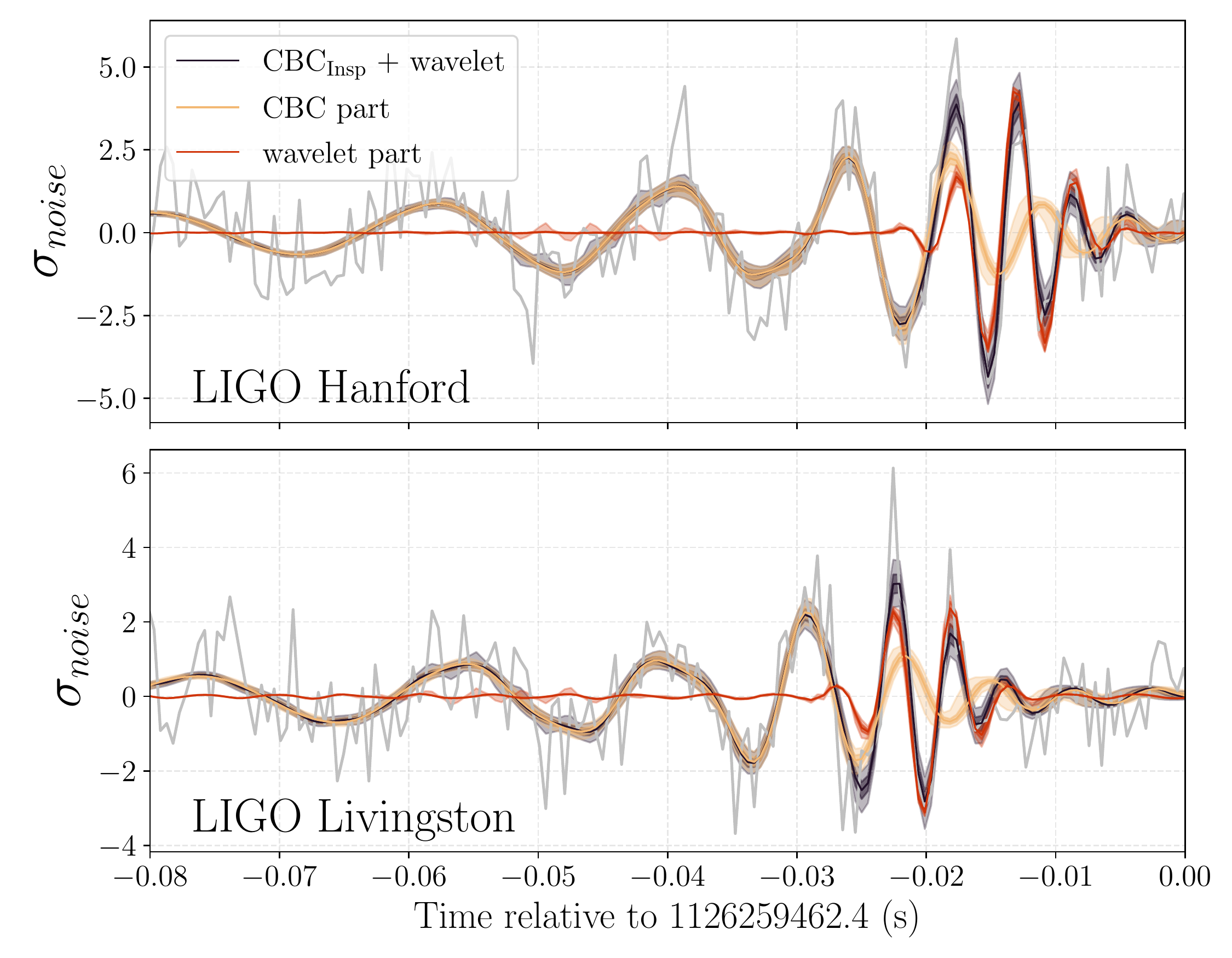}
	\caption{Breakdown of the various components of the hybrid CBC$_{\mathrm{Insp}}$+wavelets analysis of GW150914. We present 90\% credible intervals for the whitened signal reconstruction of the CBC model (yellow), the wavelet model (red), and the sum of the models (black). The sum is identical to the black lines from Fig.~\ref{fig:GW150914wf_multi}. The gray dashed line indicate the data whitened with a fair draw PSD from the noise model posterior.}
	\label{fig:GW150914CBCwavelet}
\end{figure}

Figure~\ref{fig:GW150914CBCwavelet} examines the CBC$_{\mathrm{Insp}}$+wavelets analysis in more detail and the complementary roles of the components in the hybrid model. We plot the signal reconstruction from each submodel as well as their sum. The sum of the individual model components is the same as the CBC$_{\mathrm{Insp}}$+wavelets model in Fig.~\ref{fig:GW150914wf_multi}. Figure~\ref{fig:GW150914CBCwavelet} shows that as expected the BBH inspiral phase is primarily reconstructed by the CBC PhenomDNRT model. As we move from the inspiral toward the merger time, the CBC model starts tapering off and deviating from the full GW signal. At the same time,  the wavelet model captures the part of data that is not covered by the CBC model, such that the combination of CBC and wavelets accurately models the full GW signal. This toy model for GW150914 demonstrates the main concept of our hybrid analysis: the full GW signal is modeled with a well-understood component, i.e. the CBC waveform, and a model-agnostic component, i.e. the wavelet model, that covers any parts of the signal that are not included in the CBC model.

\section{Constraints on the GW170817 postmerger}
\label{sec:170817}

We turn our attention to GW170817, the first GW detection of a BNS coalescence~\cite{TheLIGOScientific:2017qsa,Abbott:2018wiz,Abbott:2018exr}. Both the low-frequency inspiral phase and a potential short-duration high-frequency postmerger phase of the signal have been studied separately in detail, with the latter remaining undetected~\cite{Abbott:2017dke,Abbott:2018wiz}. Here we apply our hybrid CBC+wavelet model to study both parts of the GW170817 signal simultaneously and compare 
our results with separate analyses of the inspiral and postmerger signals. 

\begin{table*}
\centering
\begin{tabular}{c|c|c|c|c|c|c|c|c} 
Label  &\begin{tabular}[c]{@{}c@{}}Segment \\length (s) \end{tabular} &  \begin{tabular}[c]{@{}c@{}}Sampling \\ rate (Hz)\end{tabular}&  $f_{\textrm{low}}$(Hz) 
& Model & Waveform &\begin{tabular}[c]{@{}c@{}}Wavelet \\ window (s) \end{tabular}&\begin{tabular}[c]{@{}c@{}}Wavelet \\ bandwidth (Hz) \end{tabular} & $D_{\mathrm{min}}$ \\ 
\hline
\rule{0pt}{10pt}CBC$_{\mathrm{Insp}}$  & 64 &  4096 &  32 & CBC & PhenomDNRT & N/A & N/A & N/A \\
\rule{0pt}{10pt}CBC$_{\mathrm{Insp}}$+wavelet  & 64 & 8196  & 32 & CBC/wavelet & PhenomDNRT & (-0.125, 0.125) & (1024, 4096) & 2\\
\rule{0pt}{10pt}wavelet-only  & 64  & 8196  & 1024 & wavelet & N/A & (-0.125, 0.125) & (1024, 4096) & 2\\
\hline
\end{tabular}
\caption{
Settings for the runs of Sec.~\ref{sec:170817} on GW170817. All runs are relative to GPS time 1187008882.446. From left to right, columns correspond to the run label in Fig.~\ref{fig:GW170817_multi} and Fig.~\ref{fig:GW170817_spectrum}, the segment length, the sampling rate, 
the low frequency cut off, the models active, the CBC waveform, the wavelet window around the trigger time, the wavelet bandwidth, and the minimum number of wavelets. 
}
\label{tab:170817settings}
\end{table*}

We analyze 64s of data from LIGO Hanford and LIGO Livingston
around GPS time 1187008882.446~\cite{GWOSC,Abbott:2019ebz} where the prominent glitch in LIGO Livingston has already been subtracted \cite{GW170817Data,Pankow:2018qpo}. While Virgo data are available for that time and aided in constraining the source sky location~\cite{Abbott:2018wiz}, we do not consider them due to the lower sensitivity and the fact that we fix the sky location to the known values of $\alpha = 3.446$ rad and $\delta = -0.408$ rad \cite{2017ApJ...848L..16S,Abbott_2017}. We also use a fixed PSD rather than marginalize over the noise model for computational efficiency.
We perform three analyses with different models and data: 

\begin{enumerate}
\item  A ``CBC$_{\mathrm{Insp}}$" analysis with data in the frequency range (32,2048)Hz that are described only by the CBC model with the PhenomDRT waveform thus focusing on the inspiral signal. 
\item A ``CBC$_{\mathrm{Insp}}$+wavelet" analysis with data in the wider frequency range (32,4096)Hz that are described by a combination of the CBC model with PhenomDNRT and the wavelet model, thus targeting the full signal. The wavelets are restricted to (1024,4096)Hz targeting a potential high frequency postmerger signal.
\item A ``wavelet-only" analysis where we restrict the frequency range to (1024Hz,4096)Hz and use only the wavelet model thus focusing only on the postmerger signal.
\end{enumerate}
For both analyses with the wavelets model, we use a prior on the number of wavelets of $D \in [2, 100]$. The reason for selecting a minimum number of $2$ wavelets is different
for each analysis. In the case of the hybrid analysis, the CBC part of the model does not terminate at merger, but smoothly tapers into the postmerger phase for a few milliseconds, see
Fig~\ref{fig:wfoverlap}. We therefore need at least two wavelets, one to undo this effect of the CBC model and the other to capture the true postmerger signal. In the wavelet-only study we again need at least two wavelets such that one wavelet can capture the merger itself which extends into the analysis bandwidth and the other wavelet can capture the contribution from the postmerger signal. The full settings of our three analyses are detailed in Table~\ref{tab:170817settings}.

\begin{figure*}
	\centering
	\includegraphics[width=0.49\textwidth]{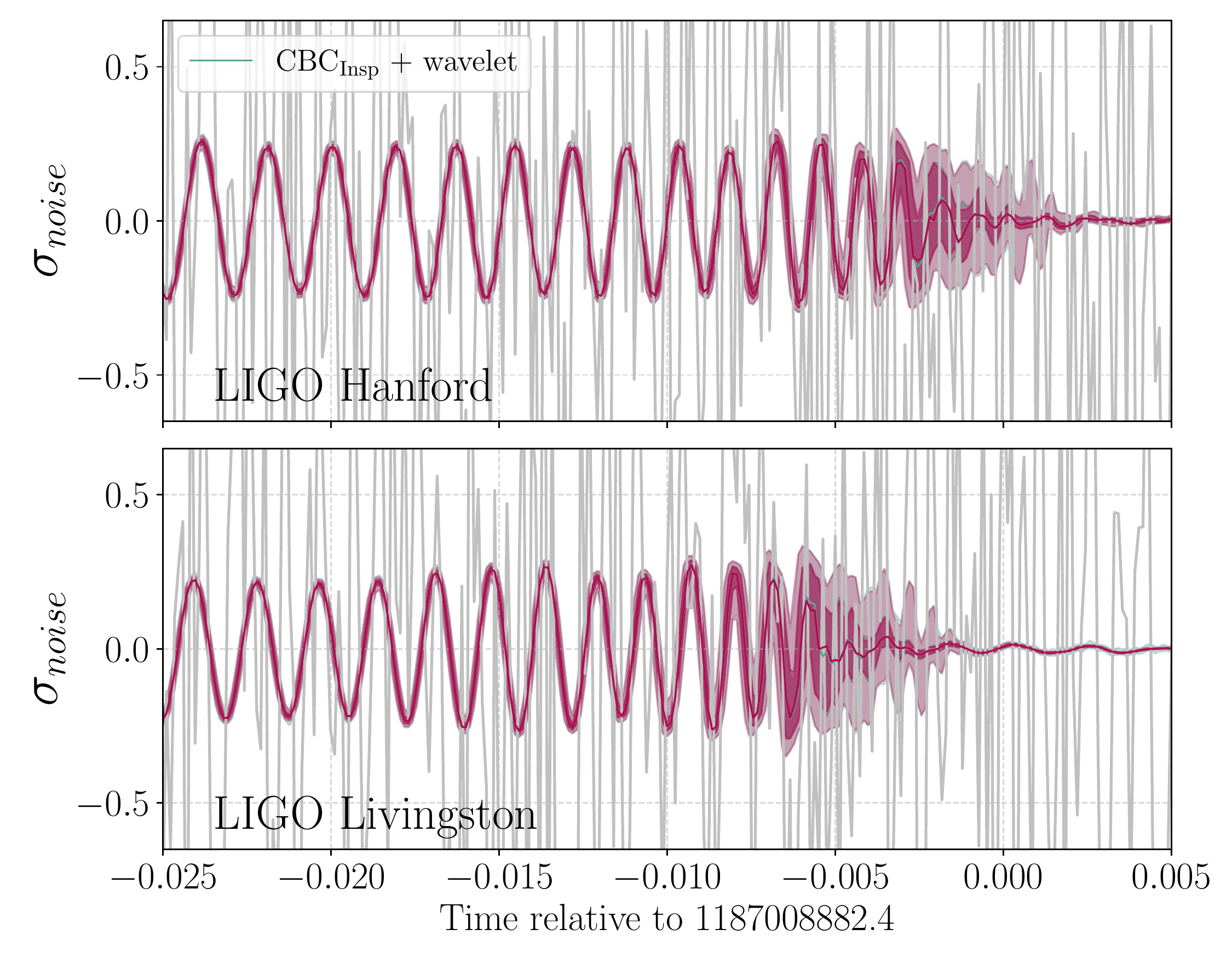}
	\includegraphics[width=0.49\textwidth]{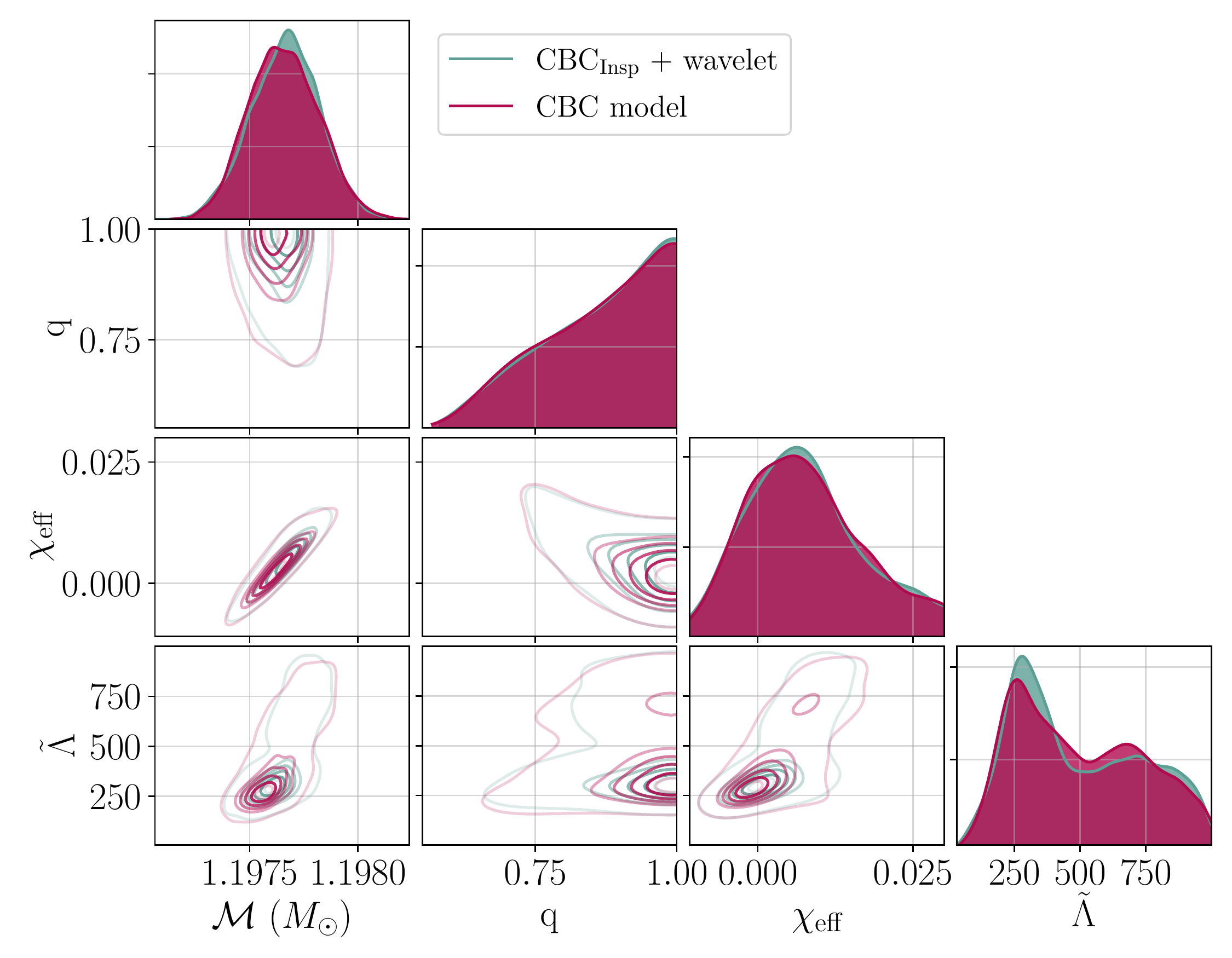}
	\caption{Results on the GW170817 data with different analyses. The left panel presents the median, 50\% and 90\% credible intervals for the whitened signal reconstructions in each detector while the right panel shows one- and two-dimensional 
		marginalized posterior distributions for selected CBC parameters: mass ratio $q$, detector frame chirp mass ${\cal{M}}$, the effective spin $\chi_\text{eff}$ and the tidal deformation parameter $\tilde{\Lambda}$.
		In both panels blue corresponds to the hybrid CBC$_{\mathrm{Insp}}$+wavelet and magenta corresponds to the CBC$_{\mathrm{Insp}}$ analysis. 
		In the left panel the gray dashed gives the whitened data. }
	\label{fig:GW170817_multi}
\end{figure*}

We compare the two analyses that include the inspiral of the signal in Fig.~\ref{fig:GW170817_multi} which shows  the signal reconstructions focused on the end of the signal (left panel) and the marginalized posterior distribution for selected source parameters (right) of GW170817. Despite the high SNR of the signal, the amplitude is relatively weak compared to the noise level, however both analyses identify the signal and lead to consistent 
reconstructions for the late inspiral phase as shown on the left panel. 
As the binary approaches merger the reconstruction uncertainty increases due to the decreasing detector sensitivity at 
increasing frequencies. This behavior is evident in both analyses, suggesting that the GW170817 postmerger signal remains undetected as expected from 
the detector sensitivity at the time of GW170817~\cite{Abbott:2018wiz}.

The right panel of Fig.~\ref{fig:GW170817_multi} examines selected CBC parameters recovered from the inspiral signal. 
The recovered parameters are consistent with previous results using the same CBC model~\cite{Abbott:2018wiz}. The only difference is that our $\tilde{\Lambda}$ posterior has more support for the higher of the two modes present in the results of~\cite{Abbott:2018wiz}. This can be attributed to different uses of prior: we use of a prior that is flat in $\tilde{\Lambda}-\delta \tilde{\Lambda}$ with the constraint that $\Lambda_1>0$ and $\Lambda_2>0$. This choice results in a marginalized prior for $\tilde{\Lambda}$ that favors larger values. In contrast,~\cite{Abbott:2018wiz} uses a flat marginalized prior for $\tilde{\Lambda}$. We have verified that reweighting our posterior to a flat marginalized prior gives consistent results with~\cite{Abbott:2018wiz}.
Furthermore, the hybrid CBC$_{\mathrm{Insp}}$+wavelet 
analysis leads to essentially identical results for the CBC parameters as the traditional CBC$_{\mathrm{Insp}}$ analysis. 
This is consistent with~\cite{Dudi:2018jzn} that
finds that failure to account for the postmerger signal will not lead to biases in the CBC parameters for typical detector sensitivities, but also serves as a sanity check of the 
CBC$_{\mathrm{Insp}}$+wavelet analysis and the fact that sampling for the joint CBC+wavelet model has converged. 

\begin{figure}
	\centering
	\includegraphics[width=0.5\textwidth]{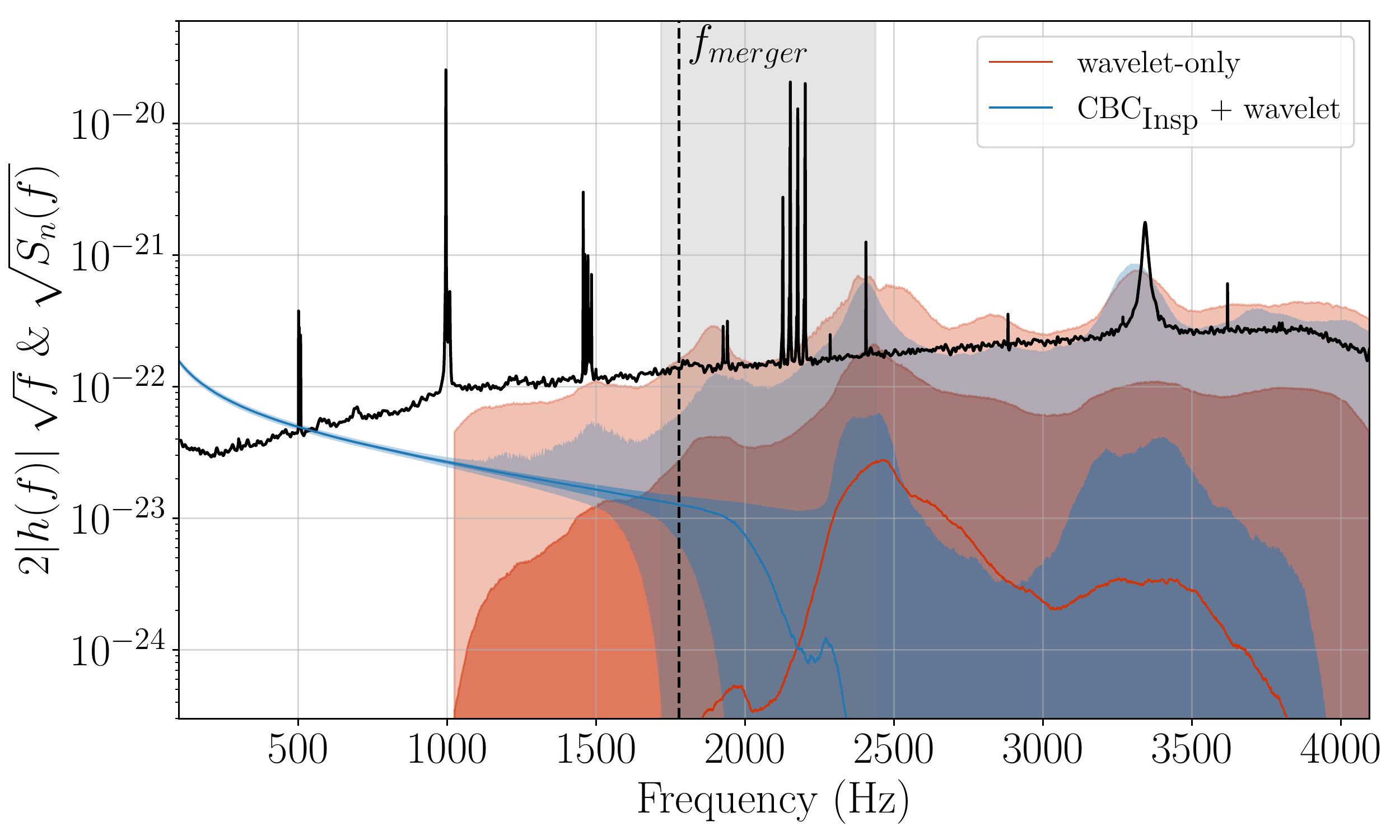}
	\caption{The median, 50$\%$ and 90$\%$ credible intervals of reconstructed GW spectrum in the LIGO Hanford detector in the CBC$_{\mathrm{Insp}}$+wavelet (blue) and the wavelet-only analysis (red). The noise amplitude spectral density is also overplotted (black dashed line). The gray band indicates the 90\% credible interval of the merger frequency, with the dashed line indicating the peak of the merger frequency posterior at $f_\text{merger}$~$\sim$~1778Hz.}
	\label{fig:GW170817_spectrum}
\end{figure}

After confirming that the CBC$_{\mathrm{Insp}}$+wavelet analysis matches the CBC$_{\mathrm{Insp}}$ analysis as far as the inspiral portion of the signal is concerned, we
switch to comparing the two analyses that include a wavelet model for a possible high-frequency postmerger signal. In Fig.~\ref{fig:GW170817_spectrum} we show the 
 reconstructed GW spectrum for CBC$_{\mathrm{Insp}}$+wavelet and wavelet-only. The vertical band gives the 90\% credible interval for the merger frequency estimated
 using Eq.~11 in \cite{2019PhRvD..99b4029D} given the binary component masses and tidal parameters from the CBC model posteriors. This merger frequency is also the location at which the PhenomDNRT template starts tapering off, so it is related to the transition between the CBC and the wavelet models.

At frequencies below $\sim 1000$Hz the CBC$_{\mathrm{Insp}}$+wavelet reconstruction is dominated by the inspiral signal which is clearly detected as the spectrum lower 
limit is nonzero. With increasing
frequency the detector sensitivity decreases, resulting in larger uncertainties in the reconstructed spectrum. Finally, at around the merger frequency, $\gtrsim1600$Hz, the
reconstruction uncertainty is large and consistent with no detected signal. A similar picture is drawn by the wavelet-only analysis which again places only upper limits
on the signal in frequencies above $1024$Hz. The upper limit from the CBC$_{\mathrm{Insp}}$+wavelet and the wavelet-only analyses is comparable, though the former is consistently lower across the frequency band, indicating more stringent constraints on the presence of a postmerger signal.
Additionally, the wavelet-only analysis does not lead to a detection of the merger signal in the frequency range (1000,1500)Hz, 
unlike the CBC$_{\mathrm{Insp}}$+wavelet case. This suggests that this high-frequency portion of the signal is not individually detectable, but only inferred coherently from the
preceding inspiral signal.

\section{Simulated signals}
\label{sec:injections}

Going beyond the GW170817 upper limits, in this section we study simulated BNSs where the postmerger emission is detectable by future GW detectors. We assume a network of two detectors at the current location of LIGO Hanford and LIGO Livingston and a zero noise realization. 
Since all signals have high SNRs we assume that an electromagnetic counterpart has been identified and the sky location of the source is known. Following~\cite{Torres-Rivas:2018svp} we 
do not explicitly select any planned GW detector such as Cosmic Explorer~\cite{Evans:2016mbw,Reitze:2019dyk,Reitze:2019iox} or the Einstein Telescope~\cite{2011CQGra..28i4013H,2010CQGra..27h4007P} and their nominal sensitivity. We instead work with the 
Advanced LIGO design sensitivity~\cite{aLIGO_design_updated} and gradually lower the noise PSD, emulating improving detector sensitivity and higher signal SNR. 

Since no NR simulation of the full BNS signal as observed by ground-based detectors exists, all simulated signals are constructed in a hybrid fashion: the postmerger
signal is obtained through NR simulations, while the premerger signal is computed with PhenomDRT using the same parameters as the NR simulation.
We choose extrinsic parameters consistent with GW170817, namely distance $D_L=40$Mpc, inclination $\iota=2.635$, polarization angle $\psi=0$, and the known sky location.
 All signals
are analyzed with the full CBC+wavelets model with run settings equivalent to the CBC$_{\mathrm{Insp}}$+wavelet analysis from Table~\ref{tab:170817settings}.

\subsection{Waveform from Kastaun and Ohme}
\label{sec:KO}

The first full BNS waveform we consider was constructed and released by~\cite{Kastaun:2021zyo}. This hybrid waveform was constructed by combining the PhenomDNRT model for the inspiral and the results of NR simulations for the postmerger GW signal. The hydrodynamical simulations assume a GW170817-like initial system with the hadronic SFHO EoS \cite{2010NuPhA.837..210H}, resulting in a short-lived hypermassive NS. We use the resulting hybrid waveform for a BNS system with mass ratio $q=0.9$, detector frame component masses of $m_1$ = 1.438$M_{\odot}$ and $m_2$ = 1.294$M_{\odot}$, and dimensionless tidal parameters $\Lambda_1$ = 280 and $\Lambda_2$ = 551, giving a binary tidal parameter $\tilde{\Lambda}$ = 396 and merger frequency $f_\mathrm{merger} = 2110$~Hz. 

 We inject the signal in a detector network with sensitivities 2xDS, 4xDS, 6xDS which respectively denote 2, 4, and 6 time improved strain sensitivity compared to the LIGO design sensitivity (i.e., the design sensitivity divided by 2, 4 and 6). In terms of amplitude spectral density, they correspond to 8.23$\times 10^{-24}~$Hz$^{-1/2}$, 4.12$\times 10^{-24}~$Hz$^{-1/2}$, and 2.72$\times 10^{-24}~$Hz$^{-1/2}$ at 3326~Hz, which is the frequency of the main postmerger mode of the signal, $\fpeak$. The resulting network SNRs are 142, 384, and 425 for the premerger signal ($f < f_\mathrm{merger}$), and 2.8, 5.6, and 8.4 for the postmerger signal ($f \geq f_\mathrm{merger}$).

\begin{figure}
	\centering
	\includegraphics[width=0.49\textwidth]{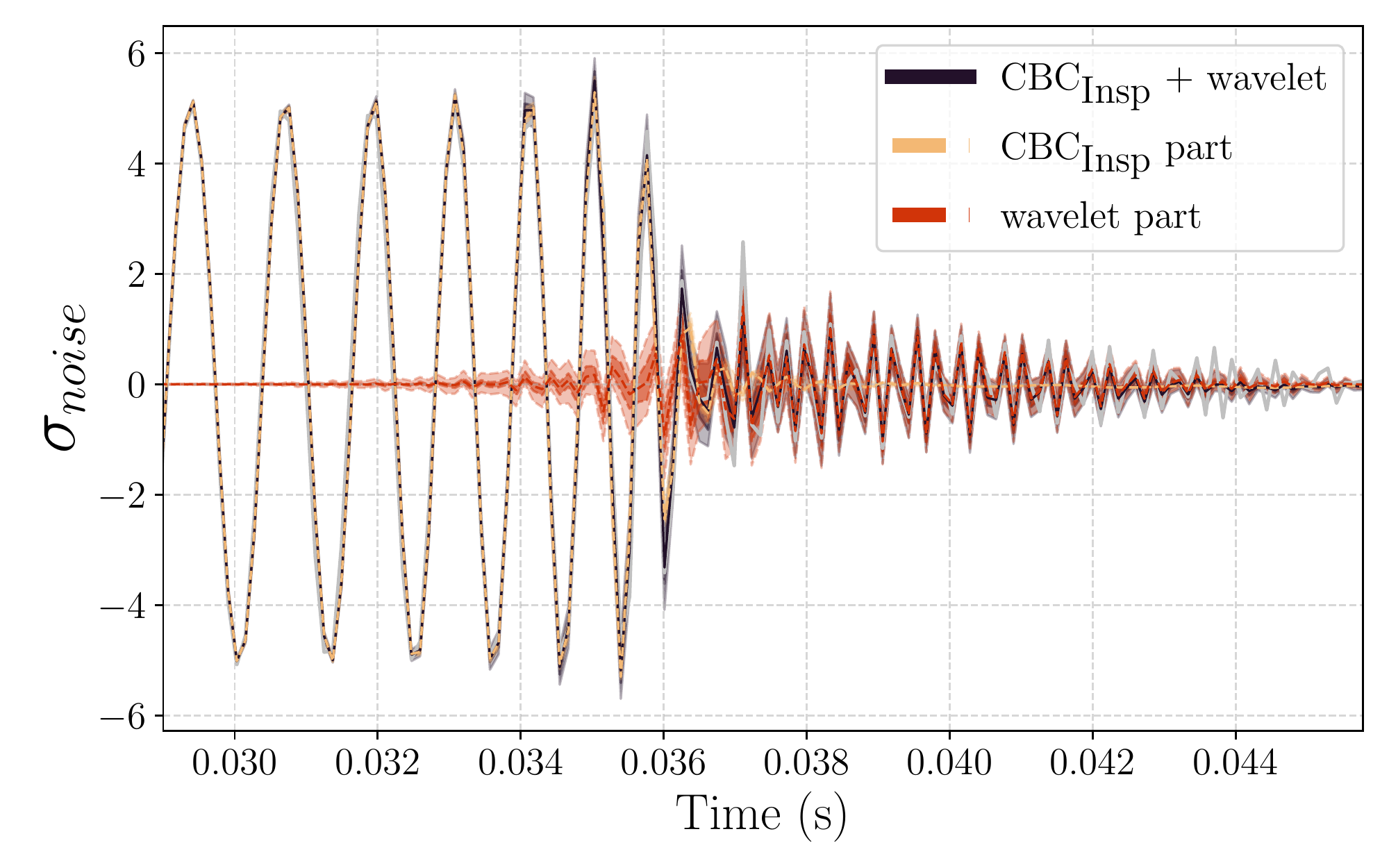}
	\caption{Whitened-time domain reconstruction of the simulated signal from~\cite{Kastaun:2021zyo} at 6xDS. The gray dashed line gives the simulated data. The shaded regions give the 50\% and 90\% credible intervals for the full CBC+wavelets reconstruction (black), as well as the CBC (yellow) and the wavelets (red) component of the full analysis. The postmerger signal is reconstructed by the wavelets model.}
	\label{fig:KO_rec}
\end{figure}

\begin{figure*}
	\centering
	\includegraphics[width=\textwidth]{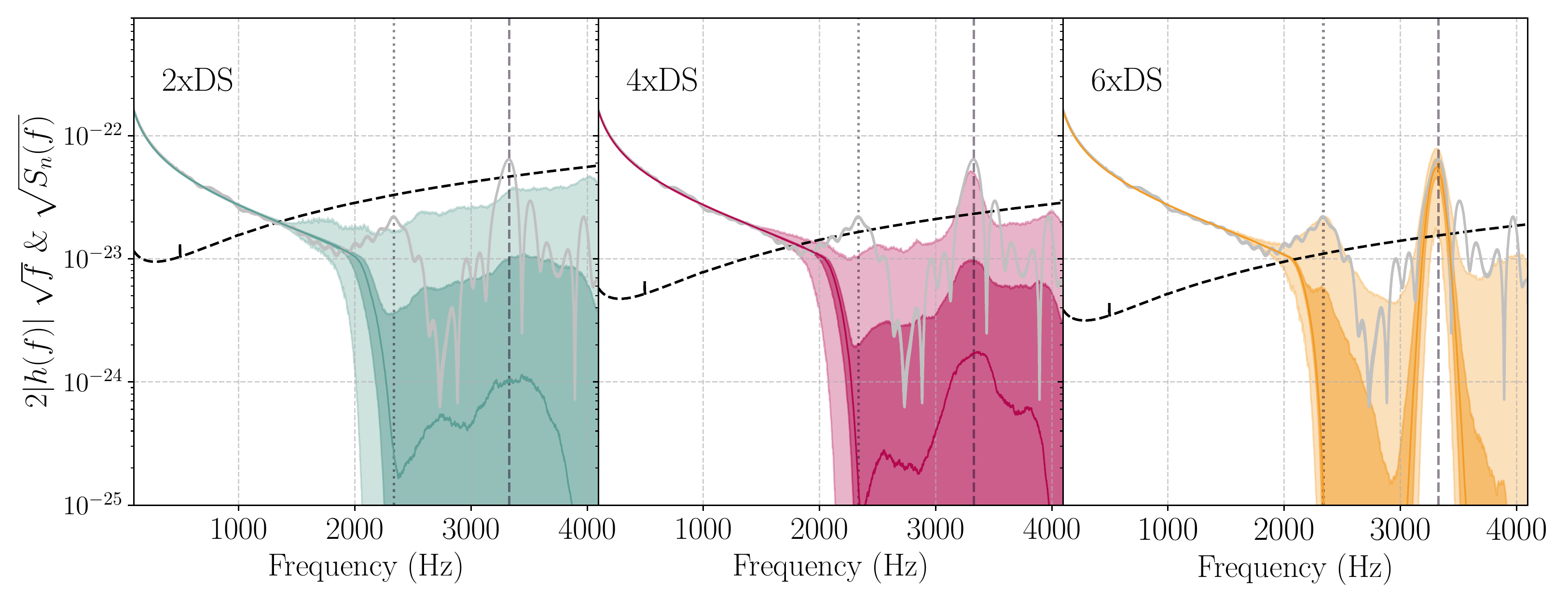}
	\caption{Median, 50\%, and 90\% credible intervals for the reconstructed spectrum from the Kastaun and Ohme~\cite{Kastaun:2021zyo} waveform with different detector sensitivities; from left to right we assume a strain sensitivity two, four, and six times better than the advanced LIGO design sensitivity. The gray line gives the simulated data, which are the same in all cases, while the black dashed line gives the detection noise amplitude spectral density. The dashed and dotted vertical lines denote, respectively, the dominant and subdominant postmerger spectrum frequency.}
	\label{fig:KO_spec}
\end{figure*}

Figure~\ref{fig:KO_rec} shows the whitened time-domain data and reconstruction for the injection at 6xDS, again focusing around the late stages of the signal. The full reconstruction with the combined CBC+wavelets model accurately captures the entire signal. We also plot the individual components of the full model, namely the CBC and the
wavelets model separately. As expected, the CBC model captures everything up to merger and then tapers off. The wavelets overlap with the CBC model in the taper region around merger, effectively canceling out the ringdown-like oscillations of the CBC model that do not match the data. At later times, the wavelet model extends to the postmerger part of the signal, capturing its main oscillatory component.

Figure~\ref{fig:KO_spec} shows the reconstructed spectrum from injections on different detectors sensitivities. As the detector sensitivity increases, the signal reconstruction becomes more accurate. The premerger signal is recovered in all analyses given its strength, and increasing sensitivity reduces the reconstruction uncertainty. The postmerger signal, on the other hand, is too weak to be detected in the 2xDS case, and therefore the reconstruction is essentially uninformative and similar to the GW170817 one. At increasing detector sensitivity, the postmerger signal emerges from the noise. At 4xDS the reconstructed spectrum starts identifying the main postmerger peak though only at the $\sim 50\%$ credible level. At 6xDS the reconstructed spectrum not only confidently identifies the main postmerger peak but also has evidence of a subdominant peak at around 2337~Hz.

\begin{figure}
	\centering
	\includegraphics[width=0.49\textwidth]{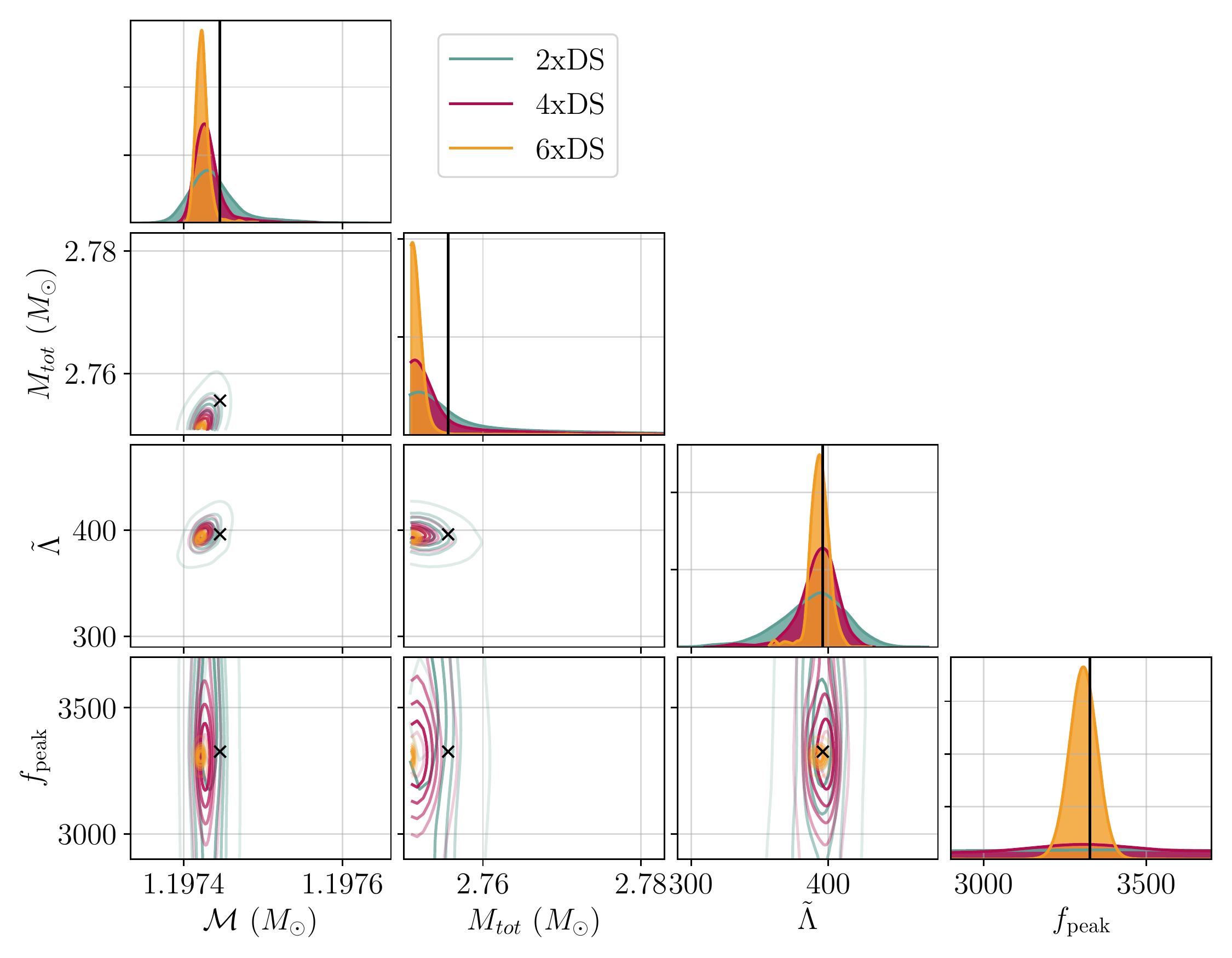}
	\caption{One- and two-dimensional marginalized posterior distributions for selected signal parameters from the analyses of Fig.~\ref{fig:KO_spec}: detector frame chirp mass $\Mc$, detector frame total mass $M$, tidal deformation parameter $\tilde{\Lambda}$, and postmerger peak frequency $\fpeak$. Vertical black lines or crosses denote the true values of each parameter. 
	}
	\label{fig:KO_params}
\end{figure}

Posteriors for select premerger and postmerger parameters are given in Fig.~\ref{fig:KO_params}. The peak frequency of the postmerger signal $\fpeak$ is defined in the same way as~\cite{Chatziioannou:2017ixj,Torres-Rivas:2018svp}: for each reconstruction posterior sample we compute the frequency at the maximum of the spectrum after the merger. If the spectrum possesses no maximum, then a sample is drawn from the $\fpeak$ prior. All posteriors are consistent with the injected values, and uncertainties decrease with increasing detector sensitivity as expected. The $\fpeak$ posterior at 2xDS is essentially the prior, consistent with the fact that the postmerger signal was not detected. At 4xDS the $\fpeak$ posterior starts exhibiting a peak at the injected value, consistent with the partial identification of the postmerger signal in the middle panel of Fig.~\ref{fig:KO_spec}. At 6xDS the peak frequency is accurately measured.

\subsection{Hadronic EoS}
\label{sec:hadronic}

We construct further simulated signals based on the postmerger NR simulations from~\cite{Torres-Rivas:2018svp}, which employ the conformal flatness approximation~\cite{Wilson1996,Isenberg1980}. Comparisons to fully relativistic studies show a very good agreement with regards to postmerger GW frequencies but an underestimation of the postmerger GW amplitude by some 10\% as a result of using the quadrupole formula for GW extraction~\cite{Bauswein:2012ya}. Since our analysis does not rely on calibration to any NR simulations, we expect our results below to be unaffected by such uncertainties. Indeed, {\tt BayesWave} has been shown to produce reliable results on simulated postmerger signals made with different NR codes~\cite{Easter:2020ifj,Chatziioannou:2017ixj}.

All simulations in~\cite{Torres-Rivas:2018svp} were at the time constructed 
to be consistent with GW170817, though in light of new data from NICER~\cite{Riley:2019yda,Miller:2019cac,Miller:2021qha,Riley:2021pdl} some of the softest EoSs
there are now disfavored. We work with EoS3 from~\cite{Torres-Rivas:2018svp}, a hadronic EoS that is consistent with radii values inferred in~\cite{Legred:2021hdx} and corresponds to $\fpeak=2880$Hz and $\Rtyp=12.6$~km. The inspiral portion of the signal is again described by PhenomDNRT for a GW170817-like system with detector frame component masses $m_1=m_2$ = 1.362$M_{\odot}$, dimensionless tidal parameters $\Lambda_{1}=\Lambda_{2} = \tilde{\Lambda}= 587$, and zero spin. 

The full waveform is constructed by aligning the projected premerger and postmerger waveforms in an overlap interval in the time domain. We use a transition window $[t_1, t_2]$ where the full waveform transitions from the PhenomDNRT template to the NR simulation data. The full waveform is then 
\begin{align}
	h(t) =  \left[1 - x(t)\right] h_\text{inspiral}(t) + x(t) A_\textrm{scale} h_\text{NR}(t),
\end{align}
where $h_\text{inspiral}$ denotes the PhenomDNRT inspiral model, $h_\text{NR}$ denotes the postmerger NR simulation, $A_\textrm{scale}$ is a scale parameter that can be varied to control the strength of the postmerger signal, and $x$ ensures a smooth transition through a Planck taper function
\begin{align}
	x(t) = 
	\begin{cases}
		0 &  t \leq t_1 \\
		\left[1 + \exp{\left( \frac{t_2 - t}{t - t_1} + \frac{t_2 - t}{t - t_2} \right)} \right]^{-1} &  t_1 < t < t_2 \\
		1 &  t \geq t_2.
	\end{cases}
\end{align}

\begin{figure}
	\centering
	\includegraphics[width=0.49\textwidth]{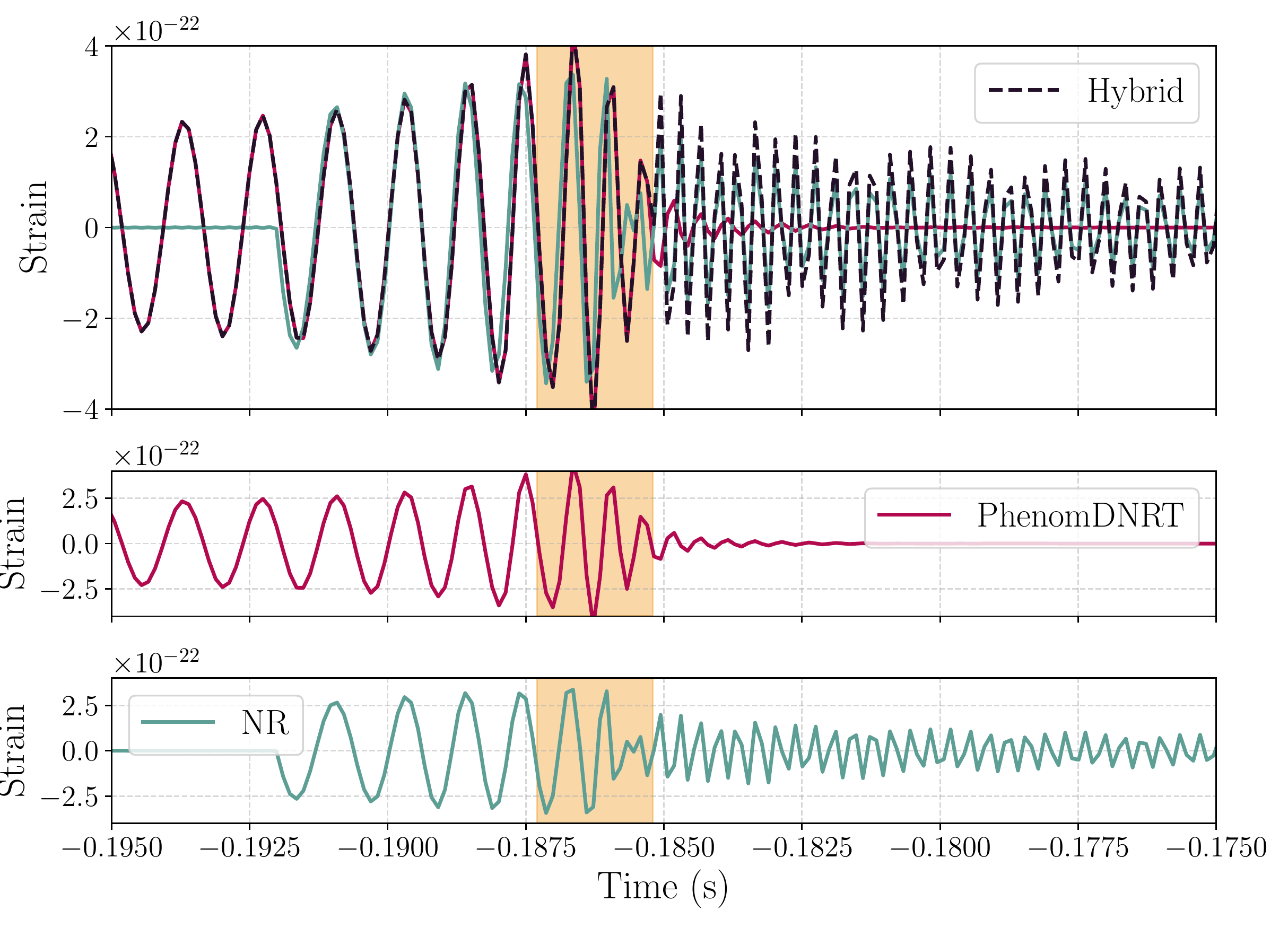}
	\caption{Comparison of the premerger (PhenomDNRT) and postmerger (NR simulation) waveform that are merged into a hybrid full waveform. The yellow band indicates the overlap region where the hybrid waveform transitions from being composed of the PhenomDNRT model to the postmerger NR simulation model. The maximum amplitude of the NR waveform used here is scaled to be 0.9 times the premerger peak amplitude.}
	\label{fig:wfoverlap}
\end{figure}

We fix the detector sensitivity to 2xDS and vary the amplitude of the postmerger signal through the scale parameter $A_\textrm{scale}$. The main motivation for this is that it allows us to vary the postmerger signal strength, while keeping the premerger SNR (and thus the computation cost) manageable. This scaling further allows us to address the underestimation of the signal amplitude within the conformal flatness approximation mentioned above. Larger sets of simulations do not find a very tight relation between the postmerger amplitude and binary parameters~\cite{Tsang:2019esi}, and simulations in general may over- or under-estimate the postmerger amplitude to some extent.

In what follows, we keep the premerger SNR constant at 112, while the postmerger SNR varies and results are presented as a function of the ratio of the premerger to the postmerger peak amplitudes. We explore three values for the ratio of the peak postmerger to the peak premerger amplitudes: 0.7, 0.9 and 1.1 (corresponding to $A_\textrm{scale}$ = 1.5, 2.0 and 2.5 and postmerger SNRs of 6.2, 8.1 and 10.0, respectively). We show an example of the above construction process in Fig.~\ref{fig:wfoverlap}.

\begin{figure*}
	\centering
	\includegraphics[width=\textwidth]{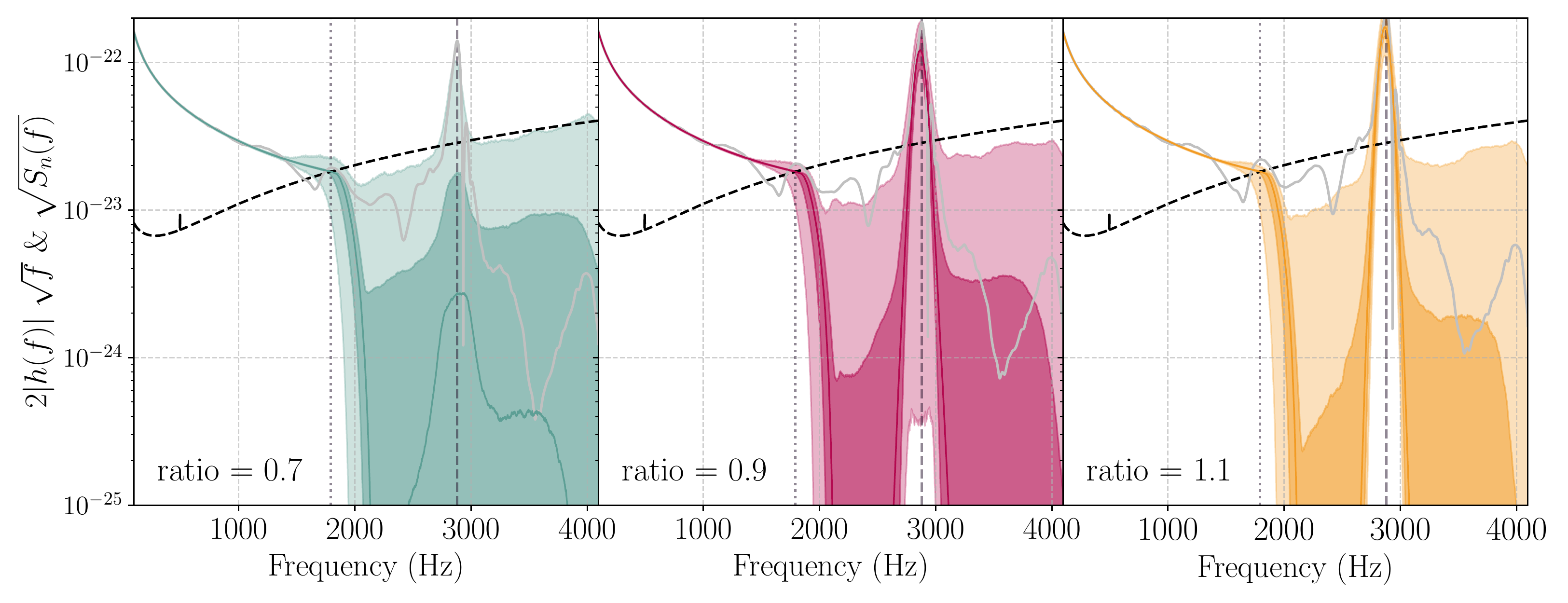}
	\caption{Median, 50$\%$ and 90$\%$ credible intervals of the reconstructed spectrum for the hadronic EoS3 signal from Sec.~\ref{sec:hadronic} and for pre/postmerger peak amplitude ratios of 0.7 (left), 0.9 (middle), 1.1 (right) in the LIGO Hanford detector. The gray line gives the simulated data. The noise amplitude spectral density is also overplotted with a black dashed line.}
	\label{fig:EoS3Spectra}
\end{figure*}

\begin{figure}
	\centering
	\includegraphics[width=0.49\textwidth]{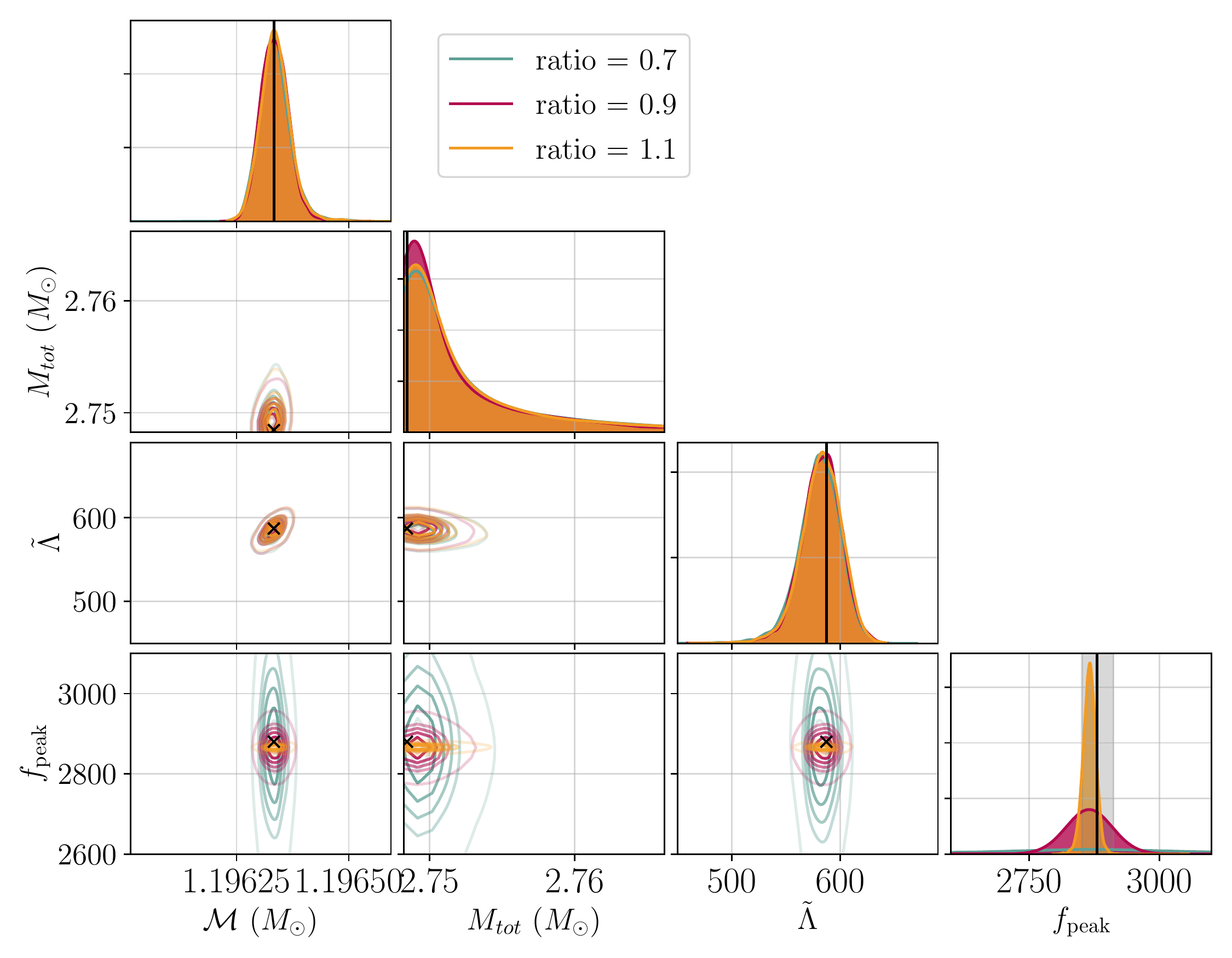}
	\caption{One- and two-dimensional marginalized posterior distributions for selected pre and postmerger parameters for the analyses of Fig.~\ref{fig:EoS3Spectra}. The parameters shown are the source-frame chirp mass ${\cal{M}}$ and total mass $M$, the tidal deformation parameter $\tilde{\Lambda}$ and the postmerger peak frequency $\fpeak$. Vertical black lines or crosses denote the true values of each parameter. The gray region is the expected value for $\fpeak$ given the premerger inferred parameters and the fit of~\cite{Chatziioannou:2017ixj} that assumes hadronic EoSs. The premerger and postmerger results are consistent with expectations for hadronic EoSs. 
}
	\label{fig:EoS3_rec}
\end{figure}

In Figure~\ref{fig:EoS3Spectra} we compare the reconstructed spectra from our analyses for the three cases of post/premerger amplitude scaling. The inspiral portion of the signal and the corresponding reconstruction are similar in the three panels. As the amplitude of the postmerger signal increases from left to right the reconstructed spectrum includes more detailed features, progressing from hints of a postmerger peak on the left panel to increasingly more confident identification in the middle and right panel. Figure~\ref{fig:EoS3_rec} again shows select recovered premerger and postmerger parameters. The gray region overlapping with the $\fpeak$ posterior is the expected value for the postmerger frequency given the premerger signal. Specifically we use the fit of~\cite{Chatziioannou:2017ixj} that holds for hadronic EoSs and the binary mass and tidal deformability as extracted from the premerger signal to compute the expected $\fpeak$. The expected and recovered posterior values for $\fpeak$ agree, showing that our analysis can correctly conclude that the premerger and postmerger signals are consistent with each other given expectations from NR simulations of hadronic EoSs.

\subsection{EoS with phase transitions}
\label{sec:PT}

\begin{figure}
	\centering
	\includegraphics[width=\columnwidth]{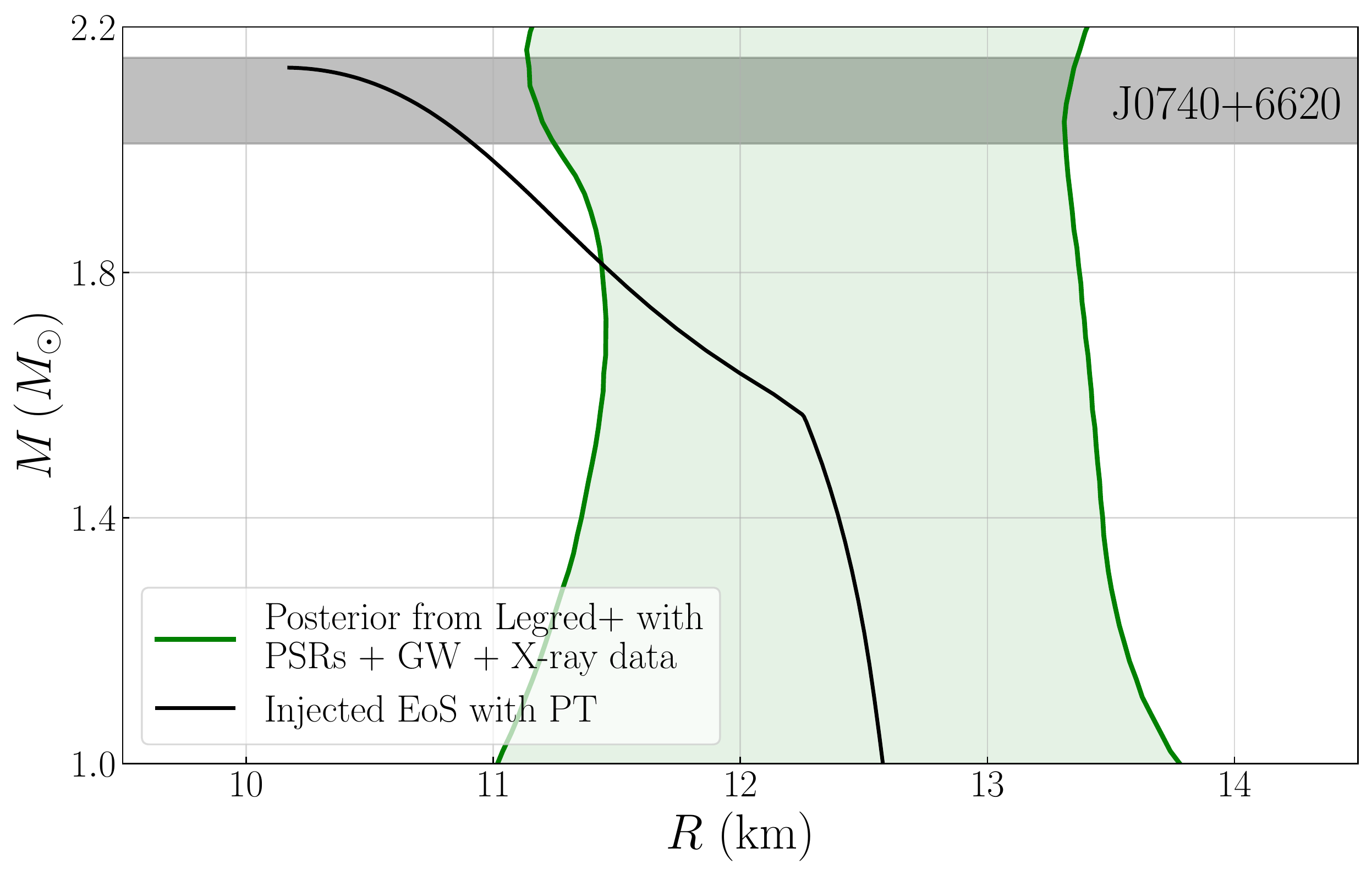}
	\caption{Mass-radius relation for EoS DD2-SF-4 used in our study of PTs (black line). The green shaded region gives the 90\% credible level of the mass-radius posterior derived in Legred et al.~\cite{Legred:2021hdx} using data from heavy pulsars, GWs, and the Miller et al.~\cite{Miller:2019cac,Miller:2021qha} radius results using X-ray data. The adopted EoS is inconsistent with this posterior at the $90\%$ level for masses above $\sim 1.8M_{\odot}$, but it is consistent to the same level with the less constraining Riley et al.~\cite{Riley:2019yda,Riley:2021pdl} results. }
	\label{fig:EoSPTMR}
\end{figure}

Since the hypermassive NS that gives rise to the postmerger signal is characterized by higher core densities and temperatures than the premerger NSs, the postmerger signal has the potential
to reveal new high-density physics. One possibility is a strong phase transition in the EoS toward quark degrees of freedom\footnote{We use the term ``strong'' phase transition to emphasize that the exact characteristics of the hadron-quark phase transition are currently unclear, and only a sufficiently strong transition, for instance a first-order phase transition with large latent heat, may impact the merger dynamics such that the occurrence of quark matter leads to an unambiguous signature. See, e.g.,~\cite{2020PhRvD.102l3023B} for a detailed discussion. In fact, quark matter may also resemble the behavior of purely hadronic matter (commonly referred to as the masquerade problem~\cite{Alford:2004pf,Alford:2015dpa}), which may not leave a very prominent imprint on the GW signal.}. Such phase transitions could occur
at lower densities and thus be detectable with premerger data only~\cite{DelPozzo:2013ala,Agathos:2015uaa,Chatziioannou:2015uea,Han:2018mtj,Chen:2019rja,Chatziioannou:2019yko,Han:2020adu,Zhang:2019fog,Pang:2020ilf,Drischler:2020fvz}. However, as already
mentioned above, the tidal deformability is a steeply decreasing function of the NS compactness, making premerger data less constraining about high-mass NSs that could 
contain quark cores. postmerger data, on the other hand, can be used to probe higher densities and NR simulation-based studies have explored the potential
signature of a high-density phase transition on the GW signal~\cite{Most:2018eaw,Bauswein:2018bma,Bauswein:2019skm,Weih:2019xvw,Bauswein:2020ggy,Liebling:2020dhf,Prakash:2021wpz}.

\begin{figure*}
	\centering
	\includegraphics[width=\textwidth]{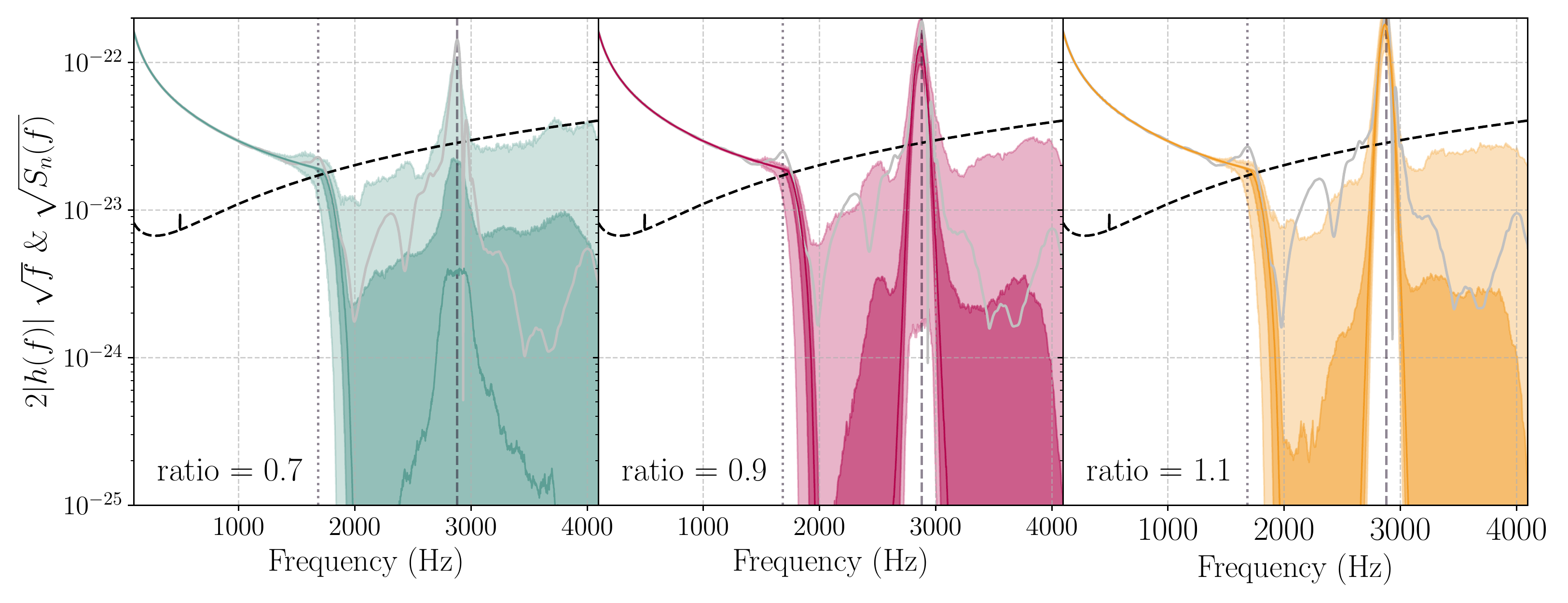}
	\caption{The median, 50$\%$ and 90$\%$ credible intervals of the reconstructed spectrum for a signal with EoS3 where the inspiral signal has been simulated with a stiffer EoS in order to mimic the phenomenology of a strong phase transitions from Sec.~\ref{sec:PT}. We show pre/postmerger peak amplitude ratios of 0.7 (left), 0.9 (middle), 1.1 (right). The gray line gives the simulated data. The noise amplitude spectral density is also overplotted with a black dashed line. }
	\label{fig:EoS3PTSpectra}
\end{figure*}

\begin{figure}
	\centering
	\includegraphics[width=0.49\textwidth]{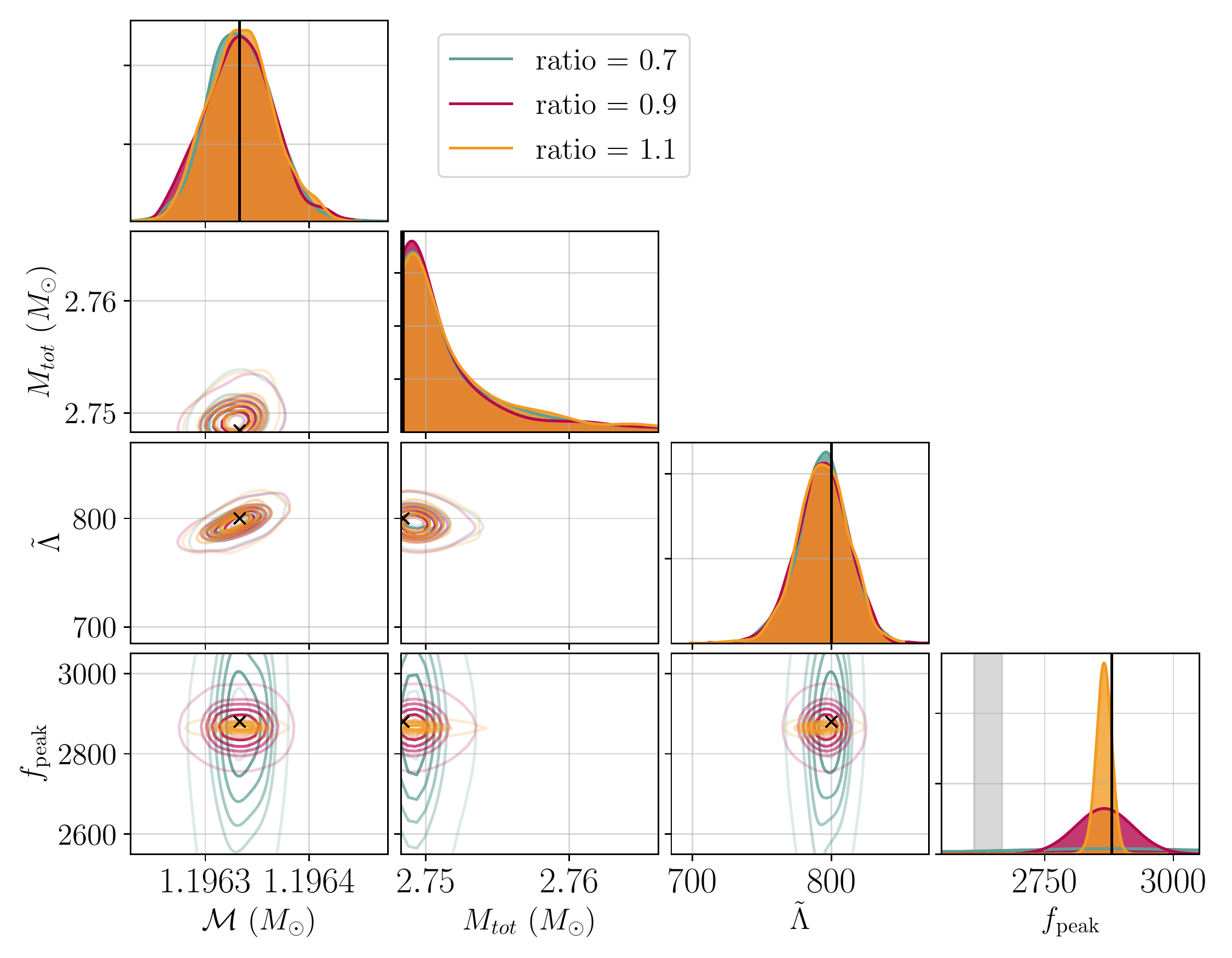}
	\caption{One- and two-dimensional marginalized posterior distributions for selected pre and postmerger parameters for the analyses of Fig.~\ref{fig:EoS3PTSpectra}. The parameters shown are the source-frame chirp mass ${\cal{M}}$ and total mass $M$, the tidal deformation parameter $\tilde{\Lambda}$ and the postmerger peak frequency $\fpeak$. Black vertical black lines or crosses denote the true values of each parameter. The gray region is the expected value for $\fpeak$ given
the premerger inferred parameters and the fit of~\cite{Chatziioannou:2017ixj} that assumes hadronic EoSs. The premerger and postmerger results are now inconsistent with expectations for hadronic EoSs.}
	\label{fig:EoS3PT_rec}
\end{figure}

The most prominent signature of a sufficiently strong phase transition would be an increase in the postmerger peak frequency as the softening of the EoS would lead to a more compact merger remnant~\cite{Bauswein:2018bma}. It has been proposed that such a frequency increase could be identified if one compares the premerger and postmerger data: for hadronic EoSs the tidal deformability and peak frequency follow the approximately EoS-insensitive relation shown in Fig.~\ref{fig:PTcomp}. This relation could be violated for EoSs with high-density phase transitions as the tidal deformability is determined by the hadronic part of the EoS alone, while the postmerger frequency is affected by the phase transition~\cite{Bauswein:2018bma,Bauswein:2020ggy}\footnote{In fact, a postmerger frequency deviation may also occur if quark matter is present before the merger~\cite{Bauswein:2020ggy}.}. The degree of deviation from the EoS-insensitive relation, and thus how detectable it is, depends on the strength of the transition~\cite{Bauswein:2018bma,2020PhRvD.102l3023B}.

As a first example of such a signal, we work again with EoS3 from~\cite{Torres-Rivas:2018svp}, but the premerger data are constructed with tidal parameters that are systematically shifted compared to their hadronic EoS value. This effectively results in a stiffer hadronic EoS that undergoes a phase transition which softens the postmerger signature to the level of EoS3. All intrinsic parameters remain to be the same as the simulations of Sec.~\ref{sec:hadronic} with the exception of $\tilde{\Lambda} = \Lambda_1 = \Lambda_2$ = 800.The postmerger SNRs remain the same as Sec.~\ref{sec:hadronic}.

Reconstructed spectra and parameter posteriors are shown in Figs.~\ref{fig:EoS3PTSpectra} and~\ref{fig:EoS3PT_rec} respectively. We obtain qualitatively similar results to Figs.~\ref{fig:EoS3Spectra} and~\ref{fig:EoS3_rec} with the main difference being that now the premerger and postmerger are now inconsistent as expected. Figure~\ref{fig:EoS3PT_rec} shows the expected $\fpeak$ value given the premerger constraints on the binary properties under the assumption of a hadronic EoS. The recovered $\fpeak$ is inconsistent with this expectation to within its measurement uncertainty, signaling the presence of additional high-density effects in the EoS that affect the postmerger signal.

\begin{figure*}
	\centering
	\includegraphics[width=\textwidth]{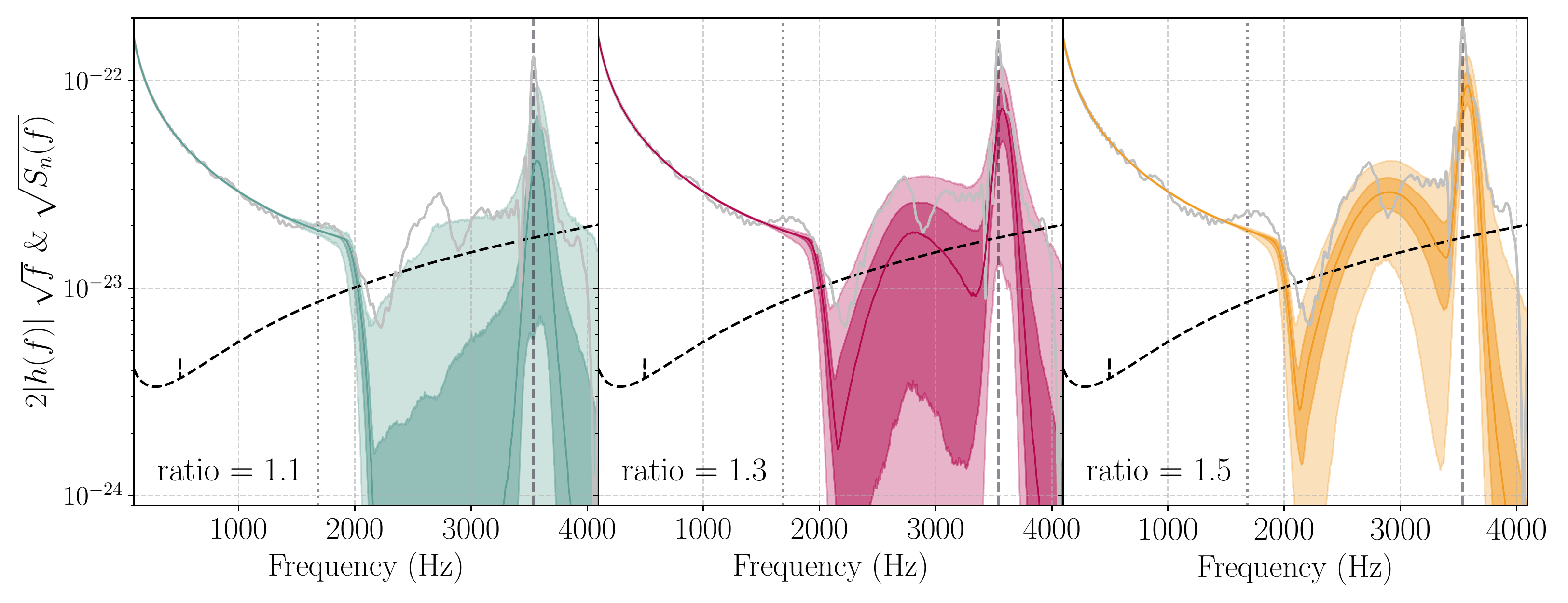}
	\caption{The median, 50$\%$ and 90$\%$ credible intervals of the reconstructed spectrum for a signal with the DD2-SF-4 EoS with phase transitions from Sec.~\ref{sec:PT} and for pre/postmerger peak amplitude ratios of 1.1 (left), 1.3 (middle), 1.5 (right). Due to the increased $\fpeak$ value toward the less sensitive detector frequency range, such a postmerger signal would require a larger strain amplitude for detection, we therefore display results with larger values of the pre/postmerger peak amplitude ratio compared to Fig.~\ref{fig:EoS3PTSpectra}. The corresponding postmerger SNR is 7.2, 8.1, 9.1 from left to right. The gray line gives the simulated data. The noise amplitude spectral density is also overplotted with a black dashed line.}
	\label{fig:DD2PTSpectra}
\end{figure*}

\begin{figure}
	\centering
	\includegraphics[width=0.49\textwidth]{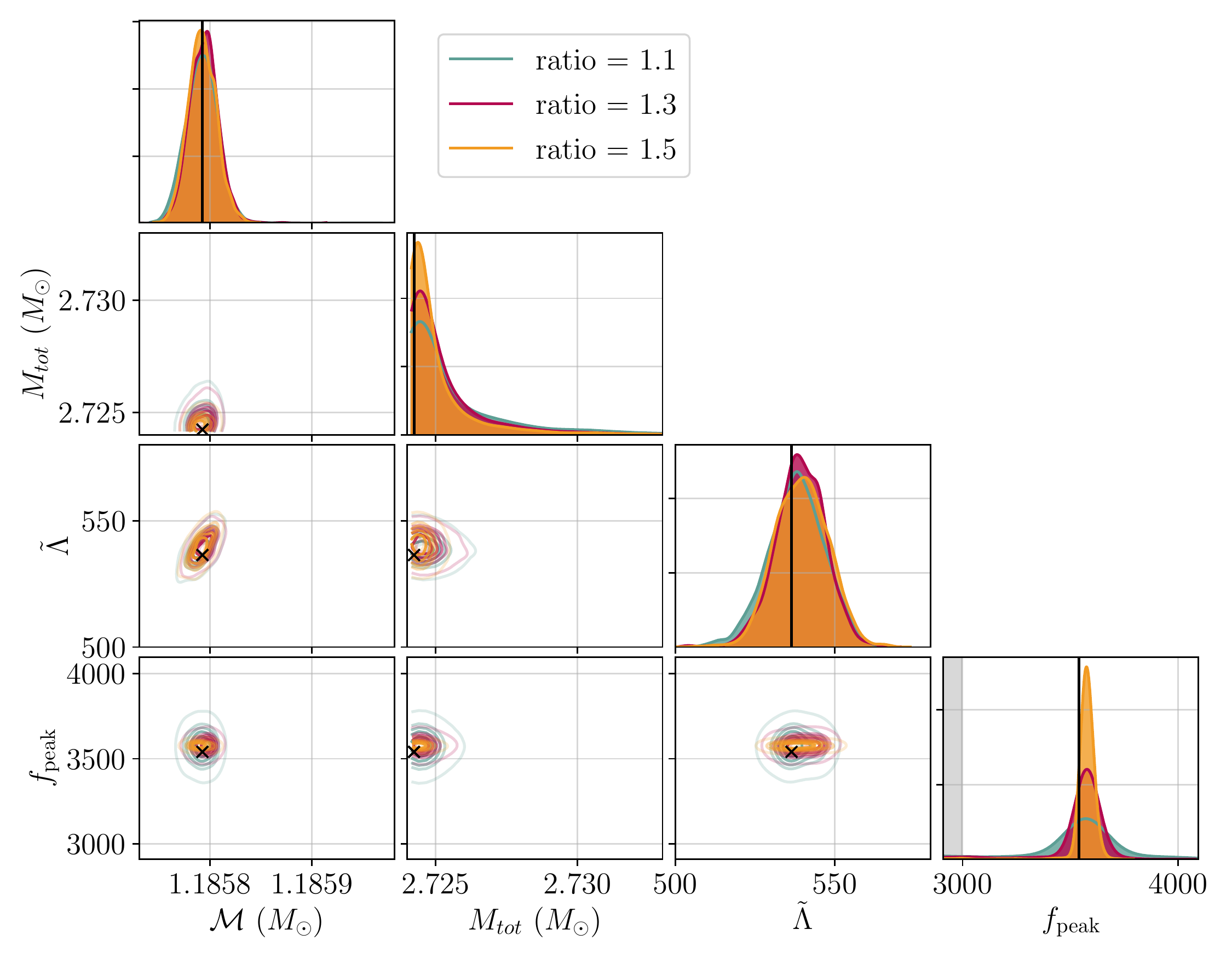}
	\caption{One- and two-dimensional marginalized posterior distributions for selected pre and postmerger parameters for the analyses of Fig.~\ref{fig:DD2PTSpectra}. The parameters shown are the detector frame chirp mass ${\cal{M}}$ and total mass $M$, the tidal deformation parameter $\tilde{\Lambda}$ and the postmerger peak frequency $\fpeak$. Vertical black lines or crosses denote the true values of each parameter. The gray region is the expected value for $\fpeak$ given
the premerger inferred parameters and the fit of~\cite{Chatziioannou:2017ixj} that assumes hadronic EoSs. The premerger and postmerger data are now as expected inconsistent with expectations for hadronic EoSs.}
	\label{fig:DD2PT_rec}
\end{figure}

We further assess how well our hybrid analysis could detect a strong phase transition by simulating a signal with the DD2-SF-4 EoS~\cite{2018NatAs...2..980F,2021PhRvD.103b3001B} and corresponding postmerger simulation from~\cite{Bauswein:2018bma}, also performed with the conformal flatness approximation. The mass-radius relation for this EoS is given in Fig.~\ref{fig:EoSPTMR} and it exhibits the characteristic radius reduction due to strong phase transitions starting at $\sim 1.5M_{\odot}$. Given the simulated binary masses of $1.35M_{\odot}$ for both components, the inspiral signal is emitted by hadronic NSs, while the postmerger signal is affected by the onset of the phase transition~\cite{Bauswein:2018bma}. For reference, we also show the 90\% symmetric credible intervals of the mass-radius posterior for the EoS derived in~\cite{Legred:2021hdx} using heavy pulsar, GW, and X-ray data. The DD2-SF-4 EoS is inconsistent with the posterior at the 90\% level for large masses as it underpredicts the radius of a $2M_\odot$ NS, though it is consistent with current data at the 95\% level. If we instead used the less constraining radius results from Riley el at.~\cite{Riley:2019yda,Riley:2021pdl}, the EoS would be consistent with the posterior at the 90\% level. Results are presented in Figs.~\ref{fig:DD2PTSpectra} and~\ref{fig:DD2PT_rec}, where again we find that the postmerger signal can be reconstructed for sufficiently loud signals, and the inconsistency between $\tilde{\Lambda}$ and $\fpeak$ can be identified.

\begin{figure}
	\centering
	\includegraphics[width=0.49\textwidth]{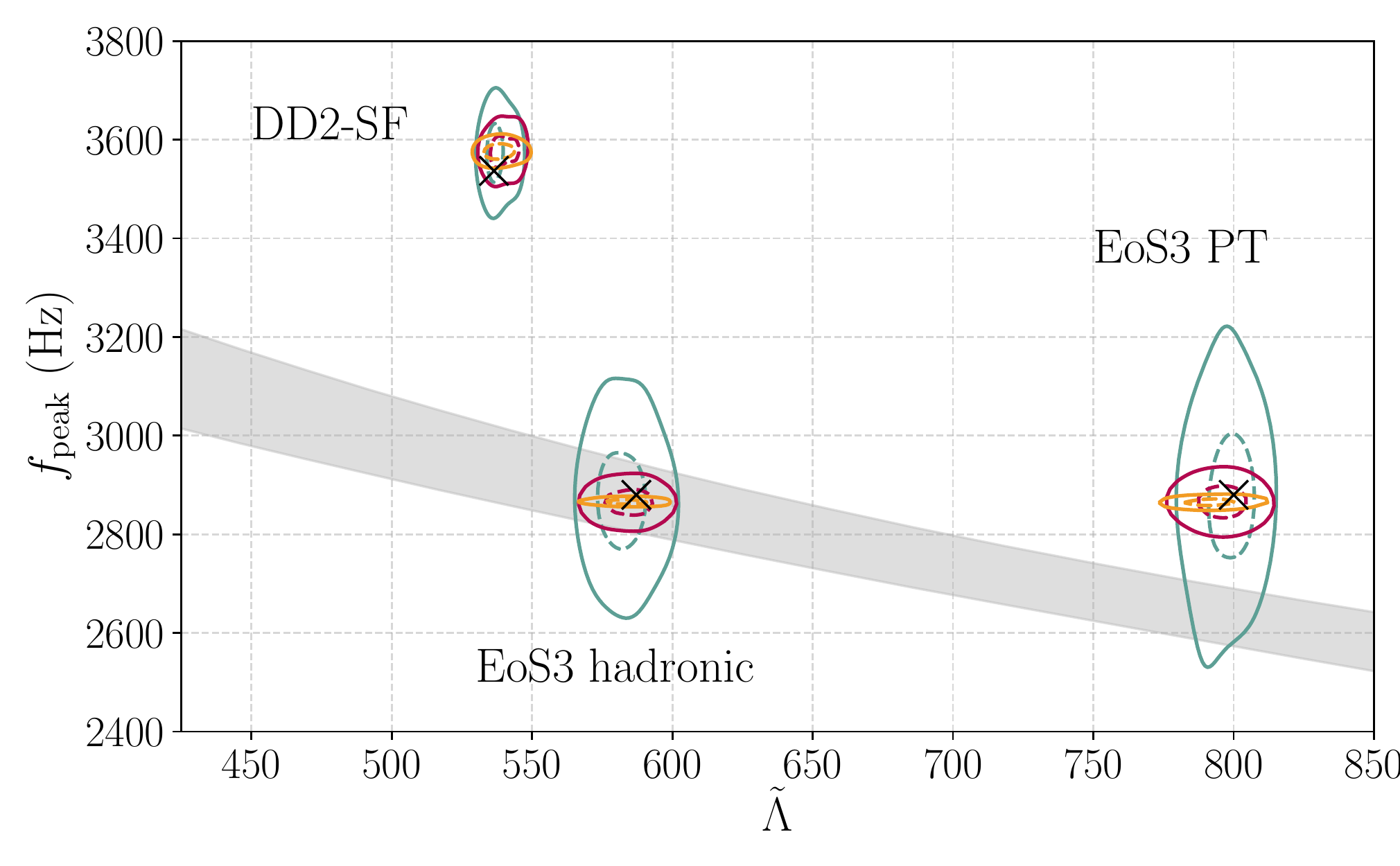}
	\caption{Comparison of the recovered $\tilde{\Lambda}$-f$_\mathrm{peak}$ posterior distribution from the EoS3 hadronic and PT data as well as the data with the DD2-SF-4 EoS. The gray bands corresponds to the EoS-independent fit relating the expected values of $\tilde{\Lambda}$-f$_\mathrm{peak}$ for hadronic NSs. We show 50\% (dashed) and 90\% (solid) contours, while colors correspond to those of Figs.~\ref{fig:EoS3_rec},~\ref{fig:EoS3PT_rec},~\ref{fig:DD2PT_rec}. In all cases, we find the correct agreement or disagreement with the expected hadronic relation for sufficiently loud signals.}
	\label{fig:PTcomp}
\end{figure}

Finally, we elaborate on the premerger and postmerger consistency in Fig.~\ref{fig:PTcomp}. The gray band shows the expected relation between $\tilde{\Lambda}$ and $\fpeak$ assuming hadronic EoSs. We compute this relation by converting the $R-\fpeak$ EoS-independent\footnote{The fit is shown in Fig. 4 of~\cite{Chatziioannou:2017ixj} and uses EoSs that predict radii between $11.5$km and $5$km approximately.} fit by~\cite{Chatziioannou:2017ixj} to a $\tilde{\Lambda}-\fpeak$ relation. We convert $R$ to $\Lambda$ using the scaling relation calibrated for GW170817-like systems for our source mass parameters $M_\text{tot}$ and $\Mc$ given by Eq. 17 in~\cite{Zhao:2018nyf}. Overplotted are the two-dimensional posteriors for $\tilde{\Lambda}-\fpeak$ for the case of a hadronic EoS (EoS3), the phase-transition version of the EoS3-based signal, and the DD2-SF-4 EoS again with a strong phase transition. The injected values for the hadronic case is consistent with expectations and this is confirmed by the recovered values to within the statistical error. On the other hand, a strong phase transition leads to a $\tilde{\Lambda}-\fpeak$ combination that is inconsistent with hadronic expectations and the extracted posteriors are able to identify this behavior.

\section{Conclusions}
\label{sec:conclusions}

We present a hybrid approach to study the full GW signal emitted during a BNS coalescence, including a possible postmerger component. Our method models the inspiral part of the coalescence with waveform templates as implemented in \code{LALSimulation} and the postmerger part of the signal with a superposition of wavelets. We do not impose phase coherence in the transition between the template and the wavelets, however, the full signal is smooth and coherent as the template model used already extends coherently past merger and into the postmerger part of the signal.

Applying our method to GW170817, we demonstrate that the inspiral parameters are consistent with previous results. We do not detect a postmerger signal for GW170817 but find that the high-frequency portion of the signal between 1000-1500 Hz is only detected using the full hybrid analysis in contrast with the traditional postmerger-only analyses. We apply our method to simulated signals with a detectable postmerger component and show that our full analysis simultaneously reconstructs both the inspiral tidal deformation and the postmerger dominant frequency peak when the SNR is sufficiently high.

In this work, we sample the inspiral tidal parameters and postmerger peak frequency independently, imposing no relation between them and no assumption on the nature of the EoS. This allows us to detect a possible signature of the hadron-quark phase transition in NS mergers, as it lead to a characteristic frequency shift of the postmerger signal. A possible extension of this analysis could use information extracted from a premerger signal to guide the analysis of the postmerger signal. If one restricts to a hadronic EoS, relations between the premerger tidal parameters and postmerger peak frequency could be used to predict the approximate location of the dominant postmerger peak from the inspiral tidal parameters. This could improve the prospects of detecting the postmerger signal in weaker signals, at the expense of assuming the EoS is similar to a hadronic one. A further possible improvement concerns the use of ``chirplets", sine-Gaussian wavelets with an evolving frequency~\cite{Millhouse:2018dgi}, which might be better suited for a time-evolving postmerger frequency mode.
 
Going beyond the main peak, numerical simulations of postmerger signals show subdominant peaks in the GW spectrum, which can also be used to characterize the properties of the merger remnant star~\cite{2011MNRAS.418..427S,2014PhRvL.113i1104T,bauswein:15}. Our analysis can identify some of these peaks for sufficiently loud signals, however they are typically less loud than the main peak and can be lost in the noise.
Numerical simulations suggest that the main and the secondary peaks in the spectrum are related to each other~\cite{bauswein:15,Clark:2015zxa,Soultanis:2021oia}. In future work, this feature can be utilized in the form of a prior that links the two peaks and is parametrized in terms of the remnant compactness. Such a prior would enhance the sensitivity of the analysis to secondary peaks as it effectively adds modeling information about the signal similar to the approach of analytic models~\cite{Hotokezaka:2013iia,Bauswein:2015vxa,Bose:2017jvk,Tsang:2019esi,Breschi:2019srl,Easter:2020ifj,Soultanis:2021oia}.

Our analysis concerns premerger SNRs in the hundreds, where systematic biases in the waveform models could be important and would have to be mitigated in the lead-up to such improved detectors~\cite{Dudi:2018jzn,Samajdar:2019ulq,Gamba:2020wgg,Chatziioannou:2021tdi,Kunert:2021hgm,Pratten:2021pro}. 
However, as we make minimal assumptions about the postmerger waveform and use sine-gaussian wavelets, we expect the reconstructed signal to be less susceptible to systematics than analytic models~\cite{Hotokezaka:2013iia,Bauswein:2015vxa,Bose:2017jvk,Tsang:2019esi,Breschi:2019srl,Easter:2020ifj,Soultanis:2021oia} that are limited by the accuracy of the NR simulations on which they are based (see for example Fig. 4 of~\cite{Breschi:2019srl}). Our analysis instead is not limited by NR accuracy and has yielded unbiased results on simulated signals form different codes~\cite{Easter:2020ifj,Chatziioannou:2017ixj}.
On the other hand, analytic models can be more straightforwardly attached to premerger waveform templates and enforce phase coherence through merger~\cite{Breschi:2019srl}. 

Next generation detectors are expected to detect thousands of BNSs~\cite{Regimbau:2012ir,Sachdev:2020bkk,LIGO:2020xsf}, the majority of which will be weak with individually undetectable postmerger signals. While we study single loud sources here, our method can also be applied to the expected numerous weaker BNS sources which might dominate the overall constraints when combined~\cite{Haster:2020sdh}. Though our analysis does not hinge on the existence of accurate models for the postmerger signal, it is possible that biases in the premerger waveform will lead to biased inferences about the postmerger when making use of EoS-independent relations connecting them. The extremely sensitive observations possible with next-generation detectors indeed require control over a wide range of potential systematic biases, and the flexible analysis presented here helps mitigate such biases from the postmerger signal.

\acknowledgments

We thank Will Farr and Wynn Ho for many useful discussions. We also thank Sophie Hourihane and Tyson Littenberg for discussions and assistance about \BayesWave.
This research has made use of data, software and/or web tools obtained from the Gravitational Wave Open Science Center (https://www.gw-openscience.org), a service of LIGO Laboratory, the LIGO Scientific Collaboration and the Virgo Collaboration.
Virgo is funded by the French Centre National de Recherche Scientifique (CNRS), the Italian Istituto Nazionale della Fisica Nucleare (INFN) and the Dutch Nikhef, with contributions by Polish and Hungarian institutes.
This material is based upon work supported by NSF's LIGO Laboratory which is a major facility fully funded by the National Science Foundation.
The authors are grateful for computational resources provided by the LIGO Laboratory and supported by National Science Foundation Grants PHY-0757058 and PHY-0823459.
M.W. gratefully acknowledges support and hospitality from the Simons Foundation through the predoctoral program at the Center for Computational Astrophysics, Flatiron Institute.
The Flatiron Institute is supported by the Simons Foundation.
K.C. was supported by NSF Grant PHY-2110111.
N.J.C. was supported by NSF Grant PHY-1912053.
AB acknowledges support by the European Research Council (ERC) under
the European Union's Horizon 2020 research and innovation programme
under grant agreement No. 759253, by Deutsche Forschungsgemeinschaft
(DFG, German Research Foundation) - Project-ID 279384907 - SFB 1245, by
DFG - Project-ID 138713538 - SFB 881 (``The Milky Way System'',
subproject A10) and by the State of Hesse within the Cluster Project
ELEMENTS.
Software: {\tt gwpy}~\cite{duncan_macleod_2020_3598469}, {\tt matplotlib}~\cite{Hunter:2007}.

\bibliography{OurRefs}

\begin{thebibliography}{174}%
\makeatletter
\providecommand \@ifxundefined [1]{%
 \@ifx{#1\undefined}
}%
\providecommand \@ifnum [1]{%
 \ifnum #1\expandafter \@firstoftwo
 \else \expandafter \@secondoftwo
 \fi
}%
\providecommand \@ifx [1]{%
 \ifx #1\expandafter \@firstoftwo
 \else \expandafter \@secondoftwo
 \fi
}%
\providecommand \natexlab [1]{#1}%
\providecommand \enquote  [1]{``#1''}%
\providecommand \bibnamefont  [1]{#1}%
\providecommand \bibfnamefont [1]{#1}%
\providecommand \citenamefont [1]{#1}%
\providecommand \href@noop [0]{\@secondoftwo}%
\providecommand \href [0]{\begingroup \@sanitize@url \@href}%
\providecommand \@href[1]{\@@startlink{#1}\@@href}%
\providecommand \@@href[1]{\endgroup#1\@@endlink}%
\providecommand \@sanitize@url [0]{\catcode `\\12\catcode `\$12\catcode
  `\&12\catcode `\#12\catcode `\^12\catcode `\_12\catcode `\%12\relax}%
\providecommand \@@startlink[1]{}%
\providecommand \@@endlink[0]{}%
\providecommand \url  [0]{\begingroup\@sanitize@url \@url }%
\providecommand \@url [1]{\endgroup\@href {#1}{\urlprefix }}%
\providecommand \urlprefix  [0]{URL }%
\providecommand \Eprint [0]{\href }%
\providecommand \doibase [0]{https://doi.org/}%
\providecommand \selectlanguage [0]{\@gobble}%
\providecommand \bibinfo  [0]{\@secondoftwo}%
\providecommand \bibfield  [0]{\@secondoftwo}%
\providecommand \translation [1]{[#1]}%
\providecommand \BibitemOpen [0]{}%
\providecommand \bibitemStop [0]{}%
\providecommand \bibitemNoStop [0]{.\EOS\space}%
\providecommand \EOS [0]{\spacefactor3000\relax}%
\providecommand \BibitemShut  [1]{\csname bibitem#1\endcsname}%
\let\auto@bib@innerbib\@empty
\bibitem [{\citenamefont {Lattimer}\ and\ \citenamefont
  {Prakash}(2016)}]{Lattimer:2015nhk}%
  \BibitemOpen
  \bibfield  {author} {\bibinfo {author} {\bibfnamefont {J.~M.}\ \bibnamefont
  {Lattimer}}\ and\ \bibinfo {author} {\bibfnamefont {M.}~\bibnamefont
  {Prakash}},\ }\bibfield  {title} {\bibinfo {title} {{The Equation of State of
  Hot, Dense Matter and Neutron Stars}},\ }\href
  {https://doi.org/10.1016/j.physrep.2015.12.005} {\bibfield  {journal}
  {\bibinfo  {journal} {Phys. Rept.}\ }\textbf {\bibinfo {volume} {621}},\
  \bibinfo {pages} {127} (\bibinfo {year} {2016})},\ \Eprint
  {https://arxiv.org/abs/1512.07820} {arXiv:1512.07820 [astro-ph.SR]}
  \BibitemShut {NoStop}%
\bibitem [{\citenamefont {{{\"O}zel}}\ and\ \citenamefont
  {{Freire}}(2016)}]{Ozel:2016oaf}%
  \BibitemOpen
  \bibfield  {author} {\bibinfo {author} {\bibfnamefont {F.}~\bibnamefont
  {{{\"O}zel}}}\ and\ \bibinfo {author} {\bibfnamefont {P.}~\bibnamefont
  {{Freire}}},\ }\bibfield  {title} {\bibinfo {title} {{Masses, Radii, and the
  Equation of State of Neutron Stars}},\ }\href
  {https://doi.org/10.1146/annurev-astro-081915-023322} {\bibfield  {journal}
  {\bibinfo  {journal} {Ann. Rev. Astron. Astrophys.}\ }\textbf {\bibinfo
  {volume} {54}},\ \bibinfo {pages} {401} (\bibinfo {year} {2016})},\ \Eprint
  {https://arxiv.org/abs/1603.02698} {arXiv:1603.02698 [astro-ph.HE]}
  \BibitemShut {NoStop}%
\bibitem [{\citenamefont {Oertel}\ \emph {et~al.}(2017)\citenamefont {Oertel},
  \citenamefont {Hempel}, \citenamefont {Klahn},\ and\ \citenamefont
  {Typel}}]{Oertel:2016bki}%
  \BibitemOpen
  \bibfield  {author} {\bibinfo {author} {\bibfnamefont {M.}~\bibnamefont
  {Oertel}}, \bibinfo {author} {\bibfnamefont {M.}~\bibnamefont {Hempel}},
  \bibinfo {author} {\bibfnamefont {T.}~\bibnamefont {Klahn}},\ and\ \bibinfo
  {author} {\bibfnamefont {S.}~\bibnamefont {Typel}},\ }\bibfield  {title}
  {\bibinfo {title} {{Equations of state for supernovae and compact stars}},\
  }\href {https://doi.org/10.1103/RevModPhys.89.015007} {\bibfield  {journal}
  {\bibinfo  {journal} {Rev. Mod. Phys.}\ }\textbf {\bibinfo {volume} {89}},\
  \bibinfo {pages} {015007} (\bibinfo {year} {2017})},\ \Eprint
  {https://arxiv.org/abs/1610.03361} {arXiv:1610.03361 [astro-ph.HE]}
  \BibitemShut {NoStop}%
\bibitem [{\citenamefont {Baym}\ \emph {et~al.}(2018)\citenamefont {Baym},
  \citenamefont {Hatsuda}, \citenamefont {Kojo}, \citenamefont {Powell},
  \citenamefont {Song},\ and\ \citenamefont {Takatsuka}}]{Baym:2017whm}%
  \BibitemOpen
  \bibfield  {author} {\bibinfo {author} {\bibfnamefont {G.}~\bibnamefont
  {Baym}}, \bibinfo {author} {\bibfnamefont {T.}~\bibnamefont {Hatsuda}},
  \bibinfo {author} {\bibfnamefont {T.}~\bibnamefont {Kojo}}, \bibinfo {author}
  {\bibfnamefont {P.~D.}\ \bibnamefont {Powell}}, \bibinfo {author}
  {\bibfnamefont {Y.}~\bibnamefont {Song}},\ and\ \bibinfo {author}
  {\bibfnamefont {T.}~\bibnamefont {Takatsuka}},\ }\bibfield  {title} {\bibinfo
  {title} {{From hadrons to quarks in neutron stars: a review}},\ }\href
  {https://doi.org/10.1088/1361-6633/aaae14} {\bibfield  {journal} {\bibinfo
  {journal} {Rept. Prog. Phys.}\ }\textbf {\bibinfo {volume} {81}},\ \bibinfo
  {pages} {056902} (\bibinfo {year} {2018})},\ \Eprint
  {https://arxiv.org/abs/1707.04966} {arXiv:1707.04966 [astro-ph.HE]}
  \BibitemShut {NoStop}%
\bibitem [{\citenamefont {Abbott}\ \emph
  {et~al.}(2017{\natexlab{a}})\citenamefont {Abbott} \emph
  {et~al.}}]{TheLIGOScientific:2017qsa}%
  \BibitemOpen
  \bibfield  {author} {\bibinfo {author} {\bibfnamefont {B.~P.}\ \bibnamefont
  {Abbott}} \emph {et~al.} (\bibinfo {collaboration} {LIGO Scientific
  Collaboration, Virgo Collaboration}),\ }\bibfield  {title} {\bibinfo {title}
  {{GW170817: Observation of Gravitational Waves from a Binary Neutron Star
  Inspiral}},\ }\href {https://doi.org/10.1103/PhysRevLett.119.161101}
  {\bibfield  {journal} {\bibinfo  {journal} {Phys. Rev. Lett.}\ }\textbf
  {\bibinfo {volume} {119}},\ \bibinfo {pages} {161101} (\bibinfo {year}
  {2017}{\natexlab{a}})},\ \Eprint {https://arxiv.org/abs/1710.05832}
  {arXiv:1710.05832 [gr-qc]} \BibitemShut {NoStop}%
\bibitem [{\citenamefont {Abbott}\ \emph {et~al.}(2020)\citenamefont {Abbott}
  \emph {et~al.}}]{Abbott:2020uma}%
  \BibitemOpen
  \bibfield  {author} {\bibinfo {author} {\bibfnamefont {B.~P.}\ \bibnamefont
  {Abbott}} \emph {et~al.} (\bibinfo {collaboration} {LIGO Scientific,
  Virgo}),\ }\bibfield  {title} {\bibinfo {title} {{GW190425: Observation of a
  Compact Binary Coalescence with Total Mass $\sim 3.4 M_{\odot}$}},\ }\href
  {https://doi.org/10.3847/2041-8213/ab75f5} {\bibfield  {journal} {\bibinfo
  {journal} {Astrophys. J. Lett.}\ }\textbf {\bibinfo {volume} {892}},\
  \bibinfo {pages} {L3} (\bibinfo {year} {2020})},\ \Eprint
  {https://arxiv.org/abs/2001.01761} {arXiv:2001.01761 [astro-ph.HE]}
  \BibitemShut {NoStop}%
\bibitem [{\citenamefont {Aasi}\ \emph {et~al.}(2015)\citenamefont {Aasi} \emph
  {et~al.}}]{TheLIGOScientific:2014jea}%
  \BibitemOpen
  \bibfield  {author} {\bibinfo {author} {\bibfnamefont {J.}~\bibnamefont
  {Aasi}} \emph {et~al.} (\bibinfo {collaboration} {LIGO Scientific}),\
  }\bibfield  {title} {\bibinfo {title} {{Advanced LIGO}},\ }\href
  {https://doi.org/10.1088/0264-9381/32/7/074001} {\bibfield  {journal}
  {\bibinfo  {journal} {Class. Quant. Grav.}\ }\textbf {\bibinfo {volume}
  {32}},\ \bibinfo {pages} {074001} (\bibinfo {year} {2015})},\ \Eprint
  {https://arxiv.org/abs/1411.4547} {arXiv:1411.4547 [gr-qc]} \BibitemShut
  {NoStop}%
\bibitem [{\citenamefont {Acernese}\ \emph {et~al.}(2015)\citenamefont
  {Acernese} \emph {et~al.}}]{TheVirgo:2014hva}%
  \BibitemOpen
  \bibfield  {author} {\bibinfo {author} {\bibfnamefont {F.}~\bibnamefont
  {Acernese}} \emph {et~al.} (\bibinfo {collaboration} {Virgo Collaboration}),\
  }\bibfield  {title} {\bibinfo {title} {{Advanced Virgo: a second-generation
  interferometric gravitational wave detector}},\ }\href
  {https://doi.org/10.1088/0264-9381/32/2/024001} {\bibfield  {journal}
  {\bibinfo  {journal} {Class. Quant. Grav.}\ }\textbf {\bibinfo {volume}
  {32}},\ \bibinfo {pages} {024001} (\bibinfo {year} {2015})},\ \Eprint
  {https://arxiv.org/abs/1408.3978} {arXiv:1408.3978 [gr-qc]} \BibitemShut
  {NoStop}%
\bibitem [{\citenamefont {Abbott}\ \emph {et~al.}(2013)\citenamefont {Abbott}
  \emph {et~al.}}]{Aasi:2013wya}%
  \BibitemOpen
  \bibfield  {author} {\bibinfo {author} {\bibfnamefont {B.~P.}\ \bibnamefont
  {Abbott}} \emph {et~al.} (\bibinfo {collaboration} {LIGO Scientific
  Collaboration, Virgo Collaboration}),\ }\bibfield  {title} {\bibinfo {title}
  {{Prospects for Observing and Localizing Gravitational-Wave Transients with
  Advanced LIGO, Advanced Virgo and KAGRA}},\ }\href
  {https://doi.org/10.1007/lrr-2016-1} {\bibfield  {journal} {\bibinfo
  {journal} {Living Rev. Rel.}\ }\textbf {\bibinfo {volume} {19}},\ \bibinfo
  {pages} {1} (\bibinfo {year} {2013})},\ \Eprint
  {https://arxiv.org/abs/1304.0670} {arXiv:1304.0670 [gr-qc]} \BibitemShut
  {NoStop}%
\bibitem [{\citenamefont {Antoniadis}\ \emph {et~al.}(2013)\citenamefont
  {Antoniadis}, \citenamefont {Freire}, \citenamefont {Wex}, \citenamefont
  {Tauris}, \citenamefont {Lynch} \emph {et~al.}}]{Antoniadis:2013pzd}%
  \BibitemOpen
  \bibfield  {author} {\bibinfo {author} {\bibfnamefont {J.}~\bibnamefont
  {Antoniadis}}, \bibinfo {author} {\bibfnamefont {P.~C.}\ \bibnamefont
  {Freire}}, \bibinfo {author} {\bibfnamefont {N.}~\bibnamefont {Wex}},
  \bibinfo {author} {\bibfnamefont {T.~M.}\ \bibnamefont {Tauris}}, \bibinfo
  {author} {\bibfnamefont {R.~S.}\ \bibnamefont {Lynch}}, \emph {et~al.},\
  }\bibfield  {title} {\bibinfo {title} {{A Massive Pulsar in a Compact
  Relativistic Binary}},\ }\href {https://doi.org/10.1126/science.1233232}
  {\bibfield  {journal} {\bibinfo  {journal} {Science}\ }\textbf {\bibinfo
  {volume} {340}},\ \bibinfo {pages} {1233232} (\bibinfo {year} {2013})},\
  \Eprint {https://arxiv.org/abs/1304.6875} {arXiv:1304.6875 [astro-ph.HE]}
  \BibitemShut {NoStop}%
\bibitem [{\citenamefont {Cromartie}\ \emph {et~al.}(2019)\citenamefont
  {Cromartie} \emph {et~al.}}]{Cromartie:2019kug}%
  \BibitemOpen
  \bibfield  {author} {\bibinfo {author} {\bibfnamefont {H.~T.}\ \bibnamefont
  {Cromartie}} \emph {et~al.},\ }\bibfield  {title} {\bibinfo {title}
  {{Relativistic Shapiro delay measurements of an extremely massive millisecond
  pulsar}},\ }\href {https://doi.org/10.1038/s41550-019-0880-2} {\bibfield
  {journal} {\bibinfo  {journal} {Nature Astron.}\ }\textbf {\bibinfo {volume}
  {4}},\ \bibinfo {pages} {72} (\bibinfo {year} {2019})},\ \Eprint
  {https://arxiv.org/abs/1904.06759} {arXiv:1904.06759} \BibitemShut {NoStop}%
\bibitem [{\citenamefont {Fonseca}\ \emph {et~al.}(2021)\citenamefont {Fonseca}
  \emph {et~al.}}]{Fonseca:2021wxt}%
  \BibitemOpen
  \bibfield  {author} {\bibinfo {author} {\bibfnamefont {E.}~\bibnamefont
  {Fonseca}} \emph {et~al.},\ }\bibfield  {title} {\bibinfo {title} {{Refined
  Mass and Geometric Measurements of the High-mass PSR J0740+6620}},\ }\href
  {https://doi.org/10.3847/2041-8213/ac03b8} {\bibfield  {journal} {\bibinfo
  {journal} {Astrophys. J. Lett.}\ }\textbf {\bibinfo {volume} {915}},\
  \bibinfo {pages} {L12} (\bibinfo {year} {2021})},\ \Eprint
  {https://arxiv.org/abs/2104.00880} {arXiv:2104.00880 [astro-ph.HE]}
  \BibitemShut {NoStop}%
\bibitem [{\citenamefont {Miller}\ \emph {et~al.}(2019)\citenamefont {Miller}
  \emph {et~al.}}]{Miller:2019cac}%
  \BibitemOpen
  \bibfield  {author} {\bibinfo {author} {\bibfnamefont {M.~C.}\ \bibnamefont
  {Miller}} \emph {et~al.},\ }\bibfield  {title} {\bibinfo {title} {{PSR
  J0030+0451 Mass and Radius from $NICER$ Data and Implications for the
  Properties of Neutron Star Matter}},\ }\href
  {https://doi.org/10.3847/2041-8213/ab50c5} {\bibfield  {journal} {\bibinfo
  {journal} {Astrophys. J. Lett.}\ }\textbf {\bibinfo {volume} {887}},\
  \bibinfo {pages} {L24} (\bibinfo {year} {2019})},\ \Eprint
  {https://arxiv.org/abs/1912.05705} {arXiv:1912.05705 [astro-ph.HE]}
  \BibitemShut {NoStop}%
\bibitem [{\citenamefont {Riley}\ \emph {et~al.}(2019)\citenamefont {Riley}
  \emph {et~al.}}]{Riley:2019yda}%
  \BibitemOpen
  \bibfield  {author} {\bibinfo {author} {\bibfnamefont {T.~E.}\ \bibnamefont
  {Riley}} \emph {et~al.},\ }\bibfield  {title} {\bibinfo {title} {{A $NICER$
  View of PSR J0030+0451: Millisecond Pulsar Parameter Estimation}},\ }\href
  {https://doi.org/10.3847/2041-8213/ab481c} {\bibfield  {journal} {\bibinfo
  {journal} {Astrophys. J. Lett.}\ }\textbf {\bibinfo {volume} {887}},\
  \bibinfo {pages} {L21} (\bibinfo {year} {2019})},\ \Eprint
  {https://arxiv.org/abs/1912.05702} {arXiv:1912.05702 [astro-ph.HE]}
  \BibitemShut {NoStop}%
\bibitem [{\citenamefont {Miller}\ \emph {et~al.}(2021)\citenamefont {Miller}
  \emph {et~al.}}]{Miller:2021qha}%
  \BibitemOpen
  \bibfield  {author} {\bibinfo {author} {\bibfnamefont {M.~C.}\ \bibnamefont
  {Miller}} \emph {et~al.},\ }\bibfield  {title} {\bibinfo {title} {{The Radius
  of PSR J0740+6620 from NICER and XMM-Newton Data}},\ }\href@noop {} {\
  (\bibinfo {year} {2021})},\ \Eprint {https://arxiv.org/abs/2105.06979}
  {arXiv:2105.06979 [astro-ph.HE]} \BibitemShut {NoStop}%
\bibitem [{\citenamefont {Riley}\ \emph {et~al.}(2021)\citenamefont {Riley}
  \emph {et~al.}}]{Riley:2021pdl}%
  \BibitemOpen
  \bibfield  {author} {\bibinfo {author} {\bibfnamefont {T.~E.}\ \bibnamefont
  {Riley}} \emph {et~al.},\ }\bibfield  {title} {\bibinfo {title} {{A NICER
  View of the Massive Pulsar PSR J0740+6620 Informed by Radio Timing and
  XMM-Newton Spectroscopy}},\ }\href@noop {} {\  (\bibinfo {year} {2021})},\
  \Eprint {https://arxiv.org/abs/2105.06980} {arXiv:2105.06980 [astro-ph.HE]}
  \BibitemShut {NoStop}%
\bibitem [{\citenamefont {Raaijmakers}\ \emph {et~al.}(2020)\citenamefont
  {Raaijmakers} \emph {et~al.}}]{Raaijmakers:2019dks}%
  \BibitemOpen
  \bibfield  {author} {\bibinfo {author} {\bibfnamefont {G.}~\bibnamefont
  {Raaijmakers}} \emph {et~al.},\ }\bibfield  {title} {\bibinfo {title}
  {{Constraining the dense matter equation of state with joint analysis of
  NICER and LIGO/Virgo measurements}},\ }\href
  {https://doi.org/10.3847/2041-8213/ab822f} {\bibfield  {journal} {\bibinfo
  {journal} {Astrophys. J. Lett.}\ }\textbf {\bibinfo {volume} {893}},\
  \bibinfo {pages} {L21} (\bibinfo {year} {2020})},\ \Eprint
  {https://arxiv.org/abs/1912.11031} {arXiv:1912.11031 [astro-ph.HE]}
  \BibitemShut {NoStop}%
\bibitem [{\citenamefont {Dietrich}\ \emph {et~al.}(2020)\citenamefont
  {Dietrich}, \citenamefont {Coughlin}, \citenamefont {Pang}, \citenamefont
  {Bulla}, \citenamefont {Heinzel}, \citenamefont {Issa}, \citenamefont
  {Tews},\ and\ \citenamefont {Antier}}]{Dietrich:2020efo}%
  \BibitemOpen
  \bibfield  {author} {\bibinfo {author} {\bibfnamefont {T.}~\bibnamefont
  {Dietrich}}, \bibinfo {author} {\bibfnamefont {M.~W.}\ \bibnamefont
  {Coughlin}}, \bibinfo {author} {\bibfnamefont {P.~T.~H.}\ \bibnamefont
  {Pang}}, \bibinfo {author} {\bibfnamefont {M.}~\bibnamefont {Bulla}},
  \bibinfo {author} {\bibfnamefont {J.}~\bibnamefont {Heinzel}}, \bibinfo
  {author} {\bibfnamefont {L.}~\bibnamefont {Issa}}, \bibinfo {author}
  {\bibfnamefont {I.}~\bibnamefont {Tews}},\ and\ \bibinfo {author}
  {\bibfnamefont {S.}~\bibnamefont {Antier}},\ }\bibfield  {title} {\bibinfo
  {title} {{Multimessenger constraints on the neutron-star equation of state
  and the Hubble constant}},\ }\href {https://doi.org/10.1126/science.abb4317}
  {\bibfield  {journal} {\bibinfo  {journal} {Science}\ }\textbf {\bibinfo
  {volume} {370}},\ \bibinfo {pages} {1450} (\bibinfo {year} {2020})},\ \Eprint
  {https://arxiv.org/abs/2002.11355} {arXiv:2002.11355 [astro-ph.HE]}
  \BibitemShut {NoStop}%
\bibitem [{\citenamefont {Landry}\ \emph {et~al.}(2020)\citenamefont {Landry},
  \citenamefont {Essick},\ and\ \citenamefont
  {Chatziioannou}}]{Landry:2020vaw}%
  \BibitemOpen
  \bibfield  {author} {\bibinfo {author} {\bibfnamefont {P.}~\bibnamefont
  {Landry}}, \bibinfo {author} {\bibfnamefont {R.}~\bibnamefont {Essick}},\
  and\ \bibinfo {author} {\bibfnamefont {K.}~\bibnamefont {Chatziioannou}},\
  }\bibfield  {title} {\bibinfo {title} {{Nonparametric constraints on neutron
  star matter with existing and upcoming gravitational wave and pulsar
  observations}},\ }\href {https://doi.org/10.1103/PhysRevD.101.123007}
  {\bibfield  {journal} {\bibinfo  {journal} {Phys. Rev. D}\ }\textbf {\bibinfo
  {volume} {101}},\ \bibinfo {pages} {123007} (\bibinfo {year} {2020})},\
  \Eprint {https://arxiv.org/abs/2003.04880} {arXiv:2003.04880 [astro-ph.HE]}
  \BibitemShut {NoStop}%
\bibitem [{\citenamefont {Al-Mamun}\ \emph {et~al.}(2021)\citenamefont
  {Al-Mamun}, \citenamefont {Steiner}, \citenamefont {N\"attil\"a},
  \citenamefont {Lange}, \citenamefont {O'Shaughnessy}, \citenamefont {Tews},
  \citenamefont {Gandolfi}, \citenamefont {Heinke},\ and\ \citenamefont
  {Han}}]{Al-Mamun:2020vzu}%
  \BibitemOpen
  \bibfield  {author} {\bibinfo {author} {\bibfnamefont {M.}~\bibnamefont
  {Al-Mamun}}, \bibinfo {author} {\bibfnamefont {A.~W.}\ \bibnamefont
  {Steiner}}, \bibinfo {author} {\bibfnamefont {J.}~\bibnamefont
  {N\"attil\"a}}, \bibinfo {author} {\bibfnamefont {J.}~\bibnamefont {Lange}},
  \bibinfo {author} {\bibfnamefont {R.}~\bibnamefont {O'Shaughnessy}}, \bibinfo
  {author} {\bibfnamefont {I.}~\bibnamefont {Tews}}, \bibinfo {author}
  {\bibfnamefont {S.}~\bibnamefont {Gandolfi}}, \bibinfo {author}
  {\bibfnamefont {C.}~\bibnamefont {Heinke}},\ and\ \bibinfo {author}
  {\bibfnamefont {S.}~\bibnamefont {Han}},\ }\bibfield  {title} {\bibinfo
  {title} {{Combining Electromagnetic and Gravitational-Wave Constraints on
  Neutron-Star Masses and Radii}},\ }\href
  {https://doi.org/10.1103/PhysRevLett.126.061101} {\bibfield  {journal}
  {\bibinfo  {journal} {Phys. Rev. Lett.}\ }\textbf {\bibinfo {volume} {126}},\
  \bibinfo {pages} {061101} (\bibinfo {year} {2021})},\ \Eprint
  {https://arxiv.org/abs/2008.12817} {arXiv:2008.12817 [astro-ph.HE]}
  \BibitemShut {NoStop}%
\bibitem [{\citenamefont {Reed}\ \emph {et~al.}(2021)\citenamefont {Reed},
  \citenamefont {Fattoyev}, \citenamefont {Horowitz},\ and\ \citenamefont
  {Piekarewicz}}]{Reed:2021nqk}%
  \BibitemOpen
  \bibfield  {author} {\bibinfo {author} {\bibfnamefont {B.~T.}\ \bibnamefont
  {Reed}}, \bibinfo {author} {\bibfnamefont {F.~J.}\ \bibnamefont {Fattoyev}},
  \bibinfo {author} {\bibfnamefont {C.~J.}\ \bibnamefont {Horowitz}},\ and\
  \bibinfo {author} {\bibfnamefont {J.}~\bibnamefont {Piekarewicz}},\
  }\bibfield  {title} {\bibinfo {title} {{Implications of PREX-2 on the
  Equation of State of Neutron-Rich Matter}},\ }\href
  {https://doi.org/10.1103/PhysRevLett.126.172503} {\bibfield  {journal}
  {\bibinfo  {journal} {Phys. Rev. Lett.}\ }\textbf {\bibinfo {volume} {126}},\
  \bibinfo {pages} {172503} (\bibinfo {year} {2021})},\ \Eprint
  {https://arxiv.org/abs/2101.03193} {arXiv:2101.03193 [nucl-th]} \BibitemShut
  {NoStop}%
\bibitem [{\citenamefont {Essick}\ \emph {et~al.}(2020)\citenamefont {Essick},
  \citenamefont {Tews}, \citenamefont {Landry}, \citenamefont {Reddy},\ and\
  \citenamefont {Holz}}]{Essick:2020flb}%
  \BibitemOpen
  \bibfield  {author} {\bibinfo {author} {\bibfnamefont {R.}~\bibnamefont
  {Essick}}, \bibinfo {author} {\bibfnamefont {I.}~\bibnamefont {Tews}},
  \bibinfo {author} {\bibfnamefont {P.}~\bibnamefont {Landry}}, \bibinfo
  {author} {\bibfnamefont {S.}~\bibnamefont {Reddy}},\ and\ \bibinfo {author}
  {\bibfnamefont {D.~E.}\ \bibnamefont {Holz}},\ }\bibfield  {title} {\bibinfo
  {title} {{Direct Astrophysical Tests of Chiral Effective Field Theory at
  Supranuclear Densities}},\ }\href
  {https://doi.org/10.1103/PhysRevC.102.055803} {\bibfield  {journal} {\bibinfo
   {journal} {Phys. Rev. C}\ }\textbf {\bibinfo {volume} {102}},\ \bibinfo
  {pages} {055803} (\bibinfo {year} {2020})},\ \Eprint
  {https://arxiv.org/abs/2004.07744} {arXiv:2004.07744 [astro-ph.HE]}
  \BibitemShut {NoStop}%
\bibitem [{\citenamefont {Raaijmakers}\ \emph {et~al.}(2021)\citenamefont
  {Raaijmakers}, \citenamefont {Greif}, \citenamefont {Hebeler}, \citenamefont
  {Hinderer}, \citenamefont {Nissanke}, \citenamefont {Schwenk}, \citenamefont
  {Riley}, \citenamefont {Watts}, \citenamefont {Lattimer},\ and\ \citenamefont
  {Ho}}]{Raaijmakers:2021uju}%
  \BibitemOpen
  \bibfield  {author} {\bibinfo {author} {\bibfnamefont {G.}~\bibnamefont
  {Raaijmakers}}, \bibinfo {author} {\bibfnamefont {S.~K.}\ \bibnamefont
  {Greif}}, \bibinfo {author} {\bibfnamefont {K.}~\bibnamefont {Hebeler}},
  \bibinfo {author} {\bibfnamefont {T.}~\bibnamefont {Hinderer}}, \bibinfo
  {author} {\bibfnamefont {S.}~\bibnamefont {Nissanke}}, \bibinfo {author}
  {\bibfnamefont {A.}~\bibnamefont {Schwenk}}, \bibinfo {author} {\bibfnamefont
  {T.~E.}\ \bibnamefont {Riley}}, \bibinfo {author} {\bibfnamefont {A.~L.}\
  \bibnamefont {Watts}}, \bibinfo {author} {\bibfnamefont {J.~M.}\ \bibnamefont
  {Lattimer}},\ and\ \bibinfo {author} {\bibfnamefont {W.~C.~G.}\ \bibnamefont
  {Ho}},\ }\bibfield  {title} {\bibinfo {title} {{Constraints on the dense
  matter equation of state and neutron star properties from NICER's mass-radius
  estimate of PSR J0740+6620 and multimessenger observations}},\ }\href@noop {}
  {\  (\bibinfo {year} {2021})},\ \Eprint {https://arxiv.org/abs/2105.06981}
  {arXiv:2105.06981 [astro-ph.HE]} \BibitemShut {NoStop}%
\bibitem [{\citenamefont {Biswas}(2021)}]{Biswas:2021yge}%
  \BibitemOpen
  \bibfield  {author} {\bibinfo {author} {\bibfnamefont {B.}~\bibnamefont
  {Biswas}},\ }\bibfield  {title} {\bibinfo {title} {{Impact of PREX-II, the
  revised mass measurement of PSRJ0740+6620, and possible NICER observation on
  the dense matter equation of state}},\ }\href@noop {} {\  (\bibinfo {year}
  {2021})},\ \Eprint {https://arxiv.org/abs/2105.02886} {arXiv:2105.02886
  [astro-ph.HE]} \BibitemShut {NoStop}%
\bibitem [{\citenamefont {Pang}\ \emph {et~al.}(2021)\citenamefont {Pang},
  \citenamefont {Tews}, \citenamefont {Coughlin}, \citenamefont {Bulla},
  \citenamefont {Van Den~Broeck},\ and\ \citenamefont
  {Dietrich}}]{Pang:2021jta}%
  \BibitemOpen
  \bibfield  {author} {\bibinfo {author} {\bibfnamefont {P.~T.~H.}\
  \bibnamefont {Pang}}, \bibinfo {author} {\bibfnamefont {I.}~\bibnamefont
  {Tews}}, \bibinfo {author} {\bibfnamefont {M.~W.}\ \bibnamefont {Coughlin}},
  \bibinfo {author} {\bibfnamefont {M.}~\bibnamefont {Bulla}}, \bibinfo
  {author} {\bibfnamefont {C.}~\bibnamefont {Van Den~Broeck}},\ and\ \bibinfo
  {author} {\bibfnamefont {T.}~\bibnamefont {Dietrich}},\ }\bibfield  {title}
  {\bibinfo {title} {{Nuclear-Physics Multi-Messenger Astrophysics Constraints
  on the Neutron-Star Equation of State: Adding NICER's PSR J0740+6620
  Measurement}},\ }\href@noop {} {\  (\bibinfo {year} {2021})},\ \Eprint
  {https://arxiv.org/abs/2105.08688} {arXiv:2105.08688 [astro-ph.HE]}
  \BibitemShut {NoStop}%
\bibitem [{\citenamefont {Legred}\ \emph {et~al.}(2021)\citenamefont {Legred},
  \citenamefont {Chatziioannou}, \citenamefont {Essick}, \citenamefont {Han},\
  and\ \citenamefont {Landry}}]{Legred:2021hdx}%
  \BibitemOpen
  \bibfield  {author} {\bibinfo {author} {\bibfnamefont {I.}~\bibnamefont
  {Legred}}, \bibinfo {author} {\bibfnamefont {K.}~\bibnamefont
  {Chatziioannou}}, \bibinfo {author} {\bibfnamefont {R.}~\bibnamefont
  {Essick}}, \bibinfo {author} {\bibfnamefont {S.}~\bibnamefont {Han}},\ and\
  \bibinfo {author} {\bibfnamefont {P.}~\bibnamefont {Landry}},\ }\bibfield
  {title} {\bibinfo {title} {{Impact of the PSR J0740+6620 radius constraint on
  the properties of high-density matter}},\ }\href
  {https://doi.org/10.1103/PhysRevD.104.063003} {\bibfield  {journal} {\bibinfo
   {journal} {Phys. Rev. D}\ }\textbf {\bibinfo {volume} {104}},\ \bibinfo
  {pages} {063003} (\bibinfo {year} {2021})},\ \Eprint
  {https://arxiv.org/abs/2106.05313} {arXiv:2106.05313 [astro-ph.HE]}
  \BibitemShut {NoStop}%
\bibitem [{\citenamefont {Essick}\ \emph
  {et~al.}(2021{\natexlab{a}})\citenamefont {Essick}, \citenamefont {Tews},
  \citenamefont {Landry},\ and\ \citenamefont {Schwenk}}]{Essick:2021kjb}%
  \BibitemOpen
  \bibfield  {author} {\bibinfo {author} {\bibfnamefont {R.}~\bibnamefont
  {Essick}}, \bibinfo {author} {\bibfnamefont {I.}~\bibnamefont {Tews}},
  \bibinfo {author} {\bibfnamefont {P.}~\bibnamefont {Landry}},\ and\ \bibinfo
  {author} {\bibfnamefont {A.}~\bibnamefont {Schwenk}},\ }\bibfield  {title}
  {\bibinfo {title} {{Astrophysical Constraints on the Symmetry Energy and the
  Neutron Skin of $^{208}$Pb with Minimal Modeling Assumptions}},\ }\href@noop
  {} {\  (\bibinfo {year} {2021}{\natexlab{a}})},\ \Eprint
  {https://arxiv.org/abs/2102.10074} {arXiv:2102.10074 [nucl-th]} \BibitemShut
  {NoStop}%
\bibitem [{\citenamefont {Essick}\ \emph
  {et~al.}(2021{\natexlab{b}})\citenamefont {Essick}, \citenamefont {Landry},
  \citenamefont {Schwenk},\ and\ \citenamefont {Tews}}]{Essick:2021ezp}%
  \BibitemOpen
  \bibfield  {author} {\bibinfo {author} {\bibfnamefont {R.}~\bibnamefont
  {Essick}}, \bibinfo {author} {\bibfnamefont {P.}~\bibnamefont {Landry}},
  \bibinfo {author} {\bibfnamefont {A.}~\bibnamefont {Schwenk}},\ and\ \bibinfo
  {author} {\bibfnamefont {I.}~\bibnamefont {Tews}},\ }\bibfield  {title}
  {\bibinfo {title} {{A Detailed Examination of Astrophysical Constraints on
  the Symmetry Energy and the Neutron Skin of $^{208}$Pb with Minimal Modeling
  Assumptions}},\ }\href@noop {} {\  (\bibinfo {year} {2021}{\natexlab{b}})},\
  \Eprint {https://arxiv.org/abs/2107.05528} {arXiv:2107.05528 [nucl-th]}
  \BibitemShut {NoStop}%
\bibitem [{\citenamefont {Blanchet}(2014)}]{Blanchet:2013haa}%
  \BibitemOpen
  \bibfield  {author} {\bibinfo {author} {\bibfnamefont {L.}~\bibnamefont
  {Blanchet}},\ }\bibfield  {title} {\bibinfo {title} {{Gravitational Radiation
  from Post-Newtonian Sources and Inspiralling Compact Binaries}},\ }\href
  {https://doi.org/10.12942/lrr-2014-2} {\bibfield  {journal} {\bibinfo
  {journal} {Living Rev. Rel.}\ }\textbf {\bibinfo {volume} {17}},\ \bibinfo
  {pages} {2} (\bibinfo {year} {2014})},\ \Eprint
  {https://arxiv.org/abs/1310.1528} {arXiv:1310.1528 [gr-qc]} \BibitemShut
  {NoStop}%
\bibitem [{\citenamefont {Abbott}\ \emph
  {et~al.}(2019{\natexlab{a}})\citenamefont {Abbott} \emph
  {et~al.}}]{Abbott:2018wiz}%
  \BibitemOpen
  \bibfield  {author} {\bibinfo {author} {\bibfnamefont {B.}~\bibnamefont
  {Abbott}} \emph {et~al.} (\bibinfo {collaboration} {LIGO Scientific,
  Virgo}),\ }\bibfield  {title} {\bibinfo {title} {{Properties of the binary
  neutron star merger GW170817}},\ }\href
  {https://doi.org/10.1103/PhysRevX.9.011001} {\bibfield  {journal} {\bibinfo
  {journal} {Phys. Rev. X}\ }\textbf {\bibinfo {volume} {9}},\ \bibinfo {pages}
  {011001} (\bibinfo {year} {2019}{\natexlab{a}})},\ \Eprint
  {https://arxiv.org/abs/1805.11579} {arXiv:1805.11579 [gr-qc]} \BibitemShut
  {NoStop}%
\bibitem [{\citenamefont {Torres-Rivas}\ \emph {et~al.}(2019)\citenamefont
  {Torres-Rivas}, \citenamefont {Chatziioannou}, \citenamefont {Bauswein},\
  and\ \citenamefont {Clark}}]{Torres-Rivas:2018svp}%
  \BibitemOpen
  \bibfield  {author} {\bibinfo {author} {\bibfnamefont {A.}~\bibnamefont
  {Torres-Rivas}}, \bibinfo {author} {\bibfnamefont {K.}~\bibnamefont
  {Chatziioannou}}, \bibinfo {author} {\bibfnamefont {A.}~\bibnamefont
  {Bauswein}},\ and\ \bibinfo {author} {\bibfnamefont {J.~A.}\ \bibnamefont
  {Clark}},\ }\bibfield  {title} {\bibinfo {title} {{Observing the post-merger
  signal of GW170817-like events with improved gravitational-wave detectors}},\
  }\href {https://doi.org/10.1103/PhysRevD.99.044014} {\bibfield  {journal}
  {\bibinfo  {journal} {Phys. Rev. D}\ }\textbf {\bibinfo {volume} {99}},\
  \bibinfo {pages} {044014} (\bibinfo {year} {2019})},\ \Eprint
  {https://arxiv.org/abs/1811.08931} {arXiv:1811.08931 [gr-qc]} \BibitemShut
  {NoStop}%
\bibitem [{\citenamefont {Chatziioannou}(2020)}]{Chatziioannou:2020pqz}%
  \BibitemOpen
  \bibfield  {author} {\bibinfo {author} {\bibfnamefont {K.}~\bibnamefont
  {Chatziioannou}},\ }\bibfield  {title} {\bibinfo {title} {{Neutron star tidal
  deformability and equation of state constraints}},\ }\href
  {https://doi.org/10.1007/s10714-020-02754-3} {\bibfield  {journal} {\bibinfo
  {journal} {Gen. Rel. Grav.}\ }\textbf {\bibinfo {volume} {52}},\ \bibinfo
  {pages} {109} (\bibinfo {year} {2020})},\ \Eprint
  {https://arxiv.org/abs/2006.03168} {arXiv:2006.03168 [gr-qc]} \BibitemShut
  {NoStop}%
\bibitem [{\citenamefont {Dietrich}\ \emph {et~al.}(2021)\citenamefont
  {Dietrich}, \citenamefont {Hinderer},\ and\ \citenamefont
  {Samajdar}}]{Dietrich:2020eud}%
  \BibitemOpen
  \bibfield  {author} {\bibinfo {author} {\bibfnamefont {T.}~\bibnamefont
  {Dietrich}}, \bibinfo {author} {\bibfnamefont {T.}~\bibnamefont {Hinderer}},\
  and\ \bibinfo {author} {\bibfnamefont {A.}~\bibnamefont {Samajdar}},\
  }\bibfield  {title} {\bibinfo {title} {{Interpreting Binary Neutron Star
  Mergers: Describing the Binary Neutron Star Dynamics, Modelling Gravitational
  Waveforms, and Analyzing Detections}},\ }\href
  {https://doi.org/10.1007/s10714-020-02751-6} {\bibfield  {journal} {\bibinfo
  {journal} {Gen. Rel. Grav.}\ }\textbf {\bibinfo {volume} {53}},\ \bibinfo
  {pages} {27} (\bibinfo {year} {2021})},\ \Eprint
  {https://arxiv.org/abs/2004.02527} {arXiv:2004.02527 [gr-qc]} \BibitemShut
  {NoStop}%
\bibitem [{\citenamefont {{Bauswein}}\ and\ \citenamefont
  {{Stergioulas}}(2019)}]{2019JPhG...46k3002B}%
  \BibitemOpen
  \bibfield  {author} {\bibinfo {author} {\bibfnamefont {A.}~\bibnamefont
  {{Bauswein}}}\ and\ \bibinfo {author} {\bibfnamefont {N.}~\bibnamefont
  {{Stergioulas}}},\ }\bibfield  {title} {\bibinfo {title} {{Spectral
  classification of gravitational-wave emission and equation of state
  constraints in binary neutron star mergers}},\ }\href
  {https://doi.org/10.1088/1361-6471/ab2b90} {\bibfield  {journal} {\bibinfo
  {journal} {Journal of Physics G Nuclear Physics}\ }\textbf {\bibinfo {volume}
  {46}},\ \bibinfo {eid} {113002} (\bibinfo {year} {2019})},\ \Eprint
  {https://arxiv.org/abs/1901.06969} {arXiv:1901.06969 [gr-qc]} \BibitemShut
  {NoStop}%
\bibitem [{\citenamefont {{Baiotti}}(2019)}]{2019PrPNP.10903714B}%
  \BibitemOpen
  \bibfield  {author} {\bibinfo {author} {\bibfnamefont {L.}~\bibnamefont
  {{Baiotti}}},\ }\bibfield  {title} {\bibinfo {title} {{Gravitational waves
  from neutron star mergers and their relation to the nuclear equation of
  state}},\ }\href {https://doi.org/10.1016/j.ppnp.2019.103714} {\bibfield
  {journal} {\bibinfo  {journal} {Progress in Particle and Nuclear Physics}\
  }\textbf {\bibinfo {volume} {109}},\ \bibinfo {eid} {103714} (\bibinfo {year}
  {2019})},\ \Eprint {https://arxiv.org/abs/1907.08534} {arXiv:1907.08534
  [astro-ph.HE]} \BibitemShut {NoStop}%
\bibitem [{\citenamefont {{Sarin}}\ and\ \citenamefont
  {{Lasky}}(2021)}]{2021GReGr..53...59S}%
  \BibitemOpen
  \bibfield  {author} {\bibinfo {author} {\bibfnamefont {N.}~\bibnamefont
  {{Sarin}}}\ and\ \bibinfo {author} {\bibfnamefont {P.~D.}\ \bibnamefont
  {{Lasky}}},\ }\bibfield  {title} {\bibinfo {title} {{The evolution of binary
  neutron star post-merger remnants: a review}},\ }\href
  {https://doi.org/10.1007/s10714-021-02831-1} {\bibfield  {journal} {\bibinfo
  {journal} {General Relativity and Gravitation}\ }\textbf {\bibinfo {volume}
  {53}},\ \bibinfo {eid} {59} (\bibinfo {year} {2021})},\ \Eprint
  {https://arxiv.org/abs/2012.08172} {arXiv:2012.08172 [astro-ph.HE]}
  \BibitemShut {NoStop}%
\bibitem [{\citenamefont {Bernuzzi}(2020)}]{Bernuzzi:2020tgt}%
  \BibitemOpen
  \bibfield  {author} {\bibinfo {author} {\bibfnamefont {S.}~\bibnamefont
  {Bernuzzi}},\ }\bibfield  {title} {\bibinfo {title} {{Neutron Star Merger
  Remnants}},\ }\href {https://doi.org/10.1007/s10714-020-02752-5} {\bibfield
  {journal} {\bibinfo  {journal} {Gen. Rel. Grav.}\ }\textbf {\bibinfo {volume}
  {52}},\ \bibinfo {pages} {108} (\bibinfo {year} {2020})},\ \Eprint
  {https://arxiv.org/abs/2004.06419} {arXiv:2004.06419 [astro-ph.HE]}
  \BibitemShut {NoStop}%
\bibitem [{\citenamefont {Baumgarte}\ \emph {et~al.}(2000)\citenamefont
  {Baumgarte}, \citenamefont {Shapiro},\ and\ \citenamefont
  {Shibata}}]{Baumgarte:1999cq}%
  \BibitemOpen
  \bibfield  {author} {\bibinfo {author} {\bibfnamefont {T.~W.}\ \bibnamefont
  {Baumgarte}}, \bibinfo {author} {\bibfnamefont {S.~L.}\ \bibnamefont
  {Shapiro}},\ and\ \bibinfo {author} {\bibfnamefont {M.}~\bibnamefont
  {Shibata}},\ }\bibfield  {title} {\bibinfo {title} {{On the maximum mass of
  differentially rotating neutron stars}},\ }\href
  {https://doi.org/10.1086/312425} {\bibfield  {journal} {\bibinfo  {journal}
  {Astrophys. J. Lett.}\ }\textbf {\bibinfo {volume} {528}},\ \bibinfo {pages}
  {L29} (\bibinfo {year} {2000})},\ \Eprint
  {https://arxiv.org/abs/astro-ph/9910565} {arXiv:astro-ph/9910565}
  \BibitemShut {NoStop}%
\bibitem [{\citenamefont {Xing}\ \emph {et~al.}(1994)\citenamefont {Xing},
  \citenamefont {Centrella},\ and\ \citenamefont {McMillan}}]{Xing:1994ak}%
  \BibitemOpen
  \bibfield  {author} {\bibinfo {author} {\bibfnamefont {Z.-G.}\ \bibnamefont
  {Xing}}, \bibinfo {author} {\bibfnamefont {J.~M.}\ \bibnamefont
  {Centrella}},\ and\ \bibinfo {author} {\bibfnamefont {S.~L.~W.}\ \bibnamefont
  {McMillan}},\ }\bibfield  {title} {\bibinfo {title} {{Gravitational radiation
  from coalescing binary neutron stars}},\ }\href
  {https://doi.org/10.1103/PhysRevD.50.6247} {\bibfield  {journal} {\bibinfo
  {journal} {Phys. Rev.}\ }\textbf {\bibinfo {volume} {D50}},\ \bibinfo {pages}
  {6247} (\bibinfo {year} {1994})},\ \Eprint
  {https://arxiv.org/abs/gr-qc/9411029} {arXiv:gr-qc/9411029 [gr-qc]}
  \BibitemShut {NoStop}%
\bibitem [{\citenamefont {Ruffert}\ \emph {et~al.}(2001)\citenamefont
  {Ruffert}, \citenamefont {Ruffert},\ and\ \citenamefont
  {Janka}}]{Ruffert:2001gf}%
  \BibitemOpen
  \bibfield  {author} {\bibinfo {author} {\bibfnamefont {M.}~\bibnamefont
  {Ruffert}}, \bibinfo {author} {\bibfnamefont {H.~T.}\ \bibnamefont
  {Ruffert}},\ and\ \bibinfo {author} {\bibfnamefont {H.~T.}\ \bibnamefont
  {Janka}},\ }\bibfield  {title} {\bibinfo {title} {{Coalescing neutron stars -
  a step towards physical models. 3. Improved numerics and different neutron
  star masses and spins}},\ }\href {https://doi.org/10.1051/0004-6361:20011453}
  {\bibfield  {journal} {\bibinfo  {journal} {Astron. Astrophys.}\ }\textbf
  {\bibinfo {volume} {380}},\ \bibinfo {pages} {544} (\bibinfo {year}
  {2001})},\ \Eprint {https://arxiv.org/abs/astro-ph/0106229}
  {arXiv:astro-ph/0106229} \BibitemShut {NoStop}%
\bibitem [{\citenamefont {Shibata}\ \emph {et~al.}(2005)\citenamefont
  {Shibata}, \citenamefont {Taniguchi},\ and\ \citenamefont
  {Uryu}}]{Shibata:2005ss}%
  \BibitemOpen
  \bibfield  {author} {\bibinfo {author} {\bibfnamefont {M.}~\bibnamefont
  {Shibata}}, \bibinfo {author} {\bibfnamefont {K.}~\bibnamefont {Taniguchi}},\
  and\ \bibinfo {author} {\bibfnamefont {K.}~\bibnamefont {Uryu}},\ }\bibfield
  {title} {\bibinfo {title} {{Merger of binary neutron stars with realistic
  equations of state in full general relativity}},\ }\href
  {https://doi.org/10.1103/PhysRevD.71.084021} {\bibfield  {journal} {\bibinfo
  {journal} {Phys.\ Rev.\ D}\ }\textbf {\bibinfo {volume} {71}},\ \bibinfo
  {pages} {084021} (\bibinfo {year} {2005})},\ \Eprint
  {https://arxiv.org/abs/gr-qc/0503119} {arXiv:gr-qc/0503119} \BibitemShut
  {NoStop}%
\bibitem [{\citenamefont {{Shibata}}(2005)}]{2005PhRvL..94t1101S}%
  \BibitemOpen
  \bibfield  {author} {\bibinfo {author} {\bibfnamefont {M.}~\bibnamefont
  {{Shibata}}},\ }\bibfield  {title} {\bibinfo {title} {{Constraining Nuclear
  Equations of State Using Gravitational Waves from Hypermassive Neutron
  Stars}},\ }\href {https://doi.org/10.1103/PhysRevLett.94.201101} {\bibfield
  {journal} {\bibinfo  {journal} {Physical Review Letters}\ }\textbf {\bibinfo
  {volume} {94}},\ \bibinfo {eid} {201101} (\bibinfo {year} {2005})},\ \Eprint
  {https://arxiv.org/abs/gr-qc/0504082} {arXiv:gr-qc/0504082 [gr-qc]}
  \BibitemShut {NoStop}%
\bibitem [{\citenamefont {{Shibata}}\ and\ \citenamefont
  {{Taniguchi}}(2006)}]{shibata:06bns}%
  \BibitemOpen
  \bibfield  {author} {\bibinfo {author} {\bibfnamefont {M.}~\bibnamefont
  {{Shibata}}}\ and\ \bibinfo {author} {\bibfnamefont {K.}~\bibnamefont
  {{Taniguchi}}},\ }\bibfield  {title} {\bibinfo {title} {{Merger of binary
  neutron stars to a black hole: Disk mass, short gamma-ray bursts, and
  quasinormal mode ringing}},\ }\href@noop {} {\bibfield  {journal} {\bibinfo
  {journal} {\prd}\ }\textbf {\bibinfo {volume} {73}},\ \bibinfo {pages}
  {064027} (\bibinfo {year} {2006})}\BibitemShut {NoStop}%
\bibitem [{\citenamefont {Oechslin}\ and\ \citenamefont
  {Janka}(2007)}]{Oechslin:2007gn}%
  \BibitemOpen
  \bibfield  {author} {\bibinfo {author} {\bibfnamefont {R.}~\bibnamefont
  {Oechslin}}\ and\ \bibinfo {author} {\bibfnamefont {H.-T.}\ \bibnamefont
  {Janka}},\ }\bibfield  {title} {\bibinfo {title} {{Gravitational waves from
  relativistic neutron star mergers with nonzero-temperature equations of
  state}},\ }\href {https://doi.org/10.1103/PhysRevLett.99.121102} {\bibfield
  {journal} {\bibinfo  {journal} {Phys.\ Rev.\ Lett.}\ }\textbf {\bibinfo
  {volume} {99}},\ \bibinfo {pages} {121102} (\bibinfo {year} {2007})},\
  \Eprint {https://arxiv.org/abs/astro-ph/0702228} {arXiv:astro-ph/0702228}
  \BibitemShut {NoStop}%
\bibitem [{\citenamefont {{Hotokezaka}}\ \emph {et~al.}(2011)\citenamefont
  {{Hotokezaka}}, \citenamefont {{Kyutoku}}, \citenamefont {{Okawa}},
  \citenamefont {{Shibata}},\ and\ \citenamefont
  {{Kiuchi}}}]{2011PhRvD..83l4008H}%
  \BibitemOpen
  \bibfield  {author} {\bibinfo {author} {\bibfnamefont {K.}~\bibnamefont
  {{Hotokezaka}}}, \bibinfo {author} {\bibfnamefont {K.}~\bibnamefont
  {{Kyutoku}}}, \bibinfo {author} {\bibfnamefont {H.}~\bibnamefont {{Okawa}}},
  \bibinfo {author} {\bibfnamefont {M.}~\bibnamefont {{Shibata}}},\ and\
  \bibinfo {author} {\bibfnamefont {K.}~\bibnamefont {{Kiuchi}}},\ }\bibfield
  {title} {\bibinfo {title} {{Binary neutron star mergers: Dependence on the
  nuclear equation of state}},\ }\href
  {https://doi.org/10.1103/PhysRevD.83.124008} {\bibfield  {journal} {\bibinfo
  {journal} {\prd}\ }\textbf {\bibinfo {volume} {83}},\ \bibinfo {eid} {124008}
  (\bibinfo {year} {2011})},\ \Eprint {https://arxiv.org/abs/1105.4370}
  {arXiv:1105.4370 [astro-ph.HE]} \BibitemShut {NoStop}%
\bibitem [{\citenamefont {{Bauswein}}\ and\ \citenamefont
  {{Janka}}(2012)}]{2012PhRvL.108a1101B}%
  \BibitemOpen
  \bibfield  {author} {\bibinfo {author} {\bibfnamefont {A.}~\bibnamefont
  {{Bauswein}}}\ and\ \bibinfo {author} {\bibfnamefont {H.~T.}\ \bibnamefont
  {{Janka}}},\ }\bibfield  {title} {\bibinfo {title} {{Measuring Neutron-Star
  Properties via Gravitational Waves from Neutron-Star Mergers}},\ }\href
  {https://doi.org/10.1103/PhysRevLett.108.011101} {\bibfield  {journal}
  {\bibinfo  {journal} {Physical Review Letters}\ }\textbf {\bibinfo {volume}
  {108}},\ \bibinfo {eid} {011101} (\bibinfo {year} {2012})},\ \Eprint
  {https://arxiv.org/abs/1106.1616} {arXiv:1106.1616 [astro-ph.SR]}
  \BibitemShut {NoStop}%
\bibitem [{\citenamefont {Bauswein}\ \emph {et~al.}(2012)\citenamefont
  {Bauswein}, \citenamefont {Janka}, \citenamefont {Hebeler},\ and\
  \citenamefont {Schwenk}}]{Bauswein:2012ya}%
  \BibitemOpen
  \bibfield  {author} {\bibinfo {author} {\bibfnamefont {A.}~\bibnamefont
  {Bauswein}}, \bibinfo {author} {\bibfnamefont {H.}~\bibnamefont {Janka}},
  \bibinfo {author} {\bibfnamefont {K.}~\bibnamefont {Hebeler}},\ and\ \bibinfo
  {author} {\bibfnamefont {A.}~\bibnamefont {Schwenk}},\ }\bibfield  {title}
  {\bibinfo {title} {{Equation-of-state dependence of the gravitational-wave
  signal from the ring-down phase of neutron-star mergers}},\ }\href
  {https://doi.org/10.1103/PhysRevD.86.063001} {\bibfield  {journal} {\bibinfo
  {journal} {Phys.\ Rev.\ D}\ }\textbf {\bibinfo {volume} {86}},\ \bibinfo
  {pages} {063001} (\bibinfo {year} {2012})},\ \Eprint
  {https://arxiv.org/abs/1204.1888} {arXiv:1204.1888 [astro-ph.SR]}
  \BibitemShut {NoStop}%
\bibitem [{\citenamefont {{Hotokezaka}}\ \emph {et~al.}(2013)\citenamefont
  {{Hotokezaka}}, \citenamefont {{Kiuchi}}, \citenamefont {{Kyutoku}},
  \citenamefont {{Okawa}}, \citenamefont {{Sekiguchi}}, \citenamefont
  {{Shibata}},\ and\ \citenamefont {{Taniguchi}}}]{hotokezaka:13}%
  \BibitemOpen
  \bibfield  {author} {\bibinfo {author} {\bibfnamefont {K.}~\bibnamefont
  {{Hotokezaka}}}, \bibinfo {author} {\bibfnamefont {K.}~\bibnamefont
  {{Kiuchi}}}, \bibinfo {author} {\bibfnamefont {K.}~\bibnamefont {{Kyutoku}}},
  \bibinfo {author} {\bibfnamefont {H.}~\bibnamefont {{Okawa}}}, \bibinfo
  {author} {\bibfnamefont {Y.-i.}\ \bibnamefont {{Sekiguchi}}}, \bibinfo
  {author} {\bibfnamefont {M.}~\bibnamefont {{Shibata}}},\ and\ \bibinfo
  {author} {\bibfnamefont {K.}~\bibnamefont {{Taniguchi}}},\ }\bibfield
  {title} {\bibinfo {title} {{Mass ejection from the merger of binary neutron
  stars}},\ }\href {https://doi.org/10.1103/PhysRevD.87.024001} {\bibfield
  {journal} {\bibinfo  {journal} {\prd}\ }\textbf {\bibinfo {volume} {87}},\
  \bibinfo {eid} {024001} (\bibinfo {year} {2013})},\ \Eprint
  {https://arxiv.org/abs/1212.0905} {arXiv:1212.0905 [astro-ph.HE]}
  \BibitemShut {NoStop}%
\bibitem [{\citenamefont {{Takami}}\ \emph {et~al.}(2014)\citenamefont
  {{Takami}}, \citenamefont {{Rezzolla}},\ and\ \citenamefont
  {{Baiotti}}}]{2014PhRvL.113i1104T}%
  \BibitemOpen
  \bibfield  {author} {\bibinfo {author} {\bibfnamefont {K.}~\bibnamefont
  {{Takami}}}, \bibinfo {author} {\bibfnamefont {L.}~\bibnamefont
  {{Rezzolla}}},\ and\ \bibinfo {author} {\bibfnamefont {L.}~\bibnamefont
  {{Baiotti}}},\ }\bibfield  {title} {\bibinfo {title} {{Constraining the
  Equation of State of Neutron Stars from Binary Mergers}},\ }\href
  {https://doi.org/10.1103/PhysRevLett.113.091104} {\bibfield  {journal}
  {\bibinfo  {journal} {Physical Review Letters}\ }\textbf {\bibinfo {volume}
  {113}},\ \bibinfo {eid} {091104} (\bibinfo {year} {2014})},\ \Eprint
  {https://arxiv.org/abs/1403.5672} {arXiv:1403.5672 [gr-qc]} \BibitemShut
  {NoStop}%
\bibitem [{\citenamefont {{Takami}}\ \emph {et~al.}(2015)\citenamefont
  {{Takami}}, \citenamefont {{Rezzolla}},\ and\ \citenamefont
  {{Baiotti}}}]{2014arXiv1412.3240T}%
  \BibitemOpen
  \bibfield  {author} {\bibinfo {author} {\bibfnamefont {K.}~\bibnamefont
  {{Takami}}}, \bibinfo {author} {\bibfnamefont {L.}~\bibnamefont
  {{Rezzolla}}},\ and\ \bibinfo {author} {\bibfnamefont {L.}~\bibnamefont
  {{Baiotti}}},\ }\bibfield  {title} {\bibinfo {title} {{Spectral properties of
  the post-merger gravitational-wave signal from binary neutron stars}},\
  }\href {https://doi.org/10.1103/PhysRevD.91.064001} {\bibfield  {journal}
  {\bibinfo  {journal} {\prd}\ }\textbf {\bibinfo {volume} {91}},\ \bibinfo
  {eid} {064001} (\bibinfo {year} {2015})}\BibitemShut {NoStop}%
\bibitem [{\citenamefont {{Kastaun}}\ and\ \citenamefont
  {{Galeazzi}}(2015)}]{2014arXiv1411.7975K}%
  \BibitemOpen
  \bibfield  {author} {\bibinfo {author} {\bibfnamefont {W.}~\bibnamefont
  {{Kastaun}}}\ and\ \bibinfo {author} {\bibfnamefont {F.}~\bibnamefont
  {{Galeazzi}}},\ }\bibfield  {title} {\bibinfo {title} {{Properties of
  hypermassive neutron stars formed in mergers of spinning binaries}},\ }\href
  {https://doi.org/10.1103/PhysRevD.91.064027} {\bibfield  {journal} {\bibinfo
  {journal} {\prd}\ }\textbf {\bibinfo {volume} {91}},\ \bibinfo {eid} {064027}
  (\bibinfo {year} {2015})}\BibitemShut {NoStop}%
\bibitem [{\citenamefont {{Bernuzzi}}\ \emph {et~al.}(2015)\citenamefont
  {{Bernuzzi}}, \citenamefont {{Dietrich}},\ and\ \citenamefont
  {{Nagar}}}]{2015arXiv150401764B}%
  \BibitemOpen
  \bibfield  {author} {\bibinfo {author} {\bibfnamefont {S.}~\bibnamefont
  {{Bernuzzi}}}, \bibinfo {author} {\bibfnamefont {T.}~\bibnamefont
  {{Dietrich}}},\ and\ \bibinfo {author} {\bibfnamefont {A.}~\bibnamefont
  {{Nagar}}},\ }\bibfield  {title} {\bibinfo {title} {{Modeling the Complete
  Gravitational Wave Spectrum of Neutron Star Mergers}},\ }\href
  {https://doi.org/10.1103/PhysRevLett.115.091101} {\bibfield  {journal}
  {\bibinfo  {journal} {Physical Review Letters}\ }\textbf {\bibinfo {volume}
  {115}},\ \bibinfo {eid} {091101} (\bibinfo {year} {2015})},\ \Eprint
  {https://arxiv.org/abs/1504.01764} {arXiv:1504.01764 [gr-qc]} \BibitemShut
  {NoStop}%
\bibitem [{\citenamefont {Bauswein}\ and\ \citenamefont
  {Stergioulas}(2015)}]{bauswein:15}%
  \BibitemOpen
  \bibfield  {author} {\bibinfo {author} {\bibfnamefont {A.}~\bibnamefont
  {Bauswein}}\ and\ \bibinfo {author} {\bibfnamefont {N.}~\bibnamefont
  {Stergioulas}},\ }\bibfield  {title} {\bibinfo {title} {Unified picture of
  the post-merger dynamics and gravitational wave emission in neutron star
  mergers},\ }\href {https://doi.org/10.1103/PhysRevD.91.124056} {\bibfield
  {journal} {\bibinfo  {journal} {Phys. Rev. D}\ }\textbf {\bibinfo {volume}
  {91}},\ \bibinfo {pages} {124056} (\bibinfo {year} {2015})}\BibitemShut
  {NoStop}%
\bibitem [{\citenamefont {{Foucart}}\ \emph {et~al.}(2016)\citenamefont
  {{Foucart}}, \citenamefont {{Haas}}, \citenamefont {{Duez}}, \citenamefont
  {{O'Connor}}, \citenamefont {{Ott}}, \citenamefont {{Roberts}}, \citenamefont
  {{Kidder}}, \citenamefont {{Lippuner}}, \citenamefont {{Pfeiffer}},\ and\
  \citenamefont {{Scheel}}}]{Foucart2016}%
  \BibitemOpen
  \bibfield  {author} {\bibinfo {author} {\bibfnamefont {F.}~\bibnamefont
  {{Foucart}}}, \bibinfo {author} {\bibfnamefont {R.}~\bibnamefont {{Haas}}},
  \bibinfo {author} {\bibfnamefont {M.~D.}\ \bibnamefont {{Duez}}}, \bibinfo
  {author} {\bibfnamefont {E.}~\bibnamefont {{O'Connor}}}, \bibinfo {author}
  {\bibfnamefont {C.~D.}\ \bibnamefont {{Ott}}}, \bibinfo {author}
  {\bibfnamefont {L.}~\bibnamefont {{Roberts}}}, \bibinfo {author}
  {\bibfnamefont {L.~E.}\ \bibnamefont {{Kidder}}}, \bibinfo {author}
  {\bibfnamefont {J.}~\bibnamefont {{Lippuner}}}, \bibinfo {author}
  {\bibfnamefont {H.~P.}\ \bibnamefont {{Pfeiffer}}},\ and\ \bibinfo {author}
  {\bibfnamefont {M.~A.}\ \bibnamefont {{Scheel}}},\ }\bibfield  {title}
  {\bibinfo {title} {{Low mass binary neutron star mergers: Gravitational waves
  and neutrino emission}},\ }\href {https://doi.org/10.1103/PhysRevD.93.044019}
  {\bibfield  {journal} {\bibinfo  {journal} {\prd}\ }\textbf {\bibinfo
  {volume} {93}},\ \bibinfo {eid} {044019} (\bibinfo {year} {2016})},\ \Eprint
  {https://arxiv.org/abs/1510.06398} {arXiv:1510.06398 [astro-ph.HE]}
  \BibitemShut {NoStop}%
\bibitem [{\citenamefont {Lehner}\ \emph {et~al.}(2016)\citenamefont {Lehner},
  \citenamefont {Liebling}, \citenamefont {Palenzuela}, \citenamefont
  {Caballero}, \citenamefont {O'Connor}, \citenamefont {Anderson},\ and\
  \citenamefont {Neilsen}}]{Lehner:2016lxy}%
  \BibitemOpen
  \bibfield  {author} {\bibinfo {author} {\bibfnamefont {L.}~\bibnamefont
  {Lehner}}, \bibinfo {author} {\bibfnamefont {S.~L.}\ \bibnamefont
  {Liebling}}, \bibinfo {author} {\bibfnamefont {C.}~\bibnamefont
  {Palenzuela}}, \bibinfo {author} {\bibfnamefont {O.~L.}\ \bibnamefont
  {Caballero}}, \bibinfo {author} {\bibfnamefont {E.}~\bibnamefont {O'Connor}},
  \bibinfo {author} {\bibfnamefont {M.}~\bibnamefont {Anderson}},\ and\
  \bibinfo {author} {\bibfnamefont {D.}~\bibnamefont {Neilsen}},\ }\bibfield
  {title} {\bibinfo {title} {{Unequal mass binary neutron star mergers and
  multimessenger signals}},\ }\href
  {https://doi.org/10.1088/0264-9381/33/18/184002} {\bibfield  {journal}
  {\bibinfo  {journal} {Class. Quant. Grav.}\ }\textbf {\bibinfo {volume}
  {33}},\ \bibinfo {pages} {184002} (\bibinfo {year} {2016})},\ \Eprint
  {https://arxiv.org/abs/1603.00501} {arXiv:1603.00501 [gr-qc]} \BibitemShut
  {NoStop}%
\bibitem [{\citenamefont {{East}}\ \emph {et~al.}(2016)\citenamefont {{East}},
  \citenamefont {{Paschalidis}},\ and\ \citenamefont
  {{Pretorius}}}]{2016CQGra..33x4004E}%
  \BibitemOpen
  \bibfield  {author} {\bibinfo {author} {\bibfnamefont {W.~E.}\ \bibnamefont
  {{East}}}, \bibinfo {author} {\bibfnamefont {V.}~\bibnamefont
  {{Paschalidis}}},\ and\ \bibinfo {author} {\bibfnamefont {F.}~\bibnamefont
  {{Pretorius}}},\ }\bibfield  {title} {\bibinfo {title} {{Equation of state
  effects and one-arm spiral instability in hypermassive neutron stars formed
  in eccentric neutron star mergers}},\ }\href
  {https://doi.org/10.1088/0264-9381/33/24/244004} {\bibfield  {journal}
  {\bibinfo  {journal} {Classical and Quantum Gravity}\ }\textbf {\bibinfo
  {volume} {33}},\ \bibinfo {eid} {244004} (\bibinfo {year} {2016})},\ \Eprint
  {https://arxiv.org/abs/1609.00725} {arXiv:1609.00725 [astro-ph.HE]}
  \BibitemShut {NoStop}%
\bibitem [{\citenamefont {{Dietrich}}\ \emph {et~al.}(2017)\citenamefont
  {{Dietrich}}, \citenamefont {{Ujevic}}, \citenamefont {{Tichy}},
  \citenamefont {{Bernuzzi}},\ and\ \citenamefont
  {{Br{\"u}gmann}}}]{Dietrich2017}%
  \BibitemOpen
  \bibfield  {author} {\bibinfo {author} {\bibfnamefont {T.}~\bibnamefont
  {{Dietrich}}}, \bibinfo {author} {\bibfnamefont {M.}~\bibnamefont
  {{Ujevic}}}, \bibinfo {author} {\bibfnamefont {W.}~\bibnamefont {{Tichy}}},
  \bibinfo {author} {\bibfnamefont {S.}~\bibnamefont {{Bernuzzi}}},\ and\
  \bibinfo {author} {\bibfnamefont {B.}~\bibnamefont {{Br{\"u}gmann}}},\
  }\bibfield  {title} {\bibinfo {title} {{Gravitational waves and mass ejecta
  from binary neutron star mergers: Effect of the mass ratio}},\ }\href
  {https://doi.org/10.1103/PhysRevD.95.024029} {\bibfield  {journal} {\bibinfo
  {journal} {\prd}\ }\textbf {\bibinfo {volume} {95}},\ \bibinfo {eid} {024029}
  (\bibinfo {year} {2017})},\ \Eprint {https://arxiv.org/abs/1607.06636}
  {arXiv:1607.06636 [gr-qc]} \BibitemShut {NoStop}%
\bibitem [{\citenamefont {{Blacker}}\ \emph {et~al.}(2020)\citenamefont
  {{Blacker}}, \citenamefont {{Bastian}}, \citenamefont {{Bauswein}},
  \citenamefont {{Blaschke}}, \citenamefont {{Fischer}}, \citenamefont
  {{Oertel}}, \citenamefont {{Soultanis}},\ and\ \citenamefont
  {{Typel}}}]{2020PhRvD.102l3023B}%
  \BibitemOpen
  \bibfield  {author} {\bibinfo {author} {\bibfnamefont {S.}~\bibnamefont
  {{Blacker}}}, \bibinfo {author} {\bibfnamefont {N.-U.~F.}\ \bibnamefont
  {{Bastian}}}, \bibinfo {author} {\bibfnamefont {A.}~\bibnamefont
  {{Bauswein}}}, \bibinfo {author} {\bibfnamefont {D.~B.}\ \bibnamefont
  {{Blaschke}}}, \bibinfo {author} {\bibfnamefont {T.}~\bibnamefont
  {{Fischer}}}, \bibinfo {author} {\bibfnamefont {M.}~\bibnamefont {{Oertel}}},
  \bibinfo {author} {\bibfnamefont {T.}~\bibnamefont {{Soultanis}}},\ and\
  \bibinfo {author} {\bibfnamefont {S.}~\bibnamefont {{Typel}}},\ }\bibfield
  {title} {\bibinfo {title} {{Constraining the onset density of the
  hadron-quark phase transition with gravitational-wave observations}},\ }\href
  {https://doi.org/10.1103/PhysRevD.102.123023} {\bibfield  {journal} {\bibinfo
   {journal} {\prd}\ }\textbf {\bibinfo {volume} {102}},\ \bibinfo {eid}
  {123023} (\bibinfo {year} {2020})},\ \Eprint
  {https://arxiv.org/abs/2006.03789} {arXiv:2006.03789 [astro-ph.HE]}
  \BibitemShut {NoStop}%
\bibitem [{\citenamefont {Raithel}\ \emph {et~al.}(2021)\citenamefont
  {Raithel}, \citenamefont {Paschalidis},\ and\ \citenamefont
  {\"Ozel}}]{Raithel:2021hye}%
  \BibitemOpen
  \bibfield  {author} {\bibinfo {author} {\bibfnamefont {C.~A.}\ \bibnamefont
  {Raithel}}, \bibinfo {author} {\bibfnamefont {V.}~\bibnamefont
  {Paschalidis}},\ and\ \bibinfo {author} {\bibfnamefont {F.}~\bibnamefont
  {\"Ozel}},\ }\bibfield  {title} {\bibinfo {title} {{Realistic
  Finite-Temperature Effects in Neutron Star Merger Simulations}},\ }\href@noop
  {} {\  (\bibinfo {year} {2021})},\ \Eprint {https://arxiv.org/abs/2104.07226}
  {arXiv:2104.07226 [astro-ph.HE]} \BibitemShut {NoStop}%
\bibitem [{\citenamefont {Hammond}\ \emph {et~al.}(2021)\citenamefont
  {Hammond}, \citenamefont {Hawke},\ and\ \citenamefont
  {Andersson}}]{Hammond:2021vtv}%
  \BibitemOpen
  \bibfield  {author} {\bibinfo {author} {\bibfnamefont {P.}~\bibnamefont
  {Hammond}}, \bibinfo {author} {\bibfnamefont {I.}~\bibnamefont {Hawke}},\
  and\ \bibinfo {author} {\bibfnamefont {N.}~\bibnamefont {Andersson}},\
  }\bibfield  {title} {\bibinfo {title} {{Thermal aspects of neutron star
  mergers}},\ }\href@noop {} {\  (\bibinfo {year} {2021})},\ \Eprint
  {https://arxiv.org/abs/2108.08649} {arXiv:2108.08649 [astro-ph.HE]}
  \BibitemShut {NoStop}%
\bibitem [{\citenamefont {Radice}\ \emph {et~al.}(2017)\citenamefont {Radice},
  \citenamefont {Bernuzzi}, \citenamefont {Del~Pozzo}, \citenamefont
  {Roberts},\ and\ \citenamefont {Ott}}]{Radice:2016rys}%
  \BibitemOpen
  \bibfield  {author} {\bibinfo {author} {\bibfnamefont {D.}~\bibnamefont
  {Radice}}, \bibinfo {author} {\bibfnamefont {S.}~\bibnamefont {Bernuzzi}},
  \bibinfo {author} {\bibfnamefont {W.}~\bibnamefont {Del~Pozzo}}, \bibinfo
  {author} {\bibfnamefont {L.~F.}\ \bibnamefont {Roberts}},\ and\ \bibinfo
  {author} {\bibfnamefont {C.~D.}\ \bibnamefont {Ott}},\ }\bibfield  {title}
  {\bibinfo {title} {{Probing Extreme-Density Matter with Gravitational Wave
  Observations of Binary Neutron Star Merger Remnants}},\ }\href
  {https://doi.org/10.3847/2041-8213/aa775f} {\bibfield  {journal} {\bibinfo
  {journal} {Astrophys. J. Lett.}\ }\textbf {\bibinfo {volume} {842}},\
  \bibinfo {pages} {L10} (\bibinfo {year} {2017})},\ \Eprint
  {https://arxiv.org/abs/1612.06429} {arXiv:1612.06429 [astro-ph.HE]}
  \BibitemShut {NoStop}%
\bibitem [{\citenamefont {Most}\ \emph {et~al.}(2019)\citenamefont {Most},
  \citenamefont {Papenfort}, \citenamefont {Dexheimer}, \citenamefont
  {Hanauske}, \citenamefont {Schramm}, \citenamefont {St\"ocker},\ and\
  \citenamefont {Rezzolla}}]{Most:2018eaw}%
  \BibitemOpen
  \bibfield  {author} {\bibinfo {author} {\bibfnamefont {E.~R.}\ \bibnamefont
  {Most}}, \bibinfo {author} {\bibfnamefont {L.~J.}\ \bibnamefont {Papenfort}},
  \bibinfo {author} {\bibfnamefont {V.}~\bibnamefont {Dexheimer}}, \bibinfo
  {author} {\bibfnamefont {M.}~\bibnamefont {Hanauske}}, \bibinfo {author}
  {\bibfnamefont {S.}~\bibnamefont {Schramm}}, \bibinfo {author} {\bibfnamefont
  {H.}~\bibnamefont {St\"ocker}},\ and\ \bibinfo {author} {\bibfnamefont
  {L.}~\bibnamefont {Rezzolla}},\ }\bibfield  {title} {\bibinfo {title}
  {{Signatures of quark-hadron phase transitions in general-relativistic
  neutron-star mergers}},\ }\href
  {https://doi.org/10.1103/PhysRevLett.122.061101} {\bibfield  {journal}
  {\bibinfo  {journal} {Phys. Rev. Lett.}\ }\textbf {\bibinfo {volume} {122}},\
  \bibinfo {pages} {061101} (\bibinfo {year} {2019})},\ \Eprint
  {https://arxiv.org/abs/1807.03684} {arXiv:1807.03684 [astro-ph.HE]}
  \BibitemShut {NoStop}%
\bibitem [{\citenamefont {Bauswein}\ \emph
  {et~al.}(2019{\natexlab{a}})\citenamefont {Bauswein}, \citenamefont
  {Bastian}, \citenamefont {Blaschke}, \citenamefont {Chatziioannou},
  \citenamefont {Clark}, \citenamefont {Fischer},\ and\ \citenamefont
  {Oertel}}]{Bauswein:2018bma}%
  \BibitemOpen
  \bibfield  {author} {\bibinfo {author} {\bibfnamefont {A.}~\bibnamefont
  {Bauswein}}, \bibinfo {author} {\bibfnamefont {N.-U.~F.}\ \bibnamefont
  {Bastian}}, \bibinfo {author} {\bibfnamefont {D.~B.}\ \bibnamefont
  {Blaschke}}, \bibinfo {author} {\bibfnamefont {K.}~\bibnamefont
  {Chatziioannou}}, \bibinfo {author} {\bibfnamefont {J.~A.}\ \bibnamefont
  {Clark}}, \bibinfo {author} {\bibfnamefont {T.}~\bibnamefont {Fischer}},\
  and\ \bibinfo {author} {\bibfnamefont {M.}~\bibnamefont {Oertel}},\
  }\bibfield  {title} {\bibinfo {title} {{Identifying a first-order phase
  transition in neutron star mergers through gravitational waves}},\ }\href
  {https://doi.org/10.1103/PhysRevLett.122.061102} {\bibfield  {journal}
  {\bibinfo  {journal} {Phys. Rev. Lett.}\ }\textbf {\bibinfo {volume} {122}},\
  \bibinfo {pages} {061102} (\bibinfo {year} {2019}{\natexlab{a}})},\ \Eprint
  {https://arxiv.org/abs/1809.01116} {arXiv:1809.01116 [astro-ph.HE]}
  \BibitemShut {NoStop}%
\bibitem [{\citenamefont {Bauswein}\ \emph
  {et~al.}(2019{\natexlab{b}})\citenamefont {Bauswein}, \citenamefont
  {Friedrich~Bastian}, \citenamefont {Blaschke}, \citenamefont {Chatziioannou},
  \citenamefont {Clark}, \citenamefont {Fischer}, \citenamefont {Janka},
  \citenamefont {Just}, \citenamefont {Oertel},\ and\ \citenamefont
  {Stergioulas}}]{Bauswein:2019skm}%
  \BibitemOpen
  \bibfield  {author} {\bibinfo {author} {\bibfnamefont {A.}~\bibnamefont
  {Bauswein}}, \bibinfo {author} {\bibfnamefont {N.-U.}\ \bibnamefont
  {Friedrich~Bastian}}, \bibinfo {author} {\bibfnamefont {D.}~\bibnamefont
  {Blaschke}}, \bibinfo {author} {\bibfnamefont {K.}~\bibnamefont
  {Chatziioannou}}, \bibinfo {author} {\bibfnamefont {J.~A.}\ \bibnamefont
  {Clark}}, \bibinfo {author} {\bibfnamefont {T.}~\bibnamefont {Fischer}},
  \bibinfo {author} {\bibfnamefont {H.-T.}\ \bibnamefont {Janka}}, \bibinfo
  {author} {\bibfnamefont {O.}~\bibnamefont {Just}}, \bibinfo {author}
  {\bibfnamefont {M.}~\bibnamefont {Oertel}},\ and\ \bibinfo {author}
  {\bibfnamefont {N.}~\bibnamefont {Stergioulas}},\ }\bibfield  {title}
  {\bibinfo {title} {{Equation-of-state Constraints and the QCD Phase
  Transition in the Era of Gravitational-Wave Astronomy}},\ }\href
  {https://doi.org/10.1063/1.5117803} {\bibfield  {journal} {\bibinfo
  {journal} {AIP Conf. Proc.}\ }\textbf {\bibinfo {volume} {2127}},\ \bibinfo
  {pages} {020013} (\bibinfo {year} {2019}{\natexlab{b}})},\ \Eprint
  {https://arxiv.org/abs/1904.01306} {arXiv:1904.01306 [astro-ph.HE]}
  \BibitemShut {NoStop}%
\bibitem [{\citenamefont {Weih}\ \emph {et~al.}(2020)\citenamefont {Weih},
  \citenamefont {Hanauske},\ and\ \citenamefont {Rezzolla}}]{Weih:2019xvw}%
  \BibitemOpen
  \bibfield  {author} {\bibinfo {author} {\bibfnamefont {L.~R.}\ \bibnamefont
  {Weih}}, \bibinfo {author} {\bibfnamefont {M.}~\bibnamefont {Hanauske}},\
  and\ \bibinfo {author} {\bibfnamefont {L.}~\bibnamefont {Rezzolla}},\
  }\bibfield  {title} {\bibinfo {title} {{Postmerger Gravitational-Wave
  Signatures of Phase Transitions in Binary Mergers}},\ }\href
  {https://doi.org/10.1103/PhysRevLett.124.171103} {\bibfield  {journal}
  {\bibinfo  {journal} {Phys. Rev. Lett.}\ }\textbf {\bibinfo {volume} {124}},\
  \bibinfo {pages} {171103} (\bibinfo {year} {2020})},\ \Eprint
  {https://arxiv.org/abs/1912.09340} {arXiv:1912.09340 [gr-qc]} \BibitemShut
  {NoStop}%
\bibitem [{\citenamefont {Bauswein}\ and\ \citenamefont
  {Blacker}(2020)}]{Bauswein:2020ggy}%
  \BibitemOpen
  \bibfield  {author} {\bibinfo {author} {\bibfnamefont {A.}~\bibnamefont
  {Bauswein}}\ and\ \bibinfo {author} {\bibfnamefont {S.}~\bibnamefont
  {Blacker}},\ }\bibfield  {title} {\bibinfo {title} {{Impact of quark
  deconfinement in neutron star mergers and hybrid star mergers}},\ }\href
  {https://doi.org/10.1140/epjst/e2020-000138-7} {\bibfield  {journal}
  {\bibinfo  {journal} {Eur. Phys. J. ST}\ }\textbf {\bibinfo {volume} {229}},\
  \bibinfo {pages} {3595} (\bibinfo {year} {2020})},\ \Eprint
  {https://arxiv.org/abs/2006.16183} {arXiv:2006.16183 [astro-ph.HE]}
  \BibitemShut {NoStop}%
\bibitem [{\citenamefont {Liebling}\ \emph {et~al.}(2021)\citenamefont
  {Liebling}, \citenamefont {Palenzuela},\ and\ \citenamefont
  {Lehner}}]{Liebling:2020dhf}%
  \BibitemOpen
  \bibfield  {author} {\bibinfo {author} {\bibfnamefont {S.~L.}\ \bibnamefont
  {Liebling}}, \bibinfo {author} {\bibfnamefont {C.}~\bibnamefont
  {Palenzuela}},\ and\ \bibinfo {author} {\bibfnamefont {L.}~\bibnamefont
  {Lehner}},\ }\bibfield  {title} {\bibinfo {title} {{Effects of High Density
  Phase Transitions on Neutron Star Dynamics}},\ }\href
  {https://doi.org/10.1088/1361-6382/abf898} {\bibfield  {journal} {\bibinfo
  {journal} {Class. Quant. Grav.}\ }\textbf {\bibinfo {volume} {38}},\ \bibinfo
  {pages} {115007} (\bibinfo {year} {2021})},\ \Eprint
  {https://arxiv.org/abs/2010.12567} {arXiv:2010.12567 [gr-qc]} \BibitemShut
  {NoStop}%
\bibitem [{\citenamefont {Prakash}\ \emph {et~al.}(2021)\citenamefont
  {Prakash}, \citenamefont {Radice}, \citenamefont {Logoteta}, \citenamefont
  {Perego}, \citenamefont {Nedora}, \citenamefont {Bombaci}, \citenamefont
  {Kashyap}, \citenamefont {Bernuzzi},\ and\ \citenamefont
  {Endrizzi}}]{Prakash:2021wpz}%
  \BibitemOpen
  \bibfield  {author} {\bibinfo {author} {\bibfnamefont {A.}~\bibnamefont
  {Prakash}}, \bibinfo {author} {\bibfnamefont {D.}~\bibnamefont {Radice}},
  \bibinfo {author} {\bibfnamefont {D.}~\bibnamefont {Logoteta}}, \bibinfo
  {author} {\bibfnamefont {A.}~\bibnamefont {Perego}}, \bibinfo {author}
  {\bibfnamefont {V.}~\bibnamefont {Nedora}}, \bibinfo {author} {\bibfnamefont
  {I.}~\bibnamefont {Bombaci}}, \bibinfo {author} {\bibfnamefont
  {R.}~\bibnamefont {Kashyap}}, \bibinfo {author} {\bibfnamefont
  {S.}~\bibnamefont {Bernuzzi}},\ and\ \bibinfo {author} {\bibfnamefont
  {A.}~\bibnamefont {Endrizzi}},\ }\bibfield  {title} {\bibinfo {title}
  {{Signatures of deconfined quark phases in binary neutron star mergers}},\
  }\href@noop {} {\  (\bibinfo {year} {2021})},\ \Eprint
  {https://arxiv.org/abs/2106.07885} {arXiv:2106.07885 [astro-ph.HE]}
  \BibitemShut {NoStop}%
\bibitem [{\citenamefont {{Lioutas}}\ \emph {et~al.}(2021)\citenamefont
  {{Lioutas}}, \citenamefont {{Bauswein}},\ and\ \citenamefont
  {{Stergioulas}}}]{2021PhRvD.104d3011L}%
  \BibitemOpen
  \bibfield  {author} {\bibinfo {author} {\bibfnamefont {G.}~\bibnamefont
  {{Lioutas}}}, \bibinfo {author} {\bibfnamefont {A.}~\bibnamefont
  {{Bauswein}}},\ and\ \bibinfo {author} {\bibfnamefont {N.}~\bibnamefont
  {{Stergioulas}}},\ }\bibfield  {title} {\bibinfo {title} {{Frequency
  deviations in universal relations of isolated neutron stars and postmerger
  remnants}},\ }\href {https://doi.org/10.1103/PhysRevD.104.043011} {\bibfield
  {journal} {\bibinfo  {journal} {\prd}\ }\textbf {\bibinfo {volume} {104}},\
  \bibinfo {eid} {043011} (\bibinfo {year} {2021})},\ \Eprint
  {https://arxiv.org/abs/2102.12455} {arXiv:2102.12455 [astro-ph.HE]}
  \BibitemShut {NoStop}%
\bibitem [{\citenamefont {Margalit}\ and\ \citenamefont
  {Metzger}(2019)}]{Margalit:2019dpi}%
  \BibitemOpen
  \bibfield  {author} {\bibinfo {author} {\bibfnamefont {B.}~\bibnamefont
  {Margalit}}\ and\ \bibinfo {author} {\bibfnamefont {B.~D.}\ \bibnamefont
  {Metzger}},\ }\bibfield  {title} {\bibinfo {title} {{The Multi-Messenger
  Matrix: the Future of Neutron Star Merger Constraints on the Nuclear Equation
  of State}},\ }\href {https://doi.org/10.3847/2041-8213/ab2ae2} {\bibfield
  {journal} {\bibinfo  {journal} {Astrophys. J. Lett.}\ }\textbf {\bibinfo
  {volume} {880}},\ \bibinfo {pages} {L15} (\bibinfo {year} {2019})},\ \Eprint
  {https://arxiv.org/abs/1904.11995} {arXiv:1904.11995 [astro-ph.HE]}
  \BibitemShut {NoStop}%
\bibitem [{\citenamefont {Bauswein}\ \emph {et~al.}(2020)\citenamefont
  {Bauswein}, \citenamefont {Blacker}, \citenamefont {Vijayan}, \citenamefont
  {Stergioulas}, \citenamefont {Chatziioannou}, \citenamefont {Clark},
  \citenamefont {Bastian}, \citenamefont {Blaschke}, \citenamefont {Cierniak},\
  and\ \citenamefont {Fischer}}]{Bauswein:2020aag}%
  \BibitemOpen
  \bibfield  {author} {\bibinfo {author} {\bibfnamefont {A.}~\bibnamefont
  {Bauswein}}, \bibinfo {author} {\bibfnamefont {S.}~\bibnamefont {Blacker}},
  \bibinfo {author} {\bibfnamefont {V.}~\bibnamefont {Vijayan}}, \bibinfo
  {author} {\bibfnamefont {N.}~\bibnamefont {Stergioulas}}, \bibinfo {author}
  {\bibfnamefont {K.}~\bibnamefont {Chatziioannou}}, \bibinfo {author}
  {\bibfnamefont {J.~A.}\ \bibnamefont {Clark}}, \bibinfo {author}
  {\bibfnamefont {N.-U.~F.}\ \bibnamefont {Bastian}}, \bibinfo {author}
  {\bibfnamefont {D.~B.}\ \bibnamefont {Blaschke}}, \bibinfo {author}
  {\bibfnamefont {M.}~\bibnamefont {Cierniak}},\ and\ \bibinfo {author}
  {\bibfnamefont {T.}~\bibnamefont {Fischer}},\ }\bibfield  {title} {\bibinfo
  {title} {{Equation of state constraints from the threshold binary mass for
  prompt collapse of neutron star mergers}},\ }\href
  {https://doi.org/10.1103/PhysRevLett.125.141103} {\bibfield  {journal}
  {\bibinfo  {journal} {Phys. Rev. Lett.}\ }\textbf {\bibinfo {volume} {125}},\
  \bibinfo {pages} {141103} (\bibinfo {year} {2020})},\ \Eprint
  {https://arxiv.org/abs/2004.00846} {arXiv:2004.00846 [astro-ph.HE]}
  \BibitemShut {NoStop}%
\bibitem [{\citenamefont {Abbott}\ \emph
  {et~al.}(2017{\natexlab{b}})\citenamefont {Abbott} \emph
  {et~al.}}]{Abbott:2017dke}%
  \BibitemOpen
  \bibfield  {author} {\bibinfo {author} {\bibfnamefont {B.~P.}\ \bibnamefont
  {Abbott}} \emph {et~al.} (\bibinfo {collaboration} {LIGO Scientific,
  Virgo}),\ }\bibfield  {title} {\bibinfo {title} {{Search for Post-merger
  Gravitational Waves from the Remnant of the Binary Neutron Star Merger
  GW170817}},\ }\href {https://doi.org/10.3847/2041-8213/aa9a35} {\bibfield
  {journal} {\bibinfo  {journal} {Astrophys. J. Lett.}\ }\textbf {\bibinfo
  {volume} {851}},\ \bibinfo {pages} {L16} (\bibinfo {year}
  {2017}{\natexlab{b}})},\ \Eprint {https://arxiv.org/abs/1710.09320}
  {arXiv:1710.09320 [astro-ph.HE]} \BibitemShut {NoStop}%
\bibitem [{\citenamefont {Abbott}\ \emph
  {et~al.}(2019{\natexlab{b}})\citenamefont {Abbott} \emph
  {et~al.}}]{LIGOScientific:2018urg}%
  \BibitemOpen
  \bibfield  {author} {\bibinfo {author} {\bibfnamefont {B.~P.}\ \bibnamefont
  {Abbott}} \emph {et~al.} (\bibinfo {collaboration} {LIGO Scientific,
  Virgo}),\ }\bibfield  {title} {\bibinfo {title} {{Search for gravitational
  waves from a long-lived remnant of the binary neutron star merger
  GW170817}},\ }\href {https://doi.org/10.3847/1538-4357/ab0f3d} {\bibfield
  {journal} {\bibinfo  {journal} {Astrophys. J.}\ }\textbf {\bibinfo {volume}
  {875}},\ \bibinfo {pages} {160} (\bibinfo {year} {2019}{\natexlab{b}})},\
  \Eprint {https://arxiv.org/abs/1810.02581} {arXiv:1810.02581 [gr-qc]}
  \BibitemShut {NoStop}%
\bibitem [{\citenamefont {Dudi}\ \emph {et~al.}(2018)\citenamefont {Dudi},
  \citenamefont {Pannarale}, \citenamefont {Dietrich}, \citenamefont {Hannam},
  \citenamefont {Bernuzzi}, \citenamefont {Ohme},\ and\ \citenamefont
  {Br\"ugmann}}]{Dudi:2018jzn}%
  \BibitemOpen
  \bibfield  {author} {\bibinfo {author} {\bibfnamefont {R.}~\bibnamefont
  {Dudi}}, \bibinfo {author} {\bibfnamefont {F.}~\bibnamefont {Pannarale}},
  \bibinfo {author} {\bibfnamefont {T.}~\bibnamefont {Dietrich}}, \bibinfo
  {author} {\bibfnamefont {M.}~\bibnamefont {Hannam}}, \bibinfo {author}
  {\bibfnamefont {S.}~\bibnamefont {Bernuzzi}}, \bibinfo {author}
  {\bibfnamefont {F.}~\bibnamefont {Ohme}},\ and\ \bibinfo {author}
  {\bibfnamefont {B.}~\bibnamefont {Br\"ugmann}},\ }\bibfield  {title}
  {\bibinfo {title} {{Relevance of tidal effects and post-merger dynamics for
  binary neutron star parameter estimation}},\ }\href
  {https://doi.org/10.1103/PhysRevD.98.084061} {\bibfield  {journal} {\bibinfo
  {journal} {Phys. Rev. D}\ }\textbf {\bibinfo {volume} {98}},\ \bibinfo
  {pages} {084061} (\bibinfo {year} {2018})},\ \Eprint
  {https://arxiv.org/abs/1808.09749} {arXiv:1808.09749 [gr-qc]} \BibitemShut
  {NoStop}%
\bibitem [{\citenamefont {{Vretinaris}}\ \emph {et~al.}(2020)\citenamefont
  {{Vretinaris}}, \citenamefont {{Stergioulas}},\ and\ \citenamefont
  {{Bauswein}}}]{2020PhRvD.101h4039V}%
  \BibitemOpen
  \bibfield  {author} {\bibinfo {author} {\bibfnamefont {S.}~\bibnamefont
  {{Vretinaris}}}, \bibinfo {author} {\bibfnamefont {N.}~\bibnamefont
  {{Stergioulas}}},\ and\ \bibinfo {author} {\bibfnamefont {A.}~\bibnamefont
  {{Bauswein}}},\ }\bibfield  {title} {\bibinfo {title} {{Empirical relations
  for gravitational-wave asteroseismology of binary neutron star mergers}},\
  }\href {https://doi.org/10.1103/PhysRevD.101.084039} {\bibfield  {journal}
  {\bibinfo  {journal} {\prd}\ }\textbf {\bibinfo {volume} {101}},\ \bibinfo
  {eid} {084039} (\bibinfo {year} {2020})},\ \Eprint
  {https://arxiv.org/abs/1910.10856} {arXiv:1910.10856 [gr-qc]} \BibitemShut
  {NoStop}%
\bibitem [{\citenamefont {Hotokezaka}\ \emph {et~al.}(2013)\citenamefont
  {Hotokezaka}, \citenamefont {Kiuchi}, \citenamefont {Kyutoku}, \citenamefont
  {Muranushi}, \citenamefont {Sekiguchi}, \citenamefont {Shibata},\ and\
  \citenamefont {Taniguchi}}]{Hotokezaka:2013iia}%
  \BibitemOpen
  \bibfield  {author} {\bibinfo {author} {\bibfnamefont {K.}~\bibnamefont
  {Hotokezaka}}, \bibinfo {author} {\bibfnamefont {K.}~\bibnamefont {Kiuchi}},
  \bibinfo {author} {\bibfnamefont {K.}~\bibnamefont {Kyutoku}}, \bibinfo
  {author} {\bibfnamefont {T.}~\bibnamefont {Muranushi}}, \bibinfo {author}
  {\bibfnamefont {Y.-i.}\ \bibnamefont {Sekiguchi}}, \bibinfo {author}
  {\bibfnamefont {M.}~\bibnamefont {Shibata}},\ and\ \bibinfo {author}
  {\bibfnamefont {K.}~\bibnamefont {Taniguchi}},\ }\bibfield  {title} {\bibinfo
  {title} {{Remnant massive neutron stars of binary neutron star mergers:
  Evolution process and gravitational waveform}},\ }\href
  {https://doi.org/10.1103/PhysRevD.88.044026} {\bibfield  {journal} {\bibinfo
  {journal} {Phys. Rev. D}\ }\textbf {\bibinfo {volume} {88}},\ \bibinfo
  {pages} {044026} (\bibinfo {year} {2013})},\ \Eprint
  {https://arxiv.org/abs/1307.5888} {arXiv:1307.5888 [astro-ph.HE]}
  \BibitemShut {NoStop}%
\bibitem [{\citenamefont {Bauswein}\ \emph {et~al.}(2016)\citenamefont
  {Bauswein}, \citenamefont {Stergioulas},\ and\ \citenamefont
  {Janka}}]{Bauswein:2015vxa}%
  \BibitemOpen
  \bibfield  {author} {\bibinfo {author} {\bibfnamefont {A.}~\bibnamefont
  {Bauswein}}, \bibinfo {author} {\bibfnamefont {N.}~\bibnamefont
  {Stergioulas}},\ and\ \bibinfo {author} {\bibfnamefont {H.-T.}\ \bibnamefont
  {Janka}},\ }\bibfield  {title} {\bibinfo {title} {{Exploring properties of
  high-density matter through remnants of neutron-star mergers}},\ }\href
  {https://doi.org/10.1140/epja/i2016-16056-7} {\bibfield  {journal} {\bibinfo
  {journal} {Eur. Phys. J. A}\ }\textbf {\bibinfo {volume} {52}},\ \bibinfo
  {pages} {56} (\bibinfo {year} {2016})},\ \Eprint
  {https://arxiv.org/abs/1508.05493} {arXiv:1508.05493 [astro-ph.HE]}
  \BibitemShut {NoStop}%
\bibitem [{\citenamefont {Bose}\ \emph {et~al.}(2018)\citenamefont {Bose},
  \citenamefont {Chakravarti}, \citenamefont {Rezzolla}, \citenamefont
  {Sathyaprakash},\ and\ \citenamefont {Takami}}]{Bose:2017jvk}%
  \BibitemOpen
  \bibfield  {author} {\bibinfo {author} {\bibfnamefont {S.}~\bibnamefont
  {Bose}}, \bibinfo {author} {\bibfnamefont {K.}~\bibnamefont {Chakravarti}},
  \bibinfo {author} {\bibfnamefont {L.}~\bibnamefont {Rezzolla}}, \bibinfo
  {author} {\bibfnamefont {B.~S.}\ \bibnamefont {Sathyaprakash}},\ and\
  \bibinfo {author} {\bibfnamefont {K.}~\bibnamefont {Takami}},\ }\bibfield
  {title} {\bibinfo {title} {{Neutron-star Radius from a Population of Binary
  Neutron Star Mergers}},\ }\href
  {https://doi.org/10.1103/PhysRevLett.120.031102} {\bibfield  {journal}
  {\bibinfo  {journal} {Phys. Rev. Lett.}\ }\textbf {\bibinfo {volume} {120}},\
  \bibinfo {pages} {031102} (\bibinfo {year} {2018})},\ \Eprint
  {https://arxiv.org/abs/1705.10850} {arXiv:1705.10850 [gr-qc]} \BibitemShut
  {NoStop}%
\bibitem [{\citenamefont {Tsang}\ \emph {et~al.}(2019)\citenamefont {Tsang},
  \citenamefont {Dietrich},\ and\ \citenamefont {Van
  Den~Broeck}}]{Tsang:2019esi}%
  \BibitemOpen
  \bibfield  {author} {\bibinfo {author} {\bibfnamefont {K.~W.}\ \bibnamefont
  {Tsang}}, \bibinfo {author} {\bibfnamefont {T.}~\bibnamefont {Dietrich}},\
  and\ \bibinfo {author} {\bibfnamefont {C.}~\bibnamefont {Van Den~Broeck}},\
  }\bibfield  {title} {\bibinfo {title} {{Modeling the postmerger gravitational
  wave signal and extracting binary properties from future binary neutron star
  detections}},\ }\href {https://doi.org/10.1103/PhysRevD.100.044047}
  {\bibfield  {journal} {\bibinfo  {journal} {Phys. Rev. D}\ }\textbf {\bibinfo
  {volume} {100}},\ \bibinfo {pages} {044047} (\bibinfo {year} {2019})},\
  \Eprint {https://arxiv.org/abs/1907.02424} {arXiv:1907.02424 [gr-qc]}
  \BibitemShut {NoStop}%
\bibitem [{\citenamefont {Breschi}\ \emph {et~al.}(2019)\citenamefont
  {Breschi}, \citenamefont {Bernuzzi}, \citenamefont {Zappa}, \citenamefont
  {Agathos}, \citenamefont {Perego}, \citenamefont {Radice},\ and\
  \citenamefont {Nagar}}]{Breschi:2019srl}%
  \BibitemOpen
  \bibfield  {author} {\bibinfo {author} {\bibfnamefont {M.}~\bibnamefont
  {Breschi}}, \bibinfo {author} {\bibfnamefont {S.}~\bibnamefont {Bernuzzi}},
  \bibinfo {author} {\bibfnamefont {F.}~\bibnamefont {Zappa}}, \bibinfo
  {author} {\bibfnamefont {M.}~\bibnamefont {Agathos}}, \bibinfo {author}
  {\bibfnamefont {A.}~\bibnamefont {Perego}}, \bibinfo {author} {\bibfnamefont
  {D.}~\bibnamefont {Radice}},\ and\ \bibinfo {author} {\bibfnamefont
  {A.}~\bibnamefont {Nagar}},\ }\bibfield  {title} {\bibinfo {title}
  {{kiloHertz gravitational waves from binary neutron star remnants:
  time-domain model and constraints on extreme matter}},\ }\href
  {https://doi.org/10.1103/PhysRevD.100.104029} {\bibfield  {journal} {\bibinfo
   {journal} {Phys. Rev. D}\ }\textbf {\bibinfo {volume} {100}},\ \bibinfo
  {pages} {104029} (\bibinfo {year} {2019})},\ \Eprint
  {https://arxiv.org/abs/1908.11418} {arXiv:1908.11418 [gr-qc]} \BibitemShut
  {NoStop}%
\bibitem [{\citenamefont {Easter}\ \emph {et~al.}(2020)\citenamefont {Easter},
  \citenamefont {Ghonge}, \citenamefont {Lasky}, \citenamefont {Casey},
  \citenamefont {Clark}, \citenamefont {Vivanco},\ and\ \citenamefont
  {Chatziioannou}}]{Easter:2020ifj}%
  \BibitemOpen
  \bibfield  {author} {\bibinfo {author} {\bibfnamefont {P.~J.}\ \bibnamefont
  {Easter}}, \bibinfo {author} {\bibfnamefont {S.}~\bibnamefont {Ghonge}},
  \bibinfo {author} {\bibfnamefont {P.~D.}\ \bibnamefont {Lasky}}, \bibinfo
  {author} {\bibfnamefont {A.~R.}\ \bibnamefont {Casey}}, \bibinfo {author}
  {\bibfnamefont {J.~A.}\ \bibnamefont {Clark}}, \bibinfo {author}
  {\bibfnamefont {F.~H.}\ \bibnamefont {Vivanco}},\ and\ \bibinfo {author}
  {\bibfnamefont {K.}~\bibnamefont {Chatziioannou}},\ }\bibfield  {title}
  {\bibinfo {title} {{Detection and parameter estimation of binary neutron star
  merger remnants}},\ }\href {https://doi.org/10.1103/PhysRevD.102.043011}
  {\bibfield  {journal} {\bibinfo  {journal} {Phys. Rev. D}\ }\textbf {\bibinfo
  {volume} {102}},\ \bibinfo {pages} {043011} (\bibinfo {year} {2020})},\
  \Eprint {https://arxiv.org/abs/2006.04396} {arXiv:2006.04396 [astro-ph.HE]}
  \BibitemShut {NoStop}%
\bibitem [{\citenamefont {Soultanis}\ \emph {et~al.}(2021)\citenamefont
  {Soultanis}, \citenamefont {Bauswein},\ and\ \citenamefont
  {Stergioulas}}]{Soultanis:2021oia}%
  \BibitemOpen
  \bibfield  {author} {\bibinfo {author} {\bibfnamefont {T.}~\bibnamefont
  {Soultanis}}, \bibinfo {author} {\bibfnamefont {A.}~\bibnamefont
  {Bauswein}},\ and\ \bibinfo {author} {\bibfnamefont {N.}~\bibnamefont
  {Stergioulas}},\ }\bibfield  {title} {\bibinfo {title} {{Analytic model of
  the spectral properties of gravitational waves from neutron star merger
  remnants}},\ }\href@noop {} {\  (\bibinfo {year} {2021})},\ \Eprint
  {https://arxiv.org/abs/2111.08353} {arXiv:2111.08353 [astro-ph.HE]}
  \BibitemShut {NoStop}%
\bibitem [{\citenamefont {Clark}\ \emph {et~al.}(2014)\citenamefont {Clark},
  \citenamefont {Bauswein}, \citenamefont {Cadonati}, \citenamefont {Janka},
  \citenamefont {Pankow},\ and\ \citenamefont {Stergioulas}}]{Clark:2014wua}%
  \BibitemOpen
  \bibfield  {author} {\bibinfo {author} {\bibfnamefont {J.}~\bibnamefont
  {Clark}}, \bibinfo {author} {\bibfnamefont {A.}~\bibnamefont {Bauswein}},
  \bibinfo {author} {\bibfnamefont {L.}~\bibnamefont {Cadonati}}, \bibinfo
  {author} {\bibfnamefont {H.~T.}\ \bibnamefont {Janka}}, \bibinfo {author}
  {\bibfnamefont {C.}~\bibnamefont {Pankow}},\ and\ \bibinfo {author}
  {\bibfnamefont {N.}~\bibnamefont {Stergioulas}},\ }\bibfield  {title}
  {\bibinfo {title} {{Prospects For High Frequency Burst Searches Following
  Binary Neutron Star Coalescence With Advanced Gravitational Wave
  Detectors}},\ }\href {https://doi.org/10.1103/PhysRevD.90.062004} {\bibfield
  {journal} {\bibinfo  {journal} {Phys. Rev. D}\ }\textbf {\bibinfo {volume}
  {90}},\ \bibinfo {pages} {062004} (\bibinfo {year} {2014})},\ \Eprint
  {https://arxiv.org/abs/1406.5444} {arXiv:1406.5444 [astro-ph.HE]}
  \BibitemShut {NoStop}%
\bibitem [{\citenamefont {Clark}\ \emph {et~al.}(2016)\citenamefont {Clark},
  \citenamefont {Bauswein}, \citenamefont {Stergioulas},\ and\ \citenamefont
  {Shoemaker}}]{Clark:2015zxa}%
  \BibitemOpen
  \bibfield  {author} {\bibinfo {author} {\bibfnamefont {J.~A.}\ \bibnamefont
  {Clark}}, \bibinfo {author} {\bibfnamefont {A.}~\bibnamefont {Bauswein}},
  \bibinfo {author} {\bibfnamefont {N.}~\bibnamefont {Stergioulas}},\ and\
  \bibinfo {author} {\bibfnamefont {D.}~\bibnamefont {Shoemaker}},\ }\bibfield
  {title} {\bibinfo {title} {{Observing Gravitational Waves From The
  Post-Merger Phase Of Binary Neutron Star Coalescence}},\ }\href
  {https://doi.org/10.1088/0264-9381/33/8/085003} {\bibfield  {journal}
  {\bibinfo  {journal} {Class. Quant. Grav.}\ }\textbf {\bibinfo {volume}
  {33}},\ \bibinfo {pages} {085003} (\bibinfo {year} {2016})},\ \Eprint
  {https://arxiv.org/abs/1509.08522} {arXiv:1509.08522 [astro-ph.HE]}
  \BibitemShut {NoStop}%
\bibitem [{\citenamefont {Chatziioannou}\ \emph {et~al.}(2017)\citenamefont
  {Chatziioannou}, \citenamefont {Clark}, \citenamefont {Bauswein},
  \citenamefont {Millhouse}, \citenamefont {Littenberg},\ and\ \citenamefont
  {Cornish}}]{Chatziioannou:2017ixj}%
  \BibitemOpen
  \bibfield  {author} {\bibinfo {author} {\bibfnamefont {K.}~\bibnamefont
  {Chatziioannou}}, \bibinfo {author} {\bibfnamefont {J.~A.}\ \bibnamefont
  {Clark}}, \bibinfo {author} {\bibfnamefont {A.}~\bibnamefont {Bauswein}},
  \bibinfo {author} {\bibfnamefont {M.}~\bibnamefont {Millhouse}}, \bibinfo
  {author} {\bibfnamefont {T.~B.}\ \bibnamefont {Littenberg}},\ and\ \bibinfo
  {author} {\bibfnamefont {N.}~\bibnamefont {Cornish}},\ }\bibfield  {title}
  {\bibinfo {title} {{Inferring the post-merger gravitational wave emission
  from binary neutron star coalescences}},\ }\href
  {https://doi.org/10.1103/PhysRevD.96.124035} {\bibfield  {journal} {\bibinfo
  {journal} {Phys. Rev. D}\ }\textbf {\bibinfo {volume} {96}},\ \bibinfo
  {pages} {124035} (\bibinfo {year} {2017})},\ \Eprint
  {https://arxiv.org/abs/1711.00040} {arXiv:1711.00040 [gr-qc]} \BibitemShut
  {NoStop}%
\bibitem [{\citenamefont {Cornish}\ and\ \citenamefont
  {Littenberg}(2015)}]{Cornish:2014kda}%
  \BibitemOpen
  \bibfield  {author} {\bibinfo {author} {\bibfnamefont {N.~J.}\ \bibnamefont
  {Cornish}}\ and\ \bibinfo {author} {\bibfnamefont {T.~B.}\ \bibnamefont
  {Littenberg}},\ }\bibfield  {title} {\bibinfo {title} {{BayesWave: Bayesian
  Inference for Gravitational Wave Bursts and Instrument Glitches}},\ }\href
  {https://doi.org/10.1088/0264-9381/32/13/135012} {\bibfield  {journal}
  {\bibinfo  {journal} {Class. Quant. Grav.}\ }\textbf {\bibinfo {volume}
  {32}},\ \bibinfo {pages} {135012} (\bibinfo {year} {2015})},\ \Eprint
  {https://arxiv.org/abs/1410.3835} {arXiv:1410.3835 [gr-qc]} \BibitemShut
  {NoStop}%
\bibitem [{\citenamefont {Cornish}\ \emph {et~al.}(2021)\citenamefont
  {Cornish}, \citenamefont {Littenberg}, \citenamefont {B\'ecsy}, \citenamefont
  {Chatziioannou}, \citenamefont {Clark}, \citenamefont {Ghonge},\ and\
  \citenamefont {Millhouse}}]{Cornish:2020dwh}%
  \BibitemOpen
  \bibfield  {author} {\bibinfo {author} {\bibfnamefont {N.~J.}\ \bibnamefont
  {Cornish}}, \bibinfo {author} {\bibfnamefont {T.~B.}\ \bibnamefont
  {Littenberg}}, \bibinfo {author} {\bibfnamefont {B.}~\bibnamefont {B\'ecsy}},
  \bibinfo {author} {\bibfnamefont {K.}~\bibnamefont {Chatziioannou}}, \bibinfo
  {author} {\bibfnamefont {J.~A.}\ \bibnamefont {Clark}}, \bibinfo {author}
  {\bibfnamefont {S.}~\bibnamefont {Ghonge}},\ and\ \bibinfo {author}
  {\bibfnamefont {M.}~\bibnamefont {Millhouse}},\ }\bibfield  {title} {\bibinfo
  {title} {{BayesWave analysis pipeline in the era of gravitational wave
  observations}},\ }\href {https://doi.org/10.1103/PhysRevD.103.044006}
  {\bibfield  {journal} {\bibinfo  {journal} {Phys. Rev. D}\ }\textbf {\bibinfo
  {volume} {103}},\ \bibinfo {pages} {044006} (\bibinfo {year} {2021})},\
  \Eprint {https://arxiv.org/abs/2011.09494} {arXiv:2011.09494 [gr-qc]}
  \BibitemShut {NoStop}%
\bibitem [{\citenamefont {{LIGO Scientific Collaboration and Virgo
  Collaboration}}(2018)}]{bayeswave}%
  \BibitemOpen
  \bibfield  {author} {\bibinfo {author} {\bibnamefont {{LIGO Scientific
  Collaboration and Virgo Collaboration}}},\ }\href
  {https://git.ligo.org/lscsoft/bayeswave} {\bibinfo {title} {{BayesWave},
  https://git.ligo.org/lscsoft/bayeswave}} (\bibinfo {year} {2018})\BibitemShut
  {NoStop}%
\bibitem [{\citenamefont {Kanner}\ \emph {et~al.}(2016)\citenamefont {Kanner},
  \citenamefont {Littenberg}, \citenamefont {Cornish}, \citenamefont
  {Millhouse}, \citenamefont {Xhakaj}, \citenamefont {Salemi}, \citenamefont
  {Drago}, \citenamefont {Vedovato},\ and\ \citenamefont
  {Klimenko}}]{Kanner:2016}%
  \BibitemOpen
  \bibfield  {author} {\bibinfo {author} {\bibfnamefont {J.~B.}\ \bibnamefont
  {Kanner}}, \bibinfo {author} {\bibfnamefont {T.~B.}\ \bibnamefont
  {Littenberg}}, \bibinfo {author} {\bibfnamefont {N.}~\bibnamefont {Cornish}},
  \bibinfo {author} {\bibfnamefont {M.}~\bibnamefont {Millhouse}}, \bibinfo
  {author} {\bibfnamefont {E.}~\bibnamefont {Xhakaj}}, \bibinfo {author}
  {\bibfnamefont {F.}~\bibnamefont {Salemi}}, \bibinfo {author} {\bibfnamefont
  {M.}~\bibnamefont {Drago}}, \bibinfo {author} {\bibfnamefont
  {G.}~\bibnamefont {Vedovato}},\ and\ \bibinfo {author} {\bibfnamefont
  {S.}~\bibnamefont {Klimenko}},\ }\bibfield  {title} {\bibinfo {title}
  {Leveraging waveform complexity for confident detection of gravitational
  waves},\ }\href {https://doi.org/10.1103/PhysRevD.93.022002} {\bibfield
  {journal} {\bibinfo  {journal} {Phys. Rev. D}\ }\textbf {\bibinfo {volume}
  {93}},\ \bibinfo {pages} {022002} (\bibinfo {year} {2016})}\BibitemShut
  {NoStop}%
\bibitem [{\citenamefont {Littenberg}\ \emph {et~al.}(2016)\citenamefont
  {Littenberg}, \citenamefont {Kanner}, \citenamefont {Cornish},\ and\
  \citenamefont {Millhouse}}]{Littenberg:2016}%
  \BibitemOpen
  \bibfield  {author} {\bibinfo {author} {\bibfnamefont {T.~B.}\ \bibnamefont
  {Littenberg}}, \bibinfo {author} {\bibfnamefont {J.~B.}\ \bibnamefont
  {Kanner}}, \bibinfo {author} {\bibfnamefont {N.~J.}\ \bibnamefont
  {Cornish}},\ and\ \bibinfo {author} {\bibfnamefont {M.}~\bibnamefont
  {Millhouse}},\ }\bibfield  {title} {\bibinfo {title} {Enabling high
  confidence detections of gravitational-wave bursts},\ }\href
  {https://doi.org/10.1103/PhysRevD.94.044050} {\bibfield  {journal} {\bibinfo
  {journal} {Phys. Rev. D}\ }\textbf {\bibinfo {volume} {94}},\ \bibinfo
  {pages} {044050} (\bibinfo {year} {2016})}\BibitemShut {NoStop}%
\bibitem [{\citenamefont {{B{\'e}csy}}\ \emph {et~al.}(2017)\citenamefont
  {{B{\'e}csy}}, \citenamefont {{Raffai}}, \citenamefont {{Cornish}},
  \citenamefont {{Essick}}, \citenamefont {{Kanner}}, \citenamefont
  {{Katsavounidis}}, \citenamefont {{Littenberg}}, \citenamefont
  {{Millhouse}},\ and\ \citenamefont {{Vitale}}}]{becsy:2017}%
  \BibitemOpen
  \bibfield  {author} {\bibinfo {author} {\bibfnamefont {B.}~\bibnamefont
  {{B{\'e}csy}}}, \bibinfo {author} {\bibfnamefont {P.}~\bibnamefont
  {{Raffai}}}, \bibinfo {author} {\bibfnamefont {N.~J.}\ \bibnamefont
  {{Cornish}}}, \bibinfo {author} {\bibfnamefont {R.}~\bibnamefont {{Essick}}},
  \bibinfo {author} {\bibfnamefont {J.}~\bibnamefont {{Kanner}}}, \bibinfo
  {author} {\bibfnamefont {E.}~\bibnamefont {{Katsavounidis}}}, \bibinfo
  {author} {\bibfnamefont {T.~B.}\ \bibnamefont {{Littenberg}}}, \bibinfo
  {author} {\bibfnamefont {M.}~\bibnamefont {{Millhouse}}},\ and\ \bibinfo
  {author} {\bibfnamefont {S.}~\bibnamefont {{Vitale}}},\ }\bibfield  {title}
  {\bibinfo {title} {{Parameter Estimation for Gravitational-wave Bursts with
  the BayesWave Pipeline}},\ }\href {https://doi.org/10.3847/1538-4357/aa63ef}
  {\bibfield  {journal} {\bibinfo  {journal} {\apj}\ }\textbf {\bibinfo
  {volume} {839}},\ \bibinfo {eid} {15} (\bibinfo {year} {2017})},\ \Eprint
  {https://arxiv.org/abs/1612.02003} {arXiv:1612.02003 [astro-ph.HE]}
  \BibitemShut {NoStop}%
\bibitem [{\citenamefont {Tsang}\ \emph {et~al.}(2018)\citenamefont {Tsang},
  \citenamefont {Rollier}, \citenamefont {Ghosh}, \citenamefont {Samajdar},
  \citenamefont {Agathos}, \citenamefont {Chatziioannou}, \citenamefont
  {Cardoso}, \citenamefont {Khanna},\ and\ \citenamefont {Van
  Den~Broeck}}]{Tsang:2018uie}%
  \BibitemOpen
  \bibfield  {author} {\bibinfo {author} {\bibfnamefont {K.~W.}\ \bibnamefont
  {Tsang}}, \bibinfo {author} {\bibfnamefont {M.}~\bibnamefont {Rollier}},
  \bibinfo {author} {\bibfnamefont {A.}~\bibnamefont {Ghosh}}, \bibinfo
  {author} {\bibfnamefont {A.}~\bibnamefont {Samajdar}}, \bibinfo {author}
  {\bibfnamefont {M.}~\bibnamefont {Agathos}}, \bibinfo {author} {\bibfnamefont
  {K.}~\bibnamefont {Chatziioannou}}, \bibinfo {author} {\bibfnamefont
  {V.}~\bibnamefont {Cardoso}}, \bibinfo {author} {\bibfnamefont
  {G.}~\bibnamefont {Khanna}},\ and\ \bibinfo {author} {\bibfnamefont
  {C.}~\bibnamefont {Van Den~Broeck}},\ }\bibfield  {title} {\bibinfo {title}
  {{A morphology-independent data analysis method for detecting and
  characterizing gravitational wave echoes}},\ }\href
  {https://doi.org/10.1103/PhysRevD.98.024023} {\bibfield  {journal} {\bibinfo
  {journal} {Phys. Rev. D}\ }\textbf {\bibinfo {volume} {98}},\ \bibinfo
  {pages} {024023} (\bibinfo {year} {2018})},\ \Eprint
  {https://arxiv.org/abs/1804.04877} {arXiv:1804.04877 [gr-qc]} \BibitemShut
  {NoStop}%
\bibitem [{\citenamefont {Tsang}\ \emph {et~al.}(2020)\citenamefont {Tsang},
  \citenamefont {Ghosh}, \citenamefont {Samajdar}, \citenamefont
  {Chatziioannou}, \citenamefont {Mastrogiovanni}, \citenamefont {Agathos},\
  and\ \citenamefont {Van Den~Broeck}}]{Tsang:2019zra}%
  \BibitemOpen
  \bibfield  {author} {\bibinfo {author} {\bibfnamefont {K.~W.}\ \bibnamefont
  {Tsang}}, \bibinfo {author} {\bibfnamefont {A.}~\bibnamefont {Ghosh}},
  \bibinfo {author} {\bibfnamefont {A.}~\bibnamefont {Samajdar}}, \bibinfo
  {author} {\bibfnamefont {K.}~\bibnamefont {Chatziioannou}}, \bibinfo {author}
  {\bibfnamefont {S.}~\bibnamefont {Mastrogiovanni}}, \bibinfo {author}
  {\bibfnamefont {M.}~\bibnamefont {Agathos}},\ and\ \bibinfo {author}
  {\bibfnamefont {C.}~\bibnamefont {Van Den~Broeck}},\ }\bibfield  {title}
  {\bibinfo {title} {{A morphology-independent search for gravitational wave
  echoes in data from the first and second observing runs of Advanced LIGO and
  Advanced Virgo}},\ }\href {https://doi.org/10.1103/PhysRevD.101.064012}
  {\bibfield  {journal} {\bibinfo  {journal} {Phys. Rev. D}\ }\textbf {\bibinfo
  {volume} {101}},\ \bibinfo {pages} {064012} (\bibinfo {year} {2020})},\
  \Eprint {https://arxiv.org/abs/1906.11168} {arXiv:1906.11168 [gr-qc]}
  \BibitemShut {NoStop}%
\bibitem [{\citenamefont {Pankow}\ \emph {et~al.}(2018)\citenamefont {Pankow}
  \emph {et~al.}}]{Pankow:2018qpo}%
  \BibitemOpen
  \bibfield  {author} {\bibinfo {author} {\bibfnamefont {C.}~\bibnamefont
  {Pankow}} \emph {et~al.},\ }\bibfield  {title} {\bibinfo {title} {{Mitigation
  of the instrumental noise transient in gravitational-wave data surrounding
  GW170817}},\ }\href {https://doi.org/10.1103/PhysRevD.98.084016} {\bibfield
  {journal} {\bibinfo  {journal} {Phys. Rev.}\ }\textbf {\bibinfo {volume}
  {D98}},\ \bibinfo {pages} {084016} (\bibinfo {year} {2018})},\ \Eprint
  {https://arxiv.org/abs/1808.03619} {arXiv:1808.03619 [gr-qc]} \BibitemShut
  {NoStop}%
\bibitem [{\citenamefont {Millhouse}\ \emph {et~al.}(2018)\citenamefont
  {Millhouse}, \citenamefont {Cornish},\ and\ \citenamefont
  {Littenberg}}]{Millhouse:2018dgi}%
  \BibitemOpen
  \bibfield  {author} {\bibinfo {author} {\bibfnamefont {M.}~\bibnamefont
  {Millhouse}}, \bibinfo {author} {\bibfnamefont {N.~J.}\ \bibnamefont
  {Cornish}},\ and\ \bibinfo {author} {\bibfnamefont {T.}~\bibnamefont
  {Littenberg}},\ }\bibfield  {title} {\bibinfo {title} {{Bayesian
  reconstruction of gravitational wave bursts using chirplets}},\ }\href
  {https://doi.org/10.1103/PhysRevD.97.104057} {\bibfield  {journal} {\bibinfo
  {journal} {Phys. Rev.}\ }\textbf {\bibinfo {volume} {D97}},\ \bibinfo {pages}
  {104057} (\bibinfo {year} {2018})},\ \Eprint
  {https://arxiv.org/abs/1804.03239} {arXiv:1804.03239 [gr-qc]} \BibitemShut
  {NoStop}%
\bibitem [{\citenamefont {Ghonge}\ \emph {et~al.}(2020)\citenamefont {Ghonge},
  \citenamefont {Chatziioannou}, \citenamefont {Clark}, \citenamefont
  {Littenberg}, \citenamefont {Millhouse}, \citenamefont {Cadonati},\ and\
  \citenamefont {Cornish}}]{Ghonge:2020suv}%
  \BibitemOpen
  \bibfield  {author} {\bibinfo {author} {\bibfnamefont {S.}~\bibnamefont
  {Ghonge}}, \bibinfo {author} {\bibfnamefont {K.}~\bibnamefont
  {Chatziioannou}}, \bibinfo {author} {\bibfnamefont {J.~A.}\ \bibnamefont
  {Clark}}, \bibinfo {author} {\bibfnamefont {T.}~\bibnamefont {Littenberg}},
  \bibinfo {author} {\bibfnamefont {M.}~\bibnamefont {Millhouse}}, \bibinfo
  {author} {\bibfnamefont {L.}~\bibnamefont {Cadonati}},\ and\ \bibinfo
  {author} {\bibfnamefont {N.}~\bibnamefont {Cornish}},\ }\bibfield  {title}
  {\bibinfo {title} {{Reconstructing gravitational wave signals from binary
  black hole mergers with minimal assumptions}},\ }\href@noop {} {\  (\bibinfo
  {year} {2020})},\ \Eprint {https://arxiv.org/abs/2003.09456}
  {arXiv:2003.09456 [gr-qc]} \BibitemShut {NoStop}%
\bibitem [{\citenamefont {D\'alya}\ \emph {et~al.}(2021)\citenamefont
  {D\'alya}, \citenamefont {Raffai},\ and\ \citenamefont
  {B\'ecsy}}]{Dalya:2020gra}%
  \BibitemOpen
  \bibfield  {author} {\bibinfo {author} {\bibfnamefont {G.}~\bibnamefont
  {D\'alya}}, \bibinfo {author} {\bibfnamefont {P.}~\bibnamefont {Raffai}},\
  and\ \bibinfo {author} {\bibfnamefont {B.}~\bibnamefont {B\'ecsy}},\
  }\bibfield  {title} {\bibinfo {title} {{Bayesian Reconstruction of
  Gravitational-wave Signals from Binary Black Holes with Nonzero
  Eccentricities}},\ }\href {https://doi.org/10.1088/1361-6382/abd7bf}
  {\bibfield  {journal} {\bibinfo  {journal} {Class. Quant. Grav.}\ }\textbf
  {\bibinfo {volume} {38}},\ \bibinfo {pages} {065002} (\bibinfo {year}
  {2021})},\ \Eprint {https://arxiv.org/abs/2006.06256} {arXiv:2006.06256
  [astro-ph.HE]} \BibitemShut {NoStop}%
\bibitem [{\citenamefont {Chatziioannou}\ \emph
  {et~al.}(2021{\natexlab{a}})\citenamefont {Chatziioannou}, \citenamefont
  {Isi}, \citenamefont {Haster},\ and\ \citenamefont
  {Littenberg}}]{Chatziioannou:2021mij}%
  \BibitemOpen
  \bibfield  {author} {\bibinfo {author} {\bibfnamefont {K.}~\bibnamefont
  {Chatziioannou}}, \bibinfo {author} {\bibfnamefont {M.}~\bibnamefont {Isi}},
  \bibinfo {author} {\bibfnamefont {C.-J.}\ \bibnamefont {Haster}},\ and\
  \bibinfo {author} {\bibfnamefont {T.~B.}\ \bibnamefont {Littenberg}},\
  }\bibfield  {title} {\bibinfo {title} {{Morphology-independent test of the
  mixed polarization content of transient gravitational wave signals}},\ }\href
  {https://doi.org/10.1103/PhysRevD.104.044005} {\bibfield  {journal} {\bibinfo
   {journal} {Phys. Rev. D}\ }\textbf {\bibinfo {volume} {104}},\ \bibinfo
  {pages} {044005} (\bibinfo {year} {2021}{\natexlab{a}})},\ \Eprint
  {https://arxiv.org/abs/2105.01521} {arXiv:2105.01521 [gr-qc]} \BibitemShut
  {NoStop}%
\bibitem [{\citenamefont {Chatziioannou}\ \emph
  {et~al.}(2021{\natexlab{b}})\citenamefont {Chatziioannou}, \citenamefont
  {Cornish}, \citenamefont {Wijngaarden},\ and\ \citenamefont
  {Littenberg}}]{Chatziioannou:2021ezd}%
  \BibitemOpen
  \bibfield  {author} {\bibinfo {author} {\bibfnamefont {K.}~\bibnamefont
  {Chatziioannou}}, \bibinfo {author} {\bibfnamefont {N.}~\bibnamefont
  {Cornish}}, \bibinfo {author} {\bibfnamefont {M.}~\bibnamefont
  {Wijngaarden}},\ and\ \bibinfo {author} {\bibfnamefont {T.~B.}\ \bibnamefont
  {Littenberg}},\ }\bibfield  {title} {\bibinfo {title} {{Modeling compact
  binary signals and instrumental glitches in gravitational wave data}},\
  }\href {https://doi.org/10.1103/PhysRevD.103.044013} {\bibfield  {journal}
  {\bibinfo  {journal} {Phys. Rev. D}\ }\textbf {\bibinfo {volume} {103}},\
  \bibinfo {pages} {044013} (\bibinfo {year} {2021}{\natexlab{b}})},\ \Eprint
  {https://arxiv.org/abs/2101.01200} {arXiv:2101.01200 [gr-qc]} \BibitemShut
  {NoStop}%
\bibitem [{\citenamefont {Green}(1995)}]{10.1093/biomet/82.4.711}%
  \BibitemOpen
  \bibfield  {author} {\bibinfo {author} {\bibfnamefont {P.~J.}\ \bibnamefont
  {Green}},\ }\bibfield  {title} {\bibinfo {title} {{Reversible jump Markov
  chain Monte Carlo computation and Bayesian model determination}},\ }\href
  {https://doi.org/10.1093/biomet/82.4.711} {\bibfield  {journal} {\bibinfo
  {journal} {Biometrika}\ }\textbf {\bibinfo {volume} {82}},\ \bibinfo {pages}
  {711} (\bibinfo {year} {1995})},\ \Eprint
  {https://arxiv.org/abs/http://oup.prod.sis.lan/biomet/article-pdf/82/4/711/699533/82-4-711.pdf}
  {http://oup.prod.sis.lan/biomet/article-pdf/82/4/711/699533/82-4-711.pdf}
  \BibitemShut {NoStop}%
\bibitem [{\citenamefont {Lartillot}\ and\ \citenamefont
  {Philippe}(2006)}]{10.1080/10635150500433722}%
  \BibitemOpen
  \bibfield  {author} {\bibinfo {author} {\bibfnamefont {N.}~\bibnamefont
  {Lartillot}}\ and\ \bibinfo {author} {\bibfnamefont {H.}~\bibnamefont
  {Philippe}},\ }\bibfield  {title} {\bibinfo {title} {{Computing Bayes Factors
  Using Thermodynamic Integration}},\ }\href
  {https://doi.org/10.1080/10635150500433722} {\bibfield  {journal} {\bibinfo
  {journal} {Systematic Biology}\ }\textbf {\bibinfo {volume} {55}},\ \bibinfo
  {pages} {195} (\bibinfo {year} {2006})},\ \Eprint
  {https://arxiv.org/abs/https://academic.oup.com/sysbio/article-pdf/55/2/195/26557316/10635150500433722.pdf}
  {https://academic.oup.com/sysbio/article-pdf/55/2/195/26557316/10635150500433722.pdf}
  \BibitemShut {NoStop}%
\bibitem [{\citenamefont {\mbox{LIGO Scientific Collaboration, Virgo
  Collaboration}}(2018)}]{lalsuite}%
  \BibitemOpen
  \bibfield  {author} {\bibinfo {author} {\bibnamefont {\mbox{LIGO Scientific
  Collaboration, Virgo Collaboration}}},\ }\href
  {https://doi.org/10.7935/GT1W-FZ16} {\bibinfo {title} {\mbox{LALSuite}}},\
  \bibinfo {howpublished}
  {\href{https://git.ligo.org/lscsoft/lalsuite}{https://git.ligo.org/lscsoft/lalsuite}}
  (\bibinfo {year} {2018})\BibitemShut {NoStop}%
\bibitem [{\citenamefont {Husa}\ \emph {et~al.}(2016)\citenamefont {Husa},
  \citenamefont {Khan}, \citenamefont {Hannam}, \citenamefont {P{\"u}rrer},
  \citenamefont {Ohme}, \citenamefont {Jim{\'e}nez~Forteza},\ and\
  \citenamefont {Boh{\'e}}}]{Husa:2015iqa}%
  \BibitemOpen
  \bibfield  {author} {\bibinfo {author} {\bibfnamefont {S.}~\bibnamefont
  {Husa}}, \bibinfo {author} {\bibfnamefont {S.}~\bibnamefont {Khan}}, \bibinfo
  {author} {\bibfnamefont {M.}~\bibnamefont {Hannam}}, \bibinfo {author}
  {\bibfnamefont {M.}~\bibnamefont {P{\"u}rrer}}, \bibinfo {author}
  {\bibfnamefont {F.}~\bibnamefont {Ohme}}, \bibinfo {author} {\bibfnamefont
  {X.}~\bibnamefont {Jim{\'e}nez~Forteza}},\ and\ \bibinfo {author}
  {\bibfnamefont {A.}~\bibnamefont {Boh{\'e}}},\ }\bibfield  {title} {\bibinfo
  {title} {{Frequency-domain gravitational waves from nonprecessing black-hole
  binaries. I. New numerical waveforms and anatomy of the signal}},\ }\href
  {https://doi.org/10.1103/PhysRevD.93.044006} {\bibfield  {journal} {\bibinfo
  {journal} {Phys. Rev. D}\ }\textbf {\bibinfo {volume} {93}},\ \bibinfo
  {pages} {044006} (\bibinfo {year} {2016})},\ \Eprint
  {https://arxiv.org/abs/1508.07250} {arXiv:1508.07250 [gr-qc]} \BibitemShut
  {NoStop}%
\bibitem [{\citenamefont {Khan}\ \emph {et~al.}(2016)\citenamefont {Khan},
  \citenamefont {Husa}, \citenamefont {Hannam}, \citenamefont {Ohme},
  \citenamefont {P{\"u}rrer}, \citenamefont {Jim{\'e}nez~Forteza},\ and\
  \citenamefont {Boh{\'e}}}]{Khan:2015jqa}%
  \BibitemOpen
  \bibfield  {author} {\bibinfo {author} {\bibfnamefont {S.}~\bibnamefont
  {Khan}}, \bibinfo {author} {\bibfnamefont {S.}~\bibnamefont {Husa}}, \bibinfo
  {author} {\bibfnamefont {M.}~\bibnamefont {Hannam}}, \bibinfo {author}
  {\bibfnamefont {F.}~\bibnamefont {Ohme}}, \bibinfo {author} {\bibfnamefont
  {M.}~\bibnamefont {P{\"u}rrer}}, \bibinfo {author} {\bibfnamefont
  {X.}~\bibnamefont {Jim{\'e}nez~Forteza}},\ and\ \bibinfo {author}
  {\bibfnamefont {A.}~\bibnamefont {Boh{\'e}}},\ }\bibfield  {title} {\bibinfo
  {title} {{Frequency-domain gravitational waves from nonprecessing black-hole
  binaries. II. A phenomenological model for the advanced detector era}},\
  }\href {https://doi.org/10.1103/PhysRevD.93.044007} {\bibfield  {journal}
  {\bibinfo  {journal} {Phys. Rev. D}\ }\textbf {\bibinfo {volume} {93}},\
  \bibinfo {pages} {044007} (\bibinfo {year} {2016})},\ \Eprint
  {https://arxiv.org/abs/1508.07253} {arXiv:1508.07253 [gr-qc]} \BibitemShut
  {NoStop}%
\bibitem [{\citenamefont {Dietrich}\ \emph {et~al.}(2017)\citenamefont
  {Dietrich}, \citenamefont {Bernuzzi},\ and\ \citenamefont
  {Tichy}}]{Dietrich:2017aum}%
  \BibitemOpen
  \bibfield  {author} {\bibinfo {author} {\bibfnamefont {T.}~\bibnamefont
  {Dietrich}}, \bibinfo {author} {\bibfnamefont {S.}~\bibnamefont {Bernuzzi}},\
  and\ \bibinfo {author} {\bibfnamefont {W.}~\bibnamefont {Tichy}},\ }\bibfield
   {title} {\bibinfo {title} {{Closed-form tidal approximants for binary
  neutron star gravitational waveforms constructed from high-resolution
  numerical relativity simulations}},\ }\href
  {https://doi.org/10.1103/PhysRevD.96.121501} {\bibfield  {journal} {\bibinfo
  {journal} {Phys. Rev. D}\ }\textbf {\bibinfo {volume} {96}},\ \bibinfo
  {pages} {121501} (\bibinfo {year} {2017})},\ \Eprint
  {https://arxiv.org/abs/1706.02969} {arXiv:1706.02969 [gr-qc]} \BibitemShut
  {NoStop}%
\bibitem [{\citenamefont {Dietrich}\ \emph
  {et~al.}(2019{\natexlab{a}})\citenamefont {Dietrich} \emph
  {et~al.}}]{Dietrich:2018uni}%
  \BibitemOpen
  \bibfield  {author} {\bibinfo {author} {\bibfnamefont {T.}~\bibnamefont
  {Dietrich}} \emph {et~al.},\ }\bibfield  {title} {\bibinfo {title} {{Matter
  imprints in waveform models for neutron star binaries: Tidal and self-spin
  effects}},\ }\href {https://doi.org/10.1103/PhysRevD.99.024029} {\bibfield
  {journal} {\bibinfo  {journal} {Phys. Rev.}\ }\textbf {\bibinfo {volume}
  {D99}},\ \bibinfo {pages} {024029} (\bibinfo {year} {2019}{\natexlab{a}})},\
  \Eprint {https://arxiv.org/abs/1804.02235} {arXiv:1804.02235 [gr-qc]}
  \BibitemShut {NoStop}%
\bibitem [{\citenamefont {Dietrich}\ \emph
  {et~al.}(2019{\natexlab{b}})\citenamefont {Dietrich}, \citenamefont
  {Samajdar}, \citenamefont {Khan}, \citenamefont {Johnson-McDaniel},
  \citenamefont {Dudi},\ and\ \citenamefont {Tichy}}]{Dietrich:2019kaq}%
  \BibitemOpen
  \bibfield  {author} {\bibinfo {author} {\bibfnamefont {T.}~\bibnamefont
  {Dietrich}}, \bibinfo {author} {\bibfnamefont {A.}~\bibnamefont {Samajdar}},
  \bibinfo {author} {\bibfnamefont {S.}~\bibnamefont {Khan}}, \bibinfo {author}
  {\bibfnamefont {N.~K.}\ \bibnamefont {Johnson-McDaniel}}, \bibinfo {author}
  {\bibfnamefont {R.}~\bibnamefont {Dudi}},\ and\ \bibinfo {author}
  {\bibfnamefont {W.}~\bibnamefont {Tichy}},\ }\bibfield  {title} {\bibinfo
  {title} {{Improving the NRTidal model for binary neutron star systems}},\
  }\href {https://doi.org/10.1103/PhysRevD.100.044003} {\bibfield  {journal}
  {\bibinfo  {journal} {Phys. Rev. D}\ }\textbf {\bibinfo {volume} {100}},\
  \bibinfo {pages} {044003} (\bibinfo {year} {2019}{\natexlab{b}})},\ \Eprint
  {https://arxiv.org/abs/1905.06011} {arXiv:1905.06011 [gr-qc]} \BibitemShut
  {NoStop}%
\bibitem [{\citenamefont {P\"urrer}\ \emph {et~al.}(2016)\citenamefont
  {P\"urrer}, \citenamefont {Hannam},\ and\ \citenamefont
  {Ohme}}]{Purrer:2015nkh}%
  \BibitemOpen
  \bibfield  {author} {\bibinfo {author} {\bibfnamefont {M.}~\bibnamefont
  {P\"urrer}}, \bibinfo {author} {\bibfnamefont {M.}~\bibnamefont {Hannam}},\
  and\ \bibinfo {author} {\bibfnamefont {F.}~\bibnamefont {Ohme}},\ }\bibfield
  {title} {\bibinfo {title} {{Can we measure individual black-hole spins from
  gravitational-wave observations?}},\ }\href
  {https://doi.org/10.1103/PhysRevD.93.084042} {\bibfield  {journal} {\bibinfo
  {journal} {Phys. Rev. D}\ }\textbf {\bibinfo {volume} {93}},\ \bibinfo
  {pages} {084042} (\bibinfo {year} {2016})},\ \Eprint
  {https://arxiv.org/abs/1512.04955} {arXiv:1512.04955 [gr-qc]} \BibitemShut
  {NoStop}%
\bibitem [{\citenamefont {Chatziioannou}\ \emph
  {et~al.}(2018{\natexlab{a}})\citenamefont {Chatziioannou}, \citenamefont
  {Lovelace}, \citenamefont {Boyle}, \citenamefont {Giesler}, \citenamefont
  {Hemberger}, \citenamefont {Katebi}, \citenamefont {Kidder}, \citenamefont
  {Pfeiffer}, \citenamefont {Scheel},\ and\ \citenamefont
  {Szil\'agyi}}]{Chatziioannou:2018wqx}%
  \BibitemOpen
  \bibfield  {author} {\bibinfo {author} {\bibfnamefont {K.}~\bibnamefont
  {Chatziioannou}}, \bibinfo {author} {\bibfnamefont {G.}~\bibnamefont
  {Lovelace}}, \bibinfo {author} {\bibfnamefont {M.}~\bibnamefont {Boyle}},
  \bibinfo {author} {\bibfnamefont {M.}~\bibnamefont {Giesler}}, \bibinfo
  {author} {\bibfnamefont {D.~A.}\ \bibnamefont {Hemberger}}, \bibinfo {author}
  {\bibfnamefont {R.}~\bibnamefont {Katebi}}, \bibinfo {author} {\bibfnamefont
  {L.~E.}\ \bibnamefont {Kidder}}, \bibinfo {author} {\bibfnamefont {H.~P.}\
  \bibnamefont {Pfeiffer}}, \bibinfo {author} {\bibfnamefont {M.~A.}\
  \bibnamefont {Scheel}},\ and\ \bibinfo {author} {\bibfnamefont
  {B.}~\bibnamefont {Szil\'agyi}},\ }\bibfield  {title} {\bibinfo {title}
  {{Measuring the properties of nearly extremal black holes with gravitational
  waves}},\ }\href {https://doi.org/10.1103/PhysRevD.98.044028} {\bibfield
  {journal} {\bibinfo  {journal} {Phys. Rev. D}\ }\textbf {\bibinfo {volume}
  {98}},\ \bibinfo {pages} {044028} (\bibinfo {year} {2018}{\natexlab{a}})},\
  \Eprint {https://arxiv.org/abs/1804.03704} {arXiv:1804.03704 [gr-qc]}
  \BibitemShut {NoStop}%
\bibitem [{\citenamefont {Racine}(2008)}]{Racine:2008qv}%
  \BibitemOpen
  \bibfield  {author} {\bibinfo {author} {\bibfnamefont {E.}~\bibnamefont
  {Racine}},\ }\bibfield  {title} {\bibinfo {title} {{Analysis of spin
  precession in binary black hole systems including quadrupole-monopole
  interaction}},\ }\href {https://doi.org/10.1103/PhysRevD.78.044021}
  {\bibfield  {journal} {\bibinfo  {journal} {Phys. Rev.}\ }\textbf {\bibinfo
  {volume} {D78}},\ \bibinfo {pages} {044021} (\bibinfo {year} {2008})},\
  \Eprint {https://arxiv.org/abs/0803.1820} {arXiv:0803.1820 [gr-qc]}
  \BibitemShut {NoStop}%
\bibitem [{\citenamefont {Favata}(2014)}]{Favata:2013rwa}%
  \BibitemOpen
  \bibfield  {author} {\bibinfo {author} {\bibfnamefont {M.}~\bibnamefont
  {Favata}},\ }\bibfield  {title} {\bibinfo {title} {{Systematic parameter
  errors in inspiraling neutron star binaries}},\ }\href
  {https://doi.org/10.1103/PhysRevLett.112.101101} {\bibfield  {journal}
  {\bibinfo  {journal} {Phys. Rev. Lett.}\ }\textbf {\bibinfo {volume} {112}},\
  \bibinfo {pages} {101101} (\bibinfo {year} {2014})},\ \Eprint
  {https://arxiv.org/abs/1310.8288} {arXiv:1310.8288 [gr-qc]} \BibitemShut
  {NoStop}%
\bibitem [{\citenamefont {Wade}\ \emph {et~al.}(2014)\citenamefont {Wade},
  \citenamefont {Creighton}, \citenamefont {Ochsner}, \citenamefont {Lackey},
  \citenamefont {Farr} \emph {et~al.}}]{Wade:2014vqa}%
  \BibitemOpen
  \bibfield  {author} {\bibinfo {author} {\bibfnamefont {L.}~\bibnamefont
  {Wade}}, \bibinfo {author} {\bibfnamefont {J.~D.}\ \bibnamefont {Creighton}},
  \bibinfo {author} {\bibfnamefont {E.}~\bibnamefont {Ochsner}}, \bibinfo
  {author} {\bibfnamefont {B.~D.}\ \bibnamefont {Lackey}}, \bibinfo {author}
  {\bibfnamefont {B.~F.}\ \bibnamefont {Farr}}, \emph {et~al.},\ }\bibfield
  {title} {\bibinfo {title} {{Systematic and statistical errors in a bayesian
  approach to the estimation of the neutron-star equation of state using
  advanced gravitational wave detectors}},\ }\href
  {https://doi.org/10.1103/PhysRevD.89.103012} {\bibfield  {journal} {\bibinfo
  {journal} {Phys. Rev. D}\ }\textbf {\bibinfo {volume} {89}},\ \bibinfo
  {pages} {103012} (\bibinfo {year} {2014})},\ \Eprint
  {https://arxiv.org/abs/1402.5156} {arXiv:1402.5156 [gr-qc]} \BibitemShut
  {NoStop}%
\bibitem [{\citenamefont {Binnington}\ and\ \citenamefont
  {Poisson}(2009)}]{Binnington:2009bb}%
  \BibitemOpen
  \bibfield  {author} {\bibinfo {author} {\bibfnamefont {T.}~\bibnamefont
  {Binnington}}\ and\ \bibinfo {author} {\bibfnamefont {E.}~\bibnamefont
  {Poisson}},\ }\bibfield  {title} {\bibinfo {title} {{Relativistic theory of
  tidal Love numbers}},\ }\href {https://doi.org/10.1103/PhysRevD.80.084018}
  {\bibfield  {journal} {\bibinfo  {journal} {Phys. Rev.}\ }\textbf {\bibinfo
  {volume} {D80}},\ \bibinfo {pages} {084018} (\bibinfo {year} {2009})},\
  \Eprint {https://arxiv.org/abs/0906.1366} {arXiv:0906.1366 [gr-qc]}
  \BibitemShut {NoStop}%
\bibitem [{\citenamefont {Poisson}(2021)}]{Poisson:2020vap}%
  \BibitemOpen
  \bibfield  {author} {\bibinfo {author} {\bibfnamefont {E.}~\bibnamefont
  {Poisson}},\ }\bibfield  {title} {\bibinfo {title} {{Compact body in a tidal
  environment: New types of relativistic Love numbers, and a post-Newtonian
  operational definition for tidally induced multipole moments}},\ }\href
  {https://doi.org/10.1103/PhysRevD.103.064023} {\bibfield  {journal} {\bibinfo
   {journal} {Phys. Rev. D}\ }\textbf {\bibinfo {volume} {103}},\ \bibinfo
  {pages} {064023} (\bibinfo {year} {2021})},\ \Eprint
  {https://arxiv.org/abs/2012.10184} {arXiv:2012.10184 [gr-qc]} \BibitemShut
  {NoStop}%
\bibitem [{\citenamefont {Cutler}\ and\ \citenamefont
  {Flanagan}(1994)}]{Cutler:1994ys}%
  \BibitemOpen
  \bibfield  {author} {\bibinfo {author} {\bibfnamefont {C.}~\bibnamefont
  {Cutler}}\ and\ \bibinfo {author} {\bibfnamefont {E.~E.}\ \bibnamefont
  {Flanagan}},\ }\bibfield  {title} {\bibinfo {title} {{Gravitational waves
  from merging compact binaries: How accurately can one extract the binary's
  parameters from the inspiral wave form?}},\ }\href
  {https://doi.org/10.1103/PhysRevD.49.2658} {\bibfield  {journal} {\bibinfo
  {journal} {Phys. Rev. D}\ }\textbf {\bibinfo {volume} {49}},\ \bibinfo
  {pages} {2658} (\bibinfo {year} {1994})},\ \Eprint
  {https://arxiv.org/abs/gr-qc/9402014} {arXiv:gr-qc/9402014} \BibitemShut
  {NoStop}%
\bibitem [{\citenamefont {Callister}(2021)}]{Callister:2021gxf}%
  \BibitemOpen
  \bibfield  {author} {\bibinfo {author} {\bibfnamefont {T.}~\bibnamefont
  {Callister}},\ }\bibfield  {title} {\bibinfo {title} {{A Thesaurus for Common
  Priors in Gravitational-Wave Astronomy}},\ }\href@noop {} {\  (\bibinfo
  {year} {2021})},\ \Eprint {https://arxiv.org/abs/2104.09508}
  {arXiv:2104.09508 [gr-qc]} \BibitemShut {NoStop}%
\bibitem [{\citenamefont {Chatziioannou}\ \emph
  {et~al.}(2018{\natexlab{b}})\citenamefont {Chatziioannou}, \citenamefont
  {Haster},\ and\ \citenamefont {Zimmerman}}]{Chatziioannou:2018vzf}%
  \BibitemOpen
  \bibfield  {author} {\bibinfo {author} {\bibfnamefont {K.}~\bibnamefont
  {Chatziioannou}}, \bibinfo {author} {\bibfnamefont {C.-J.}\ \bibnamefont
  {Haster}},\ and\ \bibinfo {author} {\bibfnamefont {A.}~\bibnamefont
  {Zimmerman}},\ }\bibfield  {title} {\bibinfo {title} {{Measuring the neutron
  star tidal deformability with equation-of-state-independent relations and
  gravitational waves}},\ }\href {https://doi.org/10.1103/PhysRevD.97.104036}
  {\bibfield  {journal} {\bibinfo  {journal} {Phys. Rev.}\ }\textbf {\bibinfo
  {volume} {D97}},\ \bibinfo {pages} {104036} (\bibinfo {year}
  {2018}{\natexlab{b}})},\ \Eprint {https://arxiv.org/abs/1804.03221}
  {arXiv:1804.03221 [gr-qc]} \BibitemShut {NoStop}%
\bibitem [{\citenamefont {Zhao}\ and\ \citenamefont
  {Lattimer}(2018)}]{Zhao:2018nyf}%
  \BibitemOpen
  \bibfield  {author} {\bibinfo {author} {\bibfnamefont {T.}~\bibnamefont
  {Zhao}}\ and\ \bibinfo {author} {\bibfnamefont {J.~M.}\ \bibnamefont
  {Lattimer}},\ }\bibfield  {title} {\bibinfo {title} {{Tidal Deformabilities
  and Neutron Star Mergers}},\ }\href
  {https://doi.org/10.1103/PhysRevD.98.063020} {\bibfield  {journal} {\bibinfo
  {journal} {Phys. Rev. D}\ }\textbf {\bibinfo {volume} {98}},\ \bibinfo
  {pages} {063020} (\bibinfo {year} {2018})},\ \Eprint
  {https://arxiv.org/abs/1808.02858} {arXiv:1808.02858 [astro-ph.HE]}
  \BibitemShut {NoStop}%
\bibitem [{\citenamefont {Carson}\ \emph {et~al.}(2019)\citenamefont {Carson},
  \citenamefont {Chatziioannou}, \citenamefont {Haster}, \citenamefont {Yagi},\
  and\ \citenamefont {Yunes}}]{Carson:2019rjx}%
  \BibitemOpen
  \bibfield  {author} {\bibinfo {author} {\bibfnamefont {Z.}~\bibnamefont
  {Carson}}, \bibinfo {author} {\bibfnamefont {K.}~\bibnamefont
  {Chatziioannou}}, \bibinfo {author} {\bibfnamefont {C.-J.}\ \bibnamefont
  {Haster}}, \bibinfo {author} {\bibfnamefont {K.}~\bibnamefont {Yagi}},\ and\
  \bibinfo {author} {\bibfnamefont {N.}~\bibnamefont {Yunes}},\ }\bibfield
  {title} {\bibinfo {title} {{Equation-of-state insensitive relations after
  GW170817}},\ }\href {https://doi.org/10.1103/PhysRevD.99.083016} {\bibfield
  {journal} {\bibinfo  {journal} {Phys. Rev. D}\ }\textbf {\bibinfo {volume}
  {99}},\ \bibinfo {pages} {083016} (\bibinfo {year} {2019})},\ \Eprint
  {https://arxiv.org/abs/1903.03909} {arXiv:1903.03909 [gr-qc]} \BibitemShut
  {NoStop}%
\bibitem [{\citenamefont {Cornish}(2021)}]{Cornish:2021wxy}%
  \BibitemOpen
  \bibfield  {author} {\bibinfo {author} {\bibfnamefont {N.~J.}\ \bibnamefont
  {Cornish}},\ }\bibfield  {title} {\bibinfo {title} {{Rapid and Robust
  Parameter Inference for Binary Mergers}},\ }\href
  {https://doi.org/10.1103/PhysRevD.103.104057} {\bibfield  {journal} {\bibinfo
   {journal} {Phys. Rev. D}\ }\textbf {\bibinfo {volume} {103}},\ \bibinfo
  {pages} {104057} (\bibinfo {year} {2021})},\ \Eprint
  {https://arxiv.org/abs/2101.01188} {arXiv:2101.01188 [gr-qc]} \BibitemShut
  {NoStop}%
\bibitem [{\citenamefont {{Ter Braak}}(2006)}]{2006S&C....16..239T}%
  \BibitemOpen
  \bibfield  {author} {\bibinfo {author} {\bibfnamefont {C.~J.~F.}\
  \bibnamefont {{Ter Braak}}},\ }\bibfield  {title} {\bibinfo {title} {{A
  Markov Chain Monte Carlo version of the genetic algorithm Differential
  Evolution: easy Bayesian computing for real parameter spaces}},\ }\href
  {https://doi.org/10.1007/s11222-006-8769-1} {\bibfield  {journal} {\bibinfo
  {journal} {Statistics and Computing}\ }\textbf {\bibinfo {volume} {16}},\
  \bibinfo {pages} {239} (\bibinfo {year} {2006})}\BibitemShut {NoStop}%
\bibitem [{\citenamefont {{Cornish}}(2010)}]{2010arXiv1007.4820C}%
  \BibitemOpen
  \bibfield  {author} {\bibinfo {author} {\bibfnamefont {N.~J.}\ \bibnamefont
  {{Cornish}}},\ }\bibfield  {title} {\bibinfo {title} {{Fast Fisher Matrices
  and Lazy Likelihoods}},\ }\href@noop {} {\bibfield  {journal} {\bibinfo
  {journal} {arXiv e-prints}\ ,\ \bibinfo {eid} {arXiv:1007.4820}} (\bibinfo
  {year} {2010})},\ \Eprint {https://arxiv.org/abs/1007.4820} {arXiv:1007.4820
  [gr-qc]} \BibitemShut {NoStop}%
\bibitem [{\citenamefont {Raymond}\ and\ \citenamefont
  {Farr}(2014)}]{Raymond:2014uha}%
  \BibitemOpen
  \bibfield  {author} {\bibinfo {author} {\bibfnamefont {V.}~\bibnamefont
  {Raymond}}\ and\ \bibinfo {author} {\bibfnamefont {W.~M.}\ \bibnamefont
  {Farr}},\ }\bibfield  {title} {\bibinfo {title} {{Physically motivated
  exploration of the extrinsic parameter space in ground-based
  gravitational-wave astronomy}},\ }\href@noop {} {\  (\bibinfo {year}
  {2014})},\ \Eprint {https://arxiv.org/abs/1402.0053} {arXiv:1402.0053
  [gr-qc]} \BibitemShut {NoStop}%
\bibitem [{\citenamefont {Veitch}\ \emph {et~al.}(2015)\citenamefont {Veitch}
  \emph {et~al.}}]{Veitch:2014wba}%
  \BibitemOpen
  \bibfield  {author} {\bibinfo {author} {\bibfnamefont {J.}~\bibnamefont
  {Veitch}} \emph {et~al.},\ }\bibfield  {title} {\bibinfo {title} {{Parameter
  estimation for compact binaries with ground-based gravitational-wave
  observations using the LALInference software library}},\ }\href
  {https://doi.org/10.1103/PhysRevD.91.042003} {\bibfield  {journal} {\bibinfo
  {journal} {Phys. Rev. D}\ }\textbf {\bibinfo {volume} {91}},\ \bibinfo
  {pages} {042003} (\bibinfo {year} {2015})},\ \Eprint
  {https://arxiv.org/abs/1409.7215} {arXiv:1409.7215 [gr-qc]} \BibitemShut
  {NoStop}%
\bibitem [{\citenamefont {Littenberg}\ and\ \citenamefont
  {Cornish}(2015)}]{Littenberg:2014oda}%
  \BibitemOpen
  \bibfield  {author} {\bibinfo {author} {\bibfnamefont {T.~B.}\ \bibnamefont
  {Littenberg}}\ and\ \bibinfo {author} {\bibfnamefont {N.~J.}\ \bibnamefont
  {Cornish}},\ }\bibfield  {title} {\bibinfo {title} {{Bayesian inference for
  spectral estimation of gravitational wave detector noise}},\ }\href
  {https://doi.org/10.1103/PhysRevD.91.084034} {\bibfield  {journal} {\bibinfo
  {journal} {Phys. Rev.}\ }\textbf {\bibinfo {volume} {D91}},\ \bibinfo {pages}
  {084034} (\bibinfo {year} {2015})},\ \Eprint
  {https://arxiv.org/abs/1410.3852} {arXiv:1410.3852 [gr-qc]} \BibitemShut
  {NoStop}%
\bibitem [{\citenamefont {Chatziioannou}\ \emph {et~al.}(2019)\citenamefont
  {Chatziioannou}, \citenamefont {Haster}, \citenamefont {Littenberg},
  \citenamefont {Farr}, \citenamefont {Ghonge}, \citenamefont {Millhouse},
  \citenamefont {Clark},\ and\ \citenamefont {Cornish}}]{Chatziioannou:2019}%
  \BibitemOpen
  \bibfield  {author} {\bibinfo {author} {\bibfnamefont {K.}~\bibnamefont
  {Chatziioannou}}, \bibinfo {author} {\bibfnamefont {C.-J.}\ \bibnamefont
  {Haster}}, \bibinfo {author} {\bibfnamefont {T.~B.}\ \bibnamefont
  {Littenberg}}, \bibinfo {author} {\bibfnamefont {W.~M.}\ \bibnamefont
  {Farr}}, \bibinfo {author} {\bibfnamefont {S.}~\bibnamefont {Ghonge}},
  \bibinfo {author} {\bibfnamefont {M.}~\bibnamefont {Millhouse}}, \bibinfo
  {author} {\bibfnamefont {J.~A.}\ \bibnamefont {Clark}},\ and\ \bibinfo
  {author} {\bibfnamefont {N.}~\bibnamefont {Cornish}},\ }\bibfield  {title}
  {\bibinfo {title} {Noise spectral estimation methods and their impact on
  gravitational wave measurement of compact binary mergers},\ }\href
  {https://doi.org/10.1103/PhysRevD.100.104004} {\bibfield  {journal} {\bibinfo
   {journal} {Phys. Rev. D}\ }\textbf {\bibinfo {volume} {100}},\ \bibinfo
  {pages} {104004} (\bibinfo {year} {2019})}\BibitemShut {NoStop}%
\bibitem [{\citenamefont {Abbott}\ \emph
  {et~al.}(2016{\natexlab{a}})\citenamefont {Abbott} \emph
  {et~al.}}]{Abbott:2016blz}%
  \BibitemOpen
  \bibfield  {author} {\bibinfo {author} {\bibfnamefont {B.}~\bibnamefont
  {Abbott}} \emph {et~al.} (\bibinfo {collaboration} {LIGO Scientific,
  Virgo}),\ }\bibfield  {title} {\bibinfo {title} {{Observation of
  Gravitational Waves from a Binary Black Hole Merger}},\ }\href
  {https://doi.org/10.1103/PhysRevLett.116.061102} {\bibfield  {journal}
  {\bibinfo  {journal} {Phys. Rev. Lett.}\ }\textbf {\bibinfo {volume} {116}},\
  \bibinfo {pages} {061102} (\bibinfo {year} {2016}{\natexlab{a}})},\ \Eprint
  {https://arxiv.org/abs/1602.03837} {arXiv:1602.03837 [gr-qc]} \BibitemShut
  {NoStop}%
\bibitem [{\citenamefont {Abbott}\ \emph
  {et~al.}(2017{\natexlab{c}})\citenamefont {Abbott} \emph
  {et~al.}}]{Abbott:2016wiq}%
  \BibitemOpen
  \bibfield  {author} {\bibinfo {author} {\bibfnamefont {B.~P.}\ \bibnamefont
  {Abbott}} \emph {et~al.} (\bibinfo {collaboration} {LIGO Scientific,
  Virgo}),\ }\bibfield  {title} {\bibinfo {title} {{Effects of waveform model
  systematics on the interpretation of GW150914}},\ }\href
  {https://doi.org/10.1088/1361-6382/aa6854} {\bibfield  {journal} {\bibinfo
  {journal} {Class. Quant. Grav.}\ }\textbf {\bibinfo {volume} {34}},\ \bibinfo
  {pages} {104002} (\bibinfo {year} {2017}{\natexlab{c}})},\ \Eprint
  {https://arxiv.org/abs/1611.07531} {arXiv:1611.07531 [gr-qc]} \BibitemShut
  {NoStop}%
\bibitem [{\citenamefont {{Gravitational Wave Open Science Center
  (GWOSC)}}()}]{GWOSC}%
  \BibitemOpen
  \bibfield  {author} {\bibinfo {author} {\bibnamefont {{Gravitational Wave
  Open Science Center (GWOSC)}}},\ }\href@noop {} {}\bibinfo {howpublished}
  {\url{https://www.gw-openscience.org/}}\BibitemShut {NoStop}%
\bibitem [{\citenamefont {Abbott}\ \emph {et~al.}(2021)\citenamefont {Abbott}
  \emph {et~al.}}]{Abbott:2019ebz}%
  \BibitemOpen
  \bibfield  {author} {\bibinfo {author} {\bibfnamefont {R.}~\bibnamefont
  {Abbott}} \emph {et~al.} (\bibinfo {collaboration} {LIGO Scientific,
  Virgo}),\ }\bibfield  {title} {\bibinfo {title} {{Open data from the first
  and second observing runs of Advanced LIGO and Advanced Virgo}},\ }\href
  {https://doi.org/10.1016/j.softx.2021.100658} {\bibfield  {journal} {\bibinfo
   {journal} {SoftwareX}\ }\textbf {\bibinfo {volume} {13}},\ \bibinfo {pages}
  {100658} (\bibinfo {year} {2021})},\ \Eprint
  {https://arxiv.org/abs/1912.11716} {arXiv:1912.11716 [gr-qc]} \BibitemShut
  {NoStop}%
\bibitem [{\citenamefont {Abbott}\ \emph
  {et~al.}(2016{\natexlab{b}})\citenamefont {Abbott} \emph
  {et~al.}}]{LIGOScientific:2016vlm}%
  \BibitemOpen
  \bibfield  {author} {\bibinfo {author} {\bibfnamefont {B.~P.}\ \bibnamefont
  {Abbott}} \emph {et~al.} (\bibinfo {collaboration} {LIGO Scientific,
  Virgo}),\ }\bibfield  {title} {\bibinfo {title} {{Properties of the Binary
  Black Hole Merger GW150914}},\ }\href
  {https://doi.org/10.1103/PhysRevLett.116.241102} {\bibfield  {journal}
  {\bibinfo  {journal} {Phys. Rev. Lett.}\ }\textbf {\bibinfo {volume} {116}},\
  \bibinfo {pages} {241102} (\bibinfo {year} {2016}{\natexlab{b}})},\ \Eprint
  {https://arxiv.org/abs/1602.03840} {arXiv:1602.03840 [gr-qc]} \BibitemShut
  {NoStop}%
\bibitem [{\citenamefont {Abbott}\ \emph
  {et~al.}(2016{\natexlab{c}})\citenamefont {Abbott} \emph
  {et~al.}}]{LIGOScientific:2016dsl}%
  \BibitemOpen
  \bibfield  {author} {\bibinfo {author} {\bibfnamefont {B.~P.}\ \bibnamefont
  {Abbott}} \emph {et~al.} (\bibinfo {collaboration} {LIGO Scientific,
  Virgo}),\ }\bibfield  {title} {\bibinfo {title} {{Binary Black Hole Mergers
  in the first Advanced LIGO Observing Run}},\ }\href
  {https://doi.org/10.1103/PhysRevX.6.041015} {\bibfield  {journal} {\bibinfo
  {journal} {Phys. Rev. X}\ }\textbf {\bibinfo {volume} {6}},\ \bibinfo {pages}
  {041015} (\bibinfo {year} {2016}{\natexlab{c}})},\ \bibinfo {note} {[Erratum:
  Phys.Rev.X 8, 039903 (2018)]},\ \Eprint {https://arxiv.org/abs/1606.04856}
  {arXiv:1606.04856 [gr-qc]} \BibitemShut {NoStop}%
\bibitem [{\citenamefont {Abbott}\ \emph
  {et~al.}(2018{\natexlab{a}})\citenamefont {Abbott} \emph
  {et~al.}}]{LIGOScientific:2018mvr}%
  \BibitemOpen
  \bibfield  {author} {\bibinfo {author} {\bibfnamefont {B.~P.}\ \bibnamefont
  {Abbott}} \emph {et~al.} (\bibinfo {collaboration} {LIGO Scientific,
  Virgo}),\ }\bibfield  {title} {\bibinfo {title} {{GWTC-1: A
  Gravitational-Wave Transient Catalog of Compact Binary Mergers Observed by
  LIGO and Virgo during the First and Second Observing Runs}},\ }\href@noop {}
  {\  (\bibinfo {year} {2018}{\natexlab{a}})},\ \Eprint
  {https://arxiv.org/abs/1811.12907} {arXiv:1811.12907 [astro-ph.HE]}
  \BibitemShut {NoStop}%
\bibitem [{\citenamefont {Abbott}\ \emph
  {et~al.}(2018{\natexlab{b}})\citenamefont {Abbott} \emph
  {et~al.}}]{Abbott:2018exr}%
  \BibitemOpen
  \bibfield  {author} {\bibinfo {author} {\bibfnamefont {B.~P.}\ \bibnamefont
  {Abbott}} \emph {et~al.} (\bibinfo {collaboration} {LIGO Scientific,
  Virgo}),\ }\bibfield  {title} {\bibinfo {title} {{GW170817: Measurements of
  neutron star radii and equation of state}},\ }\href
  {https://doi.org/10.1103/PhysRevLett.121.161101} {\bibfield  {journal}
  {\bibinfo  {journal} {Phys. Rev. Lett.}\ }\textbf {\bibinfo {volume} {121}},\
  \bibinfo {pages} {161101} (\bibinfo {year} {2018}{\natexlab{b}})},\ \Eprint
  {https://arxiv.org/abs/1805.11581} {arXiv:1805.11581 [gr-qc]} \BibitemShut
  {NoStop}%
\bibitem [{\citenamefont {{BayesWave Glitch Subtraction for
  GW170817}}()}]{GW170817Data}%
  \BibitemOpen
  \bibfield  {author} {\bibinfo {author} {\bibnamefont {{BayesWave Glitch
  Subtraction for GW170817}}},\ }\href@noop {} {}\bibinfo {howpublished}
  {\url{https://dcc.ligo.org/LIGO-T1700406/public}}\BibitemShut {NoStop}%
\bibitem [{\citenamefont {{Soares-Santos}}\ \emph {et~al.}(2017)\citenamefont
  {{Soares-Santos}}, \citenamefont {et~al.}, \citenamefont {{Dark Energy
  Survey}},\ and\ \citenamefont {{Dark Energy Camera GW-EM
  Collaboration}}}]{2017ApJ...848L..16S}%
  \BibitemOpen
  \bibfield  {author} {\bibinfo {author} {\bibfnamefont {M.}~\bibnamefont
  {{Soares-Santos}}}, \bibinfo {author} {\bibnamefont {et~al.}}, \bibinfo
  {author} {\bibnamefont {{Dark Energy Survey}}},\ and\ \bibinfo {author}
  {\bibnamefont {{Dark Energy Camera GW-EM Collaboration}}},\ }\bibfield
  {title} {\bibinfo {title} {{The Electromagnetic Counterpart of the Binary
  Neutron Star Merger LIGO/Virgo GW170817. I. Discovery of the Optical
  Counterpart Using the Dark Energy Camera}},\ }\href
  {https://doi.org/10.3847/2041-8213/aa9059} {\bibfield  {journal} {\bibinfo
  {journal} {Astrophysical Journal, Letters}\ }\textbf {\bibinfo {volume}
  {848}},\ \bibinfo {eid} {L16} (\bibinfo {year} {2017})},\ \Eprint
  {https://arxiv.org/abs/1710.05459} {arXiv:1710.05459 [astro-ph.HE]}
  \BibitemShut {NoStop}%
\bibitem [{\citenamefont {Abbott}\ \emph
  {et~al.}(2017{\natexlab{d}})\citenamefont {Abbott}, \citenamefont {Abbott},
  \citenamefont {Abbott}, \citenamefont {Acernese}, \citenamefont {Ackley},
  \citenamefont {Adams}, \citenamefont {Adams}, \citenamefont {Addesso},
  \citenamefont {Adhikari}, \citenamefont {Adya},\ and\ \citenamefont
  {et~al.}}]{Abbott_2017}%
  \BibitemOpen
  \bibfield  {author} {\bibinfo {author} {\bibfnamefont {B.~P.}\ \bibnamefont
  {Abbott}}, \bibinfo {author} {\bibfnamefont {R.}~\bibnamefont {Abbott}},
  \bibinfo {author} {\bibfnamefont {T.~D.}\ \bibnamefont {Abbott}}, \bibinfo
  {author} {\bibfnamefont {F.}~\bibnamefont {Acernese}}, \bibinfo {author}
  {\bibfnamefont {K.}~\bibnamefont {Ackley}}, \bibinfo {author} {\bibfnamefont
  {C.}~\bibnamefont {Adams}}, \bibinfo {author} {\bibfnamefont
  {T.}~\bibnamefont {Adams}}, \bibinfo {author} {\bibfnamefont
  {P.}~\bibnamefont {Addesso}}, \bibinfo {author} {\bibfnamefont {R.~X.}\
  \bibnamefont {Adhikari}}, \bibinfo {author} {\bibfnamefont {V.~B.}\
  \bibnamefont {Adya}},\ and\ \bibinfo {author} {\bibnamefont {et~al.}},\
  }\bibfield  {title} {\bibinfo {title} {Multi-messenger observations of a
  binary neutron star merger},\ }\href
  {https://doi.org/10.3847/2041-8213/aa91c9} {\bibfield  {journal} {\bibinfo
  {journal} {The Astrophysical Journal}\ }\textbf {\bibinfo {volume} {848}},\
  \bibinfo {pages} {L12} (\bibinfo {year} {2017}{\natexlab{d}})},\ \Eprint
  {https://arxiv.org/abs/1710.05833} {arXiv:1710.05833} \BibitemShut {NoStop}%
\bibitem [{\citenamefont {{Dietrich}}\ \emph {et~al.}(2019)\citenamefont
  {{Dietrich}}, \citenamefont {{Khan}}, \citenamefont {{Dudi}}, \citenamefont
  {{Kapadia}}, \citenamefont {{Kumar}}, \citenamefont {{Nagar}}, \citenamefont
  {{Ohme}}, \citenamefont {{Pannarale}}, \citenamefont {{Samajdar}},
  \citenamefont {{Bernuzzi}}, \citenamefont {{Carullo}}, \citenamefont {{Del
  Pozzo}}, \citenamefont {{Haney}}, \citenamefont {{Markakis}}, \citenamefont
  {{P{\"u}rrer}}, \citenamefont {{Riemenschneider}}, \citenamefont
  {{Setyawati}}, \citenamefont {{Tsang}},\ and\ \citenamefont {{Van Den
  Broeck}}}]{2019PhRvD..99b4029D}%
  \BibitemOpen
  \bibfield  {author} {\bibinfo {author} {\bibfnamefont {T.}~\bibnamefont
  {{Dietrich}}}, \bibinfo {author} {\bibfnamefont {S.}~\bibnamefont {{Khan}}},
  \bibinfo {author} {\bibfnamefont {R.}~\bibnamefont {{Dudi}}}, \bibinfo
  {author} {\bibfnamefont {S.~J.}\ \bibnamefont {{Kapadia}}}, \bibinfo {author}
  {\bibfnamefont {P.}~\bibnamefont {{Kumar}}}, \bibinfo {author} {\bibfnamefont
  {A.}~\bibnamefont {{Nagar}}}, \bibinfo {author} {\bibfnamefont
  {F.}~\bibnamefont {{Ohme}}}, \bibinfo {author} {\bibfnamefont
  {F.}~\bibnamefont {{Pannarale}}}, \bibinfo {author} {\bibfnamefont
  {A.}~\bibnamefont {{Samajdar}}}, \bibinfo {author} {\bibfnamefont
  {S.}~\bibnamefont {{Bernuzzi}}}, \bibinfo {author} {\bibfnamefont
  {G.}~\bibnamefont {{Carullo}}}, \bibinfo {author} {\bibfnamefont
  {W.}~\bibnamefont {{Del Pozzo}}}, \bibinfo {author} {\bibfnamefont
  {M.}~\bibnamefont {{Haney}}}, \bibinfo {author} {\bibfnamefont
  {C.}~\bibnamefont {{Markakis}}}, \bibinfo {author} {\bibfnamefont
  {M.}~\bibnamefont {{P{\"u}rrer}}}, \bibinfo {author} {\bibfnamefont
  {G.}~\bibnamefont {{Riemenschneider}}}, \bibinfo {author} {\bibfnamefont
  {Y.~E.}\ \bibnamefont {{Setyawati}}}, \bibinfo {author} {\bibfnamefont
  {K.~W.}\ \bibnamefont {{Tsang}}},\ and\ \bibinfo {author} {\bibfnamefont
  {C.}~\bibnamefont {{Van Den Broeck}}},\ }\bibfield  {title} {\bibinfo {title}
  {{Matter imprints in waveform models for neutron star binaries: Tidal and
  self-spin effects}},\ }\href {https://doi.org/10.1103/PhysRevD.99.024029}
  {\bibfield  {journal} {\bibinfo  {journal} {\prd}\ }\textbf {\bibinfo
  {volume} {99}},\ \bibinfo {eid} {024029} (\bibinfo {year} {2019})},\ \Eprint
  {https://arxiv.org/abs/1804.02235} {arXiv:1804.02235 [gr-qc]} \BibitemShut
  {NoStop}%
\bibitem [{\citenamefont {Abbott}\ \emph
  {et~al.}(2017{\natexlab{e}})\citenamefont {Abbott} \emph
  {et~al.}}]{Evans:2016mbw}%
  \BibitemOpen
  \bibfield  {author} {\bibinfo {author} {\bibfnamefont {B.~P.}\ \bibnamefont
  {Abbott}} \emph {et~al.} (\bibinfo {collaboration} {LIGO Scientific
  Collaboration}),\ }\bibfield  {title} {\bibinfo {title} {{Exploring the
  Sensitivity of Next Generation Gravitational Wave Detectors}},\ }\href
  {https://doi.org/10.1088/1361-6382/aa51f4} {\bibfield  {journal} {\bibinfo
  {journal} {Class.\ Quant.\ Grav.}\ }\textbf {\bibinfo {volume} {34}},\
  \bibinfo {pages} {044001} (\bibinfo {year} {2017}{\natexlab{e}})},\ \Eprint
  {https://arxiv.org/abs/1607.08697} {arXiv:1607.08697 [astro-ph.IM]}
  \BibitemShut {NoStop}%
\bibitem [{\citenamefont {Reitze}\ \emph
  {et~al.}(2019{\natexlab{a}})\citenamefont {Reitze} \emph
  {et~al.}}]{Reitze:2019dyk}%
  \BibitemOpen
  \bibfield  {author} {\bibinfo {author} {\bibfnamefont {D.}~\bibnamefont
  {Reitze}} \emph {et~al.},\ }\bibfield  {title} {\bibinfo {title} {{The US
  Program in Ground-Based Gravitational Wave Science: Contribution from the
  LIGO Laboratory}},\ }\href@noop {} {\bibfield  {journal} {\bibinfo  {journal}
  {Bull. Am. Astron. Soc.}\ }\textbf {\bibinfo {volume} {51}},\ \bibinfo
  {pages} {141} (\bibinfo {year} {2019}{\natexlab{a}})},\ \Eprint
  {https://arxiv.org/abs/1903.04615} {arXiv:1903.04615 [astro-ph.IM]}
  \BibitemShut {NoStop}%
\bibitem [{\citenamefont {Reitze}\ \emph
  {et~al.}(2019{\natexlab{b}})\citenamefont {Reitze} \emph
  {et~al.}}]{Reitze:2019iox}%
  \BibitemOpen
  \bibfield  {author} {\bibinfo {author} {\bibfnamefont {D.}~\bibnamefont
  {Reitze}} \emph {et~al.},\ }\bibfield  {title} {\bibinfo {title} {{Cosmic
  Explorer: The U.S. Contribution to Gravitational-Wave Astronomy beyond
  LIGO}},\ }\href@noop {} {\bibfield  {journal} {\bibinfo  {journal} {Bull. Am.
  Astron. Soc.}\ }\textbf {\bibinfo {volume} {51}},\ \bibinfo {pages} {035}
  (\bibinfo {year} {2019}{\natexlab{b}})},\ \Eprint
  {https://arxiv.org/abs/1907.04833} {arXiv:1907.04833 [astro-ph.IM]}
  \BibitemShut {NoStop}%
\bibitem [{\citenamefont {{Hild}}\ \emph {et~al.}(2011)\citenamefont {{Hild}}
  \emph {et~al.}}]{2011CQGra..28i4013H}%
  \BibitemOpen
  \bibfield  {author} {\bibinfo {author} {\bibfnamefont {S.}~\bibnamefont
  {{Hild}}} \emph {et~al.},\ }\bibfield  {title} {\bibinfo {title}
  {{Sensitivity studies for third-generation gravitational wave
  observatories}},\ }\href {https://doi.org/10.1088/0264-9381/28/9/094013}
  {\bibfield  {journal} {\bibinfo  {journal} {Classical and Quantum Gravity}\
  }\textbf {\bibinfo {volume} {28}},\ \bibinfo {eid} {094013} (\bibinfo {year}
  {2011})},\ \Eprint {https://arxiv.org/abs/1012.0908} {arXiv:1012.0908
  [gr-qc]} \BibitemShut {NoStop}%
\bibitem [{\citenamefont {{Punturo}}\ \emph {et~al.}(2010)\citenamefont
  {{Punturo}} \emph {et~al.}}]{2010CQGra..27h4007P}%
  \BibitemOpen
  \bibfield  {author} {\bibinfo {author} {\bibfnamefont {M.}~\bibnamefont
  {{Punturo}}} \emph {et~al.},\ }\bibfield  {title} {\bibinfo {title} {{The
  third generation of gravitational wave observatories and their science
  reach}},\ }\href {https://doi.org/10.1088/0264-9381/27/8/084007} {\bibfield
  {journal} {\bibinfo  {journal} {Classical and Quantum Gravity}\ }\textbf
  {\bibinfo {volume} {27}},\ \bibinfo {eid} {084007} (\bibinfo {year}
  {2010})}\BibitemShut {NoStop}%
\bibitem [{\citenamefont {Barsotti}\ \emph {et~al.}(2018)\citenamefont
  {Barsotti}, \citenamefont {Fritschel}, \citenamefont {Evans},\ and\
  \citenamefont {Gras}}]{aLIGO_design_updated}%
  \BibitemOpen
  \bibfield  {author} {\bibinfo {author} {\bibfnamefont {L.}~\bibnamefont
  {Barsotti}}, \bibinfo {author} {\bibfnamefont {P.}~\bibnamefont {Fritschel}},
  \bibinfo {author} {\bibfnamefont {M.}~\bibnamefont {Evans}},\ and\ \bibinfo
  {author} {\bibfnamefont {S.}~\bibnamefont {Gras}},\ }\href
  {https://dcc.ligo.org/LIGO-T1800044/public} {\emph {\bibinfo {title} {Updated
  Advanced LIGO sensitivity design curve}}},\ \bibinfo {type} {Tech. Rep.}\
  (\bibinfo {year} {2018})\BibitemShut {NoStop}%
\bibitem [{\citenamefont {Kastaun}\ and\ \citenamefont
  {Ohme}(2021)}]{Kastaun:2021zyo}%
  \BibitemOpen
  \bibfield  {author} {\bibinfo {author} {\bibfnamefont {W.}~\bibnamefont
  {Kastaun}}\ and\ \bibinfo {author} {\bibfnamefont {F.}~\bibnamefont {Ohme}},\
  }\bibfield  {title} {\bibinfo {title} {{Numerical inside view of hypermassive
  remnant models for GW170817}},\ }\href
  {https://doi.org/10.1103/PhysRevD.104.023001} {\bibfield  {journal} {\bibinfo
   {journal} {Phys. Rev. D}\ }\textbf {\bibinfo {volume} {104}},\ \bibinfo
  {pages} {023001} (\bibinfo {year} {2021})},\ \Eprint
  {https://arxiv.org/abs/2103.01586} {arXiv:2103.01586 [astro-ph.HE]}
  \BibitemShut {NoStop}%
\bibitem [{\citenamefont {{Hempel}}\ and\ \citenamefont
  {{Schaffner-Bielich}}(2010)}]{2010NuPhA.837..210H}%
  \BibitemOpen
  \bibfield  {author} {\bibinfo {author} {\bibfnamefont {M.}~\bibnamefont
  {{Hempel}}}\ and\ \bibinfo {author} {\bibfnamefont {J.}~\bibnamefont
  {{Schaffner-Bielich}}},\ }\bibfield  {title} {\bibinfo {title} {{A
  statistical model for a complete supernova equation of state}},\ }\href
  {https://doi.org/10.1016/j.nuclphysa.2010.02.010} {\bibfield  {journal}
  {\bibinfo  {journal} {Nuclear Physics A}\ }\textbf {\bibinfo {volume}
  {837}},\ \bibinfo {pages} {210} (\bibinfo {year} {2010})},\ \Eprint
  {https://arxiv.org/abs/0911.4073} {arXiv:0911.4073 [nucl-th]} \BibitemShut
  {NoStop}%
\bibitem [{\citenamefont {{Wilson}}\ \emph {et~al.}(1996)\citenamefont
  {{Wilson}}, \citenamefont {{Mathews}},\ and\ \citenamefont
  {{Marronetti}}}]{Wilson1996}%
  \BibitemOpen
  \bibfield  {author} {\bibinfo {author} {\bibfnamefont {J.~R.}\ \bibnamefont
  {{Wilson}}}, \bibinfo {author} {\bibfnamefont {G.~J.}\ \bibnamefont
  {{Mathews}}},\ and\ \bibinfo {author} {\bibfnamefont {P.}~\bibnamefont
  {{Marronetti}}},\ }\bibfield  {title} {\bibinfo {title} {{Relativistic
  numerical model for close neutron-star binaries}},\ }\href
  {https://doi.org/10.1103/PhysRevD.54.1317} {\bibfield  {journal} {\bibinfo
  {journal} {\prd}\ }\textbf {\bibinfo {volume} {54}},\ \bibinfo {pages} {1317}
  (\bibinfo {year} {1996})}\BibitemShut {NoStop}%
\bibitem [{\citenamefont {{Isenberg}}\ and\ \citenamefont
  {{Nester}}(1980)}]{Isenberg1980}%
  \BibitemOpen
  \bibfield  {author} {\bibinfo {author} {\bibfnamefont {J.}~\bibnamefont
  {{Isenberg}}}\ and\ \bibinfo {author} {\bibfnamefont {J.}~\bibnamefont
  {{Nester}}},\ }\bibfield  {title} {\bibinfo {title} {{Canonical Gravity}},\
  }in\ \href@noop {} {\emph {\bibinfo {booktitle} {General Relativity and
  Gravitation. Vol. 1. One hundred years after the birth of Albert Einstein.
  Edited by A. Held. New York, NY: Plenum Press, p. 23, 1980}}},\ \bibinfo
  {editor} {edited by\ \bibinfo {editor} {\bibfnamefont {A.}~\bibnamefont
  {{Held}}}}\ (\bibinfo {year} {1980})\ p.~\bibinfo {pages} {23}\BibitemShut
  {NoStop}%
\bibitem [{\citenamefont {Alford}\ \emph {et~al.}(2005)\citenamefont {Alford},
  \citenamefont {Braby}, \citenamefont {Paris},\ and\ \citenamefont
  {Reddy}}]{Alford:2004pf}%
  \BibitemOpen
  \bibfield  {author} {\bibinfo {author} {\bibfnamefont {M.}~\bibnamefont
  {Alford}}, \bibinfo {author} {\bibfnamefont {M.}~\bibnamefont {Braby}},
  \bibinfo {author} {\bibfnamefont {M.~W.}\ \bibnamefont {Paris}},\ and\
  \bibinfo {author} {\bibfnamefont {S.}~\bibnamefont {Reddy}},\ }\bibfield
  {title} {\bibinfo {title} {{Hybrid stars that masquerade as neutron stars}},\
  }\href {https://doi.org/10.1086/430902} {\bibfield  {journal} {\bibinfo
  {journal} {Astrophys. J.}\ }\textbf {\bibinfo {volume} {629}},\ \bibinfo
  {pages} {969} (\bibinfo {year} {2005})},\ \Eprint
  {https://arxiv.org/abs/nucl-th/0411016} {arXiv:nucl-th/0411016} \BibitemShut
  {NoStop}%
\bibitem [{\citenamefont {Alford}\ \emph {et~al.}(2015)\citenamefont {Alford},
  \citenamefont {Burgio}, \citenamefont {Han}, \citenamefont {Taranto},\ and\
  \citenamefont {Zappal\`a}}]{Alford:2015dpa}%
  \BibitemOpen
  \bibfield  {author} {\bibinfo {author} {\bibfnamefont {M.~G.}\ \bibnamefont
  {Alford}}, \bibinfo {author} {\bibfnamefont {G.~F.}\ \bibnamefont {Burgio}},
  \bibinfo {author} {\bibfnamefont {S.}~\bibnamefont {Han}}, \bibinfo {author}
  {\bibfnamefont {G.}~\bibnamefont {Taranto}},\ and\ \bibinfo {author}
  {\bibfnamefont {D.}~\bibnamefont {Zappal\`a}},\ }\bibfield  {title} {\bibinfo
  {title} {{Constraining and applying a generic high-density equation of
  state}},\ }\href {https://doi.org/10.1103/PhysRevD.92.083002} {\bibfield
  {journal} {\bibinfo  {journal} {Phys. Rev. D}\ }\textbf {\bibinfo {volume}
  {92}},\ \bibinfo {pages} {083002} (\bibinfo {year} {2015})},\ \Eprint
  {https://arxiv.org/abs/1501.07902} {arXiv:1501.07902 [nucl-th]} \BibitemShut
  {NoStop}%
\bibitem [{\citenamefont {Del~Pozzo}\ \emph {et~al.}(2013)\citenamefont
  {Del~Pozzo}, \citenamefont {Li}, \citenamefont {Agathos}, \citenamefont {Van
  Den~Broeck},\ and\ \citenamefont {Vitale}}]{DelPozzo:2013ala}%
  \BibitemOpen
  \bibfield  {author} {\bibinfo {author} {\bibfnamefont {W.}~\bibnamefont
  {Del~Pozzo}}, \bibinfo {author} {\bibfnamefont {T.~G.~F.}\ \bibnamefont
  {Li}}, \bibinfo {author} {\bibfnamefont {M.}~\bibnamefont {Agathos}},
  \bibinfo {author} {\bibfnamefont {C.}~\bibnamefont {Van Den~Broeck}},\ and\
  \bibinfo {author} {\bibfnamefont {S.}~\bibnamefont {Vitale}},\ }\bibfield
  {title} {\bibinfo {title} {{Demonstrating the feasibility of probing the
  neutron star equation of state with second-generation gravitational wave
  detectors}},\ }\href {https://doi.org/10.1103/PhysRevLett.111.071101}
  {\bibfield  {journal} {\bibinfo  {journal} {Phys. Rev. Lett.}\ }\textbf
  {\bibinfo {volume} {111}},\ \bibinfo {pages} {071101} (\bibinfo {year}
  {2013})},\ \Eprint {https://arxiv.org/abs/1307.8338} {arXiv:1307.8338
  [gr-qc]} \BibitemShut {NoStop}%
\bibitem [{\citenamefont {Agathos}\ \emph {et~al.}(2015)\citenamefont
  {Agathos}, \citenamefont {Meidam}, \citenamefont {Del~Pozzo}, \citenamefont
  {Li}, \citenamefont {Tompitak}, \citenamefont {Veitch}, \citenamefont
  {Vitale},\ and\ \citenamefont {Van Den~Broeck}}]{Agathos:2015uaa}%
  \BibitemOpen
  \bibfield  {author} {\bibinfo {author} {\bibfnamefont {M.}~\bibnamefont
  {Agathos}}, \bibinfo {author} {\bibfnamefont {J.}~\bibnamefont {Meidam}},
  \bibinfo {author} {\bibfnamefont {W.}~\bibnamefont {Del~Pozzo}}, \bibinfo
  {author} {\bibfnamefont {T.~G.~F.}\ \bibnamefont {Li}}, \bibinfo {author}
  {\bibfnamefont {M.}~\bibnamefont {Tompitak}}, \bibinfo {author}
  {\bibfnamefont {J.}~\bibnamefont {Veitch}}, \bibinfo {author} {\bibfnamefont
  {S.}~\bibnamefont {Vitale}},\ and\ \bibinfo {author} {\bibfnamefont
  {C.}~\bibnamefont {Van Den~Broeck}},\ }\bibfield  {title} {\bibinfo {title}
  {{Constraining the neutron star equation of state with gravitational wave
  signals from coalescing binary neutron stars}},\ }\href
  {https://doi.org/10.1103/PhysRevD.92.023012} {\bibfield  {journal} {\bibinfo
  {journal} {Phys. Rev. D}\ }\textbf {\bibinfo {volume} {92}},\ \bibinfo
  {pages} {023012} (\bibinfo {year} {2015})},\ \Eprint
  {https://arxiv.org/abs/1503.05405} {arXiv:1503.05405 [gr-qc]} \BibitemShut
  {NoStop}%
\bibitem [{\citenamefont {Chatziioannou}\ \emph {et~al.}(2015)\citenamefont
  {Chatziioannou}, \citenamefont {Yagi}, \citenamefont {Klein}, \citenamefont
  {Cornish},\ and\ \citenamefont {Yunes}}]{Chatziioannou:2015uea}%
  \BibitemOpen
  \bibfield  {author} {\bibinfo {author} {\bibfnamefont {K.}~\bibnamefont
  {Chatziioannou}}, \bibinfo {author} {\bibfnamefont {K.}~\bibnamefont {Yagi}},
  \bibinfo {author} {\bibfnamefont {A.}~\bibnamefont {Klein}}, \bibinfo
  {author} {\bibfnamefont {N.}~\bibnamefont {Cornish}},\ and\ \bibinfo {author}
  {\bibfnamefont {N.}~\bibnamefont {Yunes}},\ }\bibfield  {title} {\bibinfo
  {title} {{Probing the Internal Composition of Neutron Stars with
  Gravitational Waves}},\ }\href {https://doi.org/10.1103/PhysRevD.92.104008}
  {\bibfield  {journal} {\bibinfo  {journal} {Phys. Rev. D}\ }\textbf {\bibinfo
  {volume} {92}},\ \bibinfo {pages} {104008} (\bibinfo {year} {2015})},\
  \Eprint {https://arxiv.org/abs/1508.02062} {arXiv:1508.02062 [gr-qc]}
  \BibitemShut {NoStop}%
\bibitem [{\citenamefont {Han}\ and\ \citenamefont
  {Steiner}(2019)}]{Han:2018mtj}%
  \BibitemOpen
  \bibfield  {author} {\bibinfo {author} {\bibfnamefont {S.}~\bibnamefont
  {Han}}\ and\ \bibinfo {author} {\bibfnamefont {A.~W.}\ \bibnamefont
  {Steiner}},\ }\bibfield  {title} {\bibinfo {title} {{Tidal deformability with
  sharp phase transitions in (binary) neutron stars}},\ }\href
  {https://doi.org/10.1103/PhysRevD.99.083014} {\bibfield  {journal} {\bibinfo
  {journal} {Phys. Rev. D}\ }\textbf {\bibinfo {volume} {99}},\ \bibinfo
  {pages} {083014} (\bibinfo {year} {2019})},\ \Eprint
  {https://arxiv.org/abs/1810.10967} {arXiv:1810.10967 [nucl-th]} \BibitemShut
  {NoStop}%
\bibitem [{\citenamefont {Chen}\ \emph {et~al.}(2020)\citenamefont {Chen},
  \citenamefont {Chesler},\ and\ \citenamefont {Loeb}}]{Chen:2019rja}%
  \BibitemOpen
  \bibfield  {author} {\bibinfo {author} {\bibfnamefont {H.-Y.}\ \bibnamefont
  {Chen}}, \bibinfo {author} {\bibfnamefont {P.~M.}\ \bibnamefont {Chesler}},\
  and\ \bibinfo {author} {\bibfnamefont {A.}~\bibnamefont {Loeb}},\ }\bibfield
  {title} {\bibinfo {title} {{Searching for exotic cores with binary neutron
  star inspirals}},\ }\href {https://doi.org/10.3847/2041-8213/ab830f}
  {\bibfield  {journal} {\bibinfo  {journal} {Astrophys. J. Lett.}\ }\textbf
  {\bibinfo {volume} {893}},\ \bibinfo {pages} {L4} (\bibinfo {year} {2020})},\
  \Eprint {https://arxiv.org/abs/1909.04096} {arXiv:1909.04096 [astro-ph.HE]}
  \BibitemShut {NoStop}%
\bibitem [{\citenamefont {Chatziioannou}\ and\ \citenamefont
  {Han}(2020)}]{Chatziioannou:2019yko}%
  \BibitemOpen
  \bibfield  {author} {\bibinfo {author} {\bibfnamefont {K.}~\bibnamefont
  {Chatziioannou}}\ and\ \bibinfo {author} {\bibfnamefont {S.}~\bibnamefont
  {Han}},\ }\bibfield  {title} {\bibinfo {title} {{Studying strong phase
  transitions in neutron stars with gravitational waves}},\ }\href
  {https://doi.org/10.1103/PhysRevD.101.044019} {\bibfield  {journal} {\bibinfo
   {journal} {Phys. Rev. D}\ }\textbf {\bibinfo {volume} {101}},\ \bibinfo
  {pages} {044019} (\bibinfo {year} {2020})},\ \Eprint
  {https://arxiv.org/abs/1911.07091} {arXiv:1911.07091 [gr-qc]} \BibitemShut
  {NoStop}%
\bibitem [{\citenamefont {Han}\ and\ \citenamefont
  {Prakash}(2020)}]{Han:2020adu}%
  \BibitemOpen
  \bibfield  {author} {\bibinfo {author} {\bibfnamefont {S.}~\bibnamefont
  {Han}}\ and\ \bibinfo {author} {\bibfnamefont {M.}~\bibnamefont {Prakash}},\
  }\bibfield  {title} {\bibinfo {title} {{On the Minimum Radius of Very Massive
  Neutron Stars}},\ }\href {https://doi.org/10.3847/1538-4357/aba3c7}
  {\bibfield  {journal} {\bibinfo  {journal} {Astrophys. J.}\ }\textbf
  {\bibinfo {volume} {899}},\ \bibinfo {pages} {164} (\bibinfo {year}
  {2020})},\ \Eprint {https://arxiv.org/abs/2006.02207} {arXiv:2006.02207
  [astro-ph.HE]} \BibitemShut {NoStop}%
\bibitem [{\citenamefont {Zhang}\ and\ \citenamefont
  {Li}(2019)}]{Zhang:2019fog}%
  \BibitemOpen
  \bibfield  {author} {\bibinfo {author} {\bibfnamefont {N.-B.}\ \bibnamefont
  {Zhang}}\ and\ \bibinfo {author} {\bibfnamefont {B.-A.}\ \bibnamefont {Li}},\
  }\bibfield  {title} {\bibinfo {title} {{Implications of the Mass
  $M=2.17^{+0.11}_{-0.10}$M$_\odot$ of PSR J0740+6620 on the Equation of State
  of Super-dense Neutron-rich Nuclear Matter}},\ }\href
  {https://doi.org/10.3847/1538-4357/ab24cb} {\bibfield  {journal} {\bibinfo
  {journal} {Astrophys. J.}\ }\textbf {\bibinfo {volume} {879}},\ \bibinfo
  {pages} {99} (\bibinfo {year} {2019})},\ \Eprint
  {https://arxiv.org/abs/1904.10998} {arXiv:1904.10998 [nucl-th]} \BibitemShut
  {NoStop}%
\bibitem [{\citenamefont {Pang}\ \emph {et~al.}(2020)\citenamefont {Pang},
  \citenamefont {Dietrich}, \citenamefont {Tews},\ and\ \citenamefont {Van
  Den~Broeck}}]{Pang:2020ilf}%
  \BibitemOpen
  \bibfield  {author} {\bibinfo {author} {\bibfnamefont {P.~T.~H.}\
  \bibnamefont {Pang}}, \bibinfo {author} {\bibfnamefont {T.}~\bibnamefont
  {Dietrich}}, \bibinfo {author} {\bibfnamefont {I.}~\bibnamefont {Tews}},\
  and\ \bibinfo {author} {\bibfnamefont {C.}~\bibnamefont {Van Den~Broeck}},\
  }\bibfield  {title} {\bibinfo {title} {{Parameter estimation for strong phase
  transitions in supranuclear matter using gravitational-wave astronomy}},\
  }\href {https://doi.org/10.1103/PhysRevResearch.2.033514} {\bibfield
  {journal} {\bibinfo  {journal} {Phys. Rev. Res.}\ }\textbf {\bibinfo {volume}
  {2}},\ \bibinfo {pages} {033514} (\bibinfo {year} {2020})},\ \Eprint
  {https://arxiv.org/abs/2006.14936} {arXiv:2006.14936 [astro-ph.HE]}
  \BibitemShut {NoStop}%
\bibitem [{\citenamefont {Drischler}\ \emph {et~al.}(2021)\citenamefont
  {Drischler}, \citenamefont {Han}, \citenamefont {Lattimer}, \citenamefont
  {Prakash}, \citenamefont {Reddy},\ and\ \citenamefont
  {Zhao}}]{Drischler:2020fvz}%
  \BibitemOpen
  \bibfield  {author} {\bibinfo {author} {\bibfnamefont {C.}~\bibnamefont
  {Drischler}}, \bibinfo {author} {\bibfnamefont {S.}~\bibnamefont {Han}},
  \bibinfo {author} {\bibfnamefont {J.~M.}\ \bibnamefont {Lattimer}}, \bibinfo
  {author} {\bibfnamefont {M.}~\bibnamefont {Prakash}}, \bibinfo {author}
  {\bibfnamefont {S.}~\bibnamefont {Reddy}},\ and\ \bibinfo {author}
  {\bibfnamefont {T.}~\bibnamefont {Zhao}},\ }\bibfield  {title} {\bibinfo
  {title} {{Limiting masses and radii of neutron stars and their
  implications}},\ }\href {https://doi.org/10.1103/PhysRevC.103.045808}
  {\bibfield  {journal} {\bibinfo  {journal} {Phys. Rev. C}\ }\textbf {\bibinfo
  {volume} {103}},\ \bibinfo {pages} {045808} (\bibinfo {year} {2021})},\
  \Eprint {https://arxiv.org/abs/2009.06441} {arXiv:2009.06441 [nucl-th]}
  \BibitemShut {NoStop}%
\bibitem [{\citenamefont {{Fischer}}\ \emph {et~al.}(2018)\citenamefont
  {{Fischer}}, \citenamefont {{Bastian}}, \citenamefont {{Wu}}, \citenamefont
  {{Baklanov}}, \citenamefont {{Sorokina}}, \citenamefont {{Blinnikov}},
  \citenamefont {{Typel}}, \citenamefont {{Kl{\"a}hn}},\ and\ \citenamefont
  {{Blaschke}}}]{2018NatAs...2..980F}%
  \BibitemOpen
  \bibfield  {author} {\bibinfo {author} {\bibfnamefont {T.}~\bibnamefont
  {{Fischer}}}, \bibinfo {author} {\bibfnamefont {N.-U.~F.}\ \bibnamefont
  {{Bastian}}}, \bibinfo {author} {\bibfnamefont {M.-R.}\ \bibnamefont {{Wu}}},
  \bibinfo {author} {\bibfnamefont {P.}~\bibnamefont {{Baklanov}}}, \bibinfo
  {author} {\bibfnamefont {E.}~\bibnamefont {{Sorokina}}}, \bibinfo {author}
  {\bibfnamefont {S.}~\bibnamefont {{Blinnikov}}}, \bibinfo {author}
  {\bibfnamefont {S.}~\bibnamefont {{Typel}}}, \bibinfo {author} {\bibfnamefont
  {T.}~\bibnamefont {{Kl{\"a}hn}}},\ and\ \bibinfo {author} {\bibfnamefont
  {D.~B.}\ \bibnamefont {{Blaschke}}},\ }\bibfield  {title} {\bibinfo {title}
  {{Quark deconfinement as a supernova explosion engine for massive blue
  supergiant stars}},\ }\href {https://doi.org/10.1038/s41550-018-0583-0}
  {\bibfield  {journal} {\bibinfo  {journal} {Nature Astronomy}\ }\textbf
  {\bibinfo {volume} {2}},\ \bibinfo {pages} {980} (\bibinfo {year} {2018})},\
  \Eprint {https://arxiv.org/abs/1712.08788} {arXiv:1712.08788 [astro-ph.HE]}
  \BibitemShut {NoStop}%
\bibitem [{\citenamefont {{Bastian}}(2021)}]{2021PhRvD.103b3001B}%
  \BibitemOpen
  \bibfield  {author} {\bibinfo {author} {\bibfnamefont {N.-U.~F.}\
  \bibnamefont {{Bastian}}},\ }\bibfield  {title} {\bibinfo {title}
  {{Phenomenological quark-hadron equations of state with first-order phase
  transitions for astrophysical applications}},\ }\href
  {https://doi.org/10.1103/PhysRevD.103.023001} {\bibfield  {journal} {\bibinfo
   {journal} {\prd}\ }\textbf {\bibinfo {volume} {103}},\ \bibinfo {eid}
  {023001} (\bibinfo {year} {2021})},\ \Eprint
  {https://arxiv.org/abs/2009.10846} {arXiv:2009.10846 [nucl-th]} \BibitemShut
  {NoStop}%
\bibitem [{\citenamefont {{Stergioulas}}\ \emph {et~al.}(2011)\citenamefont
  {{Stergioulas}}, \citenamefont {{Bauswein}}, \citenamefont {{Zagkouris}},\
  and\ \citenamefont {{Janka}}}]{2011MNRAS.418..427S}%
  \BibitemOpen
  \bibfield  {author} {\bibinfo {author} {\bibfnamefont {N.}~\bibnamefont
  {{Stergioulas}}}, \bibinfo {author} {\bibfnamefont {A.}~\bibnamefont
  {{Bauswein}}}, \bibinfo {author} {\bibfnamefont {K.}~\bibnamefont
  {{Zagkouris}}},\ and\ \bibinfo {author} {\bibfnamefont {H.-T.}\ \bibnamefont
  {{Janka}}},\ }\bibfield  {title} {\bibinfo {title} {{Gravitational waves and
  non-axisymmetric oscillation modes in mergers of compact object binaries}},\
  }\href {https://doi.org/10.1111/j.1365-2966.2011.19493.x} {\bibfield
  {journal} {\bibinfo  {journal} {Monthly Notices of the Royal Astronomical
  Society}\ }\textbf {\bibinfo {volume} {418}},\ \bibinfo {pages} {427}
  (\bibinfo {year} {2011})},\ \Eprint {https://arxiv.org/abs/1105.0368}
  {arXiv:1105.0368 [gr-qc]} \BibitemShut {NoStop}%
\bibitem [{\citenamefont {Samajdar}\ and\ \citenamefont
  {Dietrich}(2019)}]{Samajdar:2019ulq}%
  \BibitemOpen
  \bibfield  {author} {\bibinfo {author} {\bibfnamefont {A.}~\bibnamefont
  {Samajdar}}\ and\ \bibinfo {author} {\bibfnamefont {T.}~\bibnamefont
  {Dietrich}},\ }\bibfield  {title} {\bibinfo {title} {{Waveform systematics
  for binary neutron star gravitational wave signals: Effects of spin,
  precession, and the observation of electromagnetic counterparts}},\ }\href
  {https://doi.org/10.1103/PhysRevD.100.024046} {\bibfield  {journal} {\bibinfo
   {journal} {Phys. Rev. D}\ }\textbf {\bibinfo {volume} {100}},\ \bibinfo
  {pages} {024046} (\bibinfo {year} {2019})},\ \Eprint
  {https://arxiv.org/abs/1905.03118} {arXiv:1905.03118 [gr-qc]} \BibitemShut
  {NoStop}%
\bibitem [{\citenamefont {Gamba}\ \emph {et~al.}(2021)\citenamefont {Gamba},
  \citenamefont {Breschi}, \citenamefont {Bernuzzi}, \citenamefont {Agathos},\
  and\ \citenamefont {Nagar}}]{Gamba:2020wgg}%
  \BibitemOpen
  \bibfield  {author} {\bibinfo {author} {\bibfnamefont {R.}~\bibnamefont
  {Gamba}}, \bibinfo {author} {\bibfnamefont {M.}~\bibnamefont {Breschi}},
  \bibinfo {author} {\bibfnamefont {S.}~\bibnamefont {Bernuzzi}}, \bibinfo
  {author} {\bibfnamefont {M.}~\bibnamefont {Agathos}},\ and\ \bibinfo {author}
  {\bibfnamefont {A.}~\bibnamefont {Nagar}},\ }\bibfield  {title} {\bibinfo
  {title} {{Waveform systematics in the gravitational-wave inference of tidal
  parameters and equation of state from binary neutron star signals}},\ }\href
  {https://doi.org/10.1103/PhysRevD.103.124015} {\bibfield  {journal} {\bibinfo
   {journal} {Phys. Rev. D}\ }\textbf {\bibinfo {volume} {103}},\ \bibinfo
  {pages} {124015} (\bibinfo {year} {2021})},\ \Eprint
  {https://arxiv.org/abs/2009.08467} {arXiv:2009.08467 [gr-qc]} \BibitemShut
  {NoStop}%
\bibitem [{\citenamefont {Chatziioannou}(2021)}]{Chatziioannou:2021tdi}%
  \BibitemOpen
  \bibfield  {author} {\bibinfo {author} {\bibfnamefont {K.}~\bibnamefont
  {Chatziioannou}},\ }\bibfield  {title} {\bibinfo {title} {{Uncertainty limits
  on neutron star radius measurements with gravitational waves}},\ }\href@noop
  {} {\  (\bibinfo {year} {2021})},\ \Eprint {https://arxiv.org/abs/2108.12368}
  {arXiv:2108.12368 [gr-qc]} \BibitemShut {NoStop}%
\bibitem [{\citenamefont {Kunert}\ \emph {et~al.}(2021)\citenamefont {Kunert},
  \citenamefont {Pang}, \citenamefont {Tews}, \citenamefont {Coughlin},\ and\
  \citenamefont {Dietrich}}]{Kunert:2021hgm}%
  \BibitemOpen
  \bibfield  {author} {\bibinfo {author} {\bibfnamefont {N.}~\bibnamefont
  {Kunert}}, \bibinfo {author} {\bibfnamefont {P.~T.~H.}\ \bibnamefont {Pang}},
  \bibinfo {author} {\bibfnamefont {I.}~\bibnamefont {Tews}}, \bibinfo {author}
  {\bibfnamefont {M.~W.}\ \bibnamefont {Coughlin}},\ and\ \bibinfo {author}
  {\bibfnamefont {T.}~\bibnamefont {Dietrich}},\ }\bibfield  {title} {\bibinfo
  {title} {{Quantifying modelling uncertainties when combining multiple
  gravitational-wave detections from binary neutron star sources}},\
  }\href@noop {} {\  (\bibinfo {year} {2021})},\ \Eprint
  {https://arxiv.org/abs/2110.11835} {arXiv:2110.11835 [astro-ph.HE]}
  \BibitemShut {NoStop}%
\bibitem [{\citenamefont {Pratten}\ \emph {et~al.}(2021)\citenamefont
  {Pratten}, \citenamefont {Schmidt},\ and\ \citenamefont
  {Williams}}]{Pratten:2021pro}%
  \BibitemOpen
  \bibfield  {author} {\bibinfo {author} {\bibfnamefont {G.}~\bibnamefont
  {Pratten}}, \bibinfo {author} {\bibfnamefont {P.}~\bibnamefont {Schmidt}},\
  and\ \bibinfo {author} {\bibfnamefont {N.}~\bibnamefont {Williams}},\
  }\bibfield  {title} {\bibinfo {title} {{Impact of Dynamical Tides on the
  Reconstruction of the Neutron Star Equation of State}},\ }\href@noop {} {\
  (\bibinfo {year} {2021})},\ \Eprint {https://arxiv.org/abs/2109.07566}
  {arXiv:2109.07566 [astro-ph.HE]} \BibitemShut {NoStop}%
\bibitem [{\citenamefont {Regimbau}\ \emph {et~al.}(2012)\citenamefont
  {Regimbau} \emph {et~al.}}]{Regimbau:2012ir}%
  \BibitemOpen
  \bibfield  {author} {\bibinfo {author} {\bibfnamefont {T.}~\bibnamefont
  {Regimbau}} \emph {et~al.},\ }\bibfield  {title} {\bibinfo {title} {{A Mock
  Data Challenge for the Einstein Gravitational-Wave Telescope}},\ }\href
  {https://doi.org/10.1103/PhysRevD.86.122001} {\bibfield  {journal} {\bibinfo
  {journal} {Phys. Rev. D}\ }\textbf {\bibinfo {volume} {86}},\ \bibinfo
  {pages} {122001} (\bibinfo {year} {2012})},\ \Eprint
  {https://arxiv.org/abs/1201.3563} {arXiv:1201.3563 [gr-qc]} \BibitemShut
  {NoStop}%
\bibitem [{\citenamefont {Sachdev}\ \emph {et~al.}(2020)\citenamefont
  {Sachdev}, \citenamefont {Regimbau},\ and\ \citenamefont
  {Sathyaprakash}}]{Sachdev:2020bkk}%
  \BibitemOpen
  \bibfield  {author} {\bibinfo {author} {\bibfnamefont {S.}~\bibnamefont
  {Sachdev}}, \bibinfo {author} {\bibfnamefont {T.}~\bibnamefont {Regimbau}},\
  and\ \bibinfo {author} {\bibfnamefont {B.~S.}\ \bibnamefont
  {Sathyaprakash}},\ }\bibfield  {title} {\bibinfo {title} {{Subtracting
  compact binary foreground sources to reveal primordial gravitational-wave
  backgrounds}},\ }\href {https://doi.org/10.1103/PhysRevD.102.024051}
  {\bibfield  {journal} {\bibinfo  {journal} {Phys. Rev. D}\ }\textbf {\bibinfo
  {volume} {102}},\ \bibinfo {pages} {024051} (\bibinfo {year} {2020})},\
  \Eprint {https://arxiv.org/abs/2002.05365} {arXiv:2002.05365 [gr-qc]}
  \BibitemShut {NoStop}%
\bibitem [{\citenamefont {Adhikari}\ \emph {et~al.}(2020)\citenamefont
  {Adhikari} \emph {et~al.}}]{LIGO:2020xsf}%
  \BibitemOpen
  \bibfield  {author} {\bibinfo {author} {\bibfnamefont {R.~X.}\ \bibnamefont
  {Adhikari}} \emph {et~al.} (\bibinfo {collaboration} {LIGO}),\ }\bibfield
  {title} {\bibinfo {title} {{A cryogenic silicon interferometer for
  gravitational-wave detection}},\ }\href
  {https://doi.org/10.1088/1361-6382/ab9143} {\bibfield  {journal} {\bibinfo
  {journal} {Class. Quant. Grav.}\ }\textbf {\bibinfo {volume} {37}},\ \bibinfo
  {pages} {165003} (\bibinfo {year} {2020})},\ \Eprint
  {https://arxiv.org/abs/2001.11173} {arXiv:2001.11173 [astro-ph.IM]}
  \BibitemShut {NoStop}%
\bibitem [{\citenamefont {Haster}\ \emph {et~al.}(2020)\citenamefont {Haster},
  \citenamefont {Chatziioannou}, \citenamefont {Bauswein},\ and\ \citenamefont
  {Clark}}]{Haster:2020sdh}%
  \BibitemOpen
  \bibfield  {author} {\bibinfo {author} {\bibfnamefont {C.-J.}\ \bibnamefont
  {Haster}}, \bibinfo {author} {\bibfnamefont {K.}~\bibnamefont
  {Chatziioannou}}, \bibinfo {author} {\bibfnamefont {A.}~\bibnamefont
  {Bauswein}},\ and\ \bibinfo {author} {\bibfnamefont {J.~A.}\ \bibnamefont
  {Clark}},\ }\bibfield  {title} {\bibinfo {title} {{Inference of the neutron
  star equation of state from cosmological distances}},\ }\href
  {https://doi.org/10.1103/PhysRevLett.125.261101} {\bibfield  {journal}
  {\bibinfo  {journal} {Phys. Rev. Lett.}\ }\textbf {\bibinfo {volume} {125}},\
  \bibinfo {pages} {261101} (\bibinfo {year} {2020})},\ \Eprint
  {https://arxiv.org/abs/2004.11334} {arXiv:2004.11334 [gr-qc]} \BibitemShut
  {NoStop}%
\bibitem [{\citenamefont {Macleod}\ \emph {et~al.}(2020)\citenamefont
  {Macleod}, \citenamefont {Urban}, \citenamefont {Coughlin}, \citenamefont
  {Massinger}, \citenamefont {Pitkin}, \citenamefont {paulaltin}, \citenamefont
  {Areeda}, \citenamefont {Quintero}, \citenamefont {Badger}, \citenamefont
  {Singer},\ and\ \citenamefont {Leinweber}}]{duncan_macleod_2020_3598469}%
  \BibitemOpen
  \bibfield  {author} {\bibinfo {author} {\bibfnamefont {D.}~\bibnamefont
  {Macleod}}, \bibinfo {author} {\bibfnamefont {A.~L.}\ \bibnamefont {Urban}},
  \bibinfo {author} {\bibfnamefont {S.}~\bibnamefont {Coughlin}}, \bibinfo
  {author} {\bibfnamefont {T.}~\bibnamefont {Massinger}}, \bibinfo {author}
  {\bibfnamefont {M.}~\bibnamefont {Pitkin}}, \bibinfo {author} {\bibnamefont
  {paulaltin}}, \bibinfo {author} {\bibfnamefont {J.}~\bibnamefont {Areeda}},
  \bibinfo {author} {\bibfnamefont {E.}~\bibnamefont {Quintero}}, \bibinfo
  {author} {\bibfnamefont {T.~G.}\ \bibnamefont {Badger}}, \bibinfo {author}
  {\bibfnamefont {L.}~\bibnamefont {Singer}},\ and\ \bibinfo {author}
  {\bibfnamefont {K.}~\bibnamefont {Leinweber}},\ }\href
  {https://doi.org/10.5281/zenodo.3598469} {\bibinfo {title} {gwpy/gwpy:
  1.0.1}} (\bibinfo {year} {2020})\BibitemShut {NoStop}%
\bibitem [{\citenamefont {Hunter}(2007)}]{Hunter:2007}%
  \BibitemOpen
  \bibfield  {author} {\bibinfo {author} {\bibfnamefont {J.~D.}\ \bibnamefont
  {Hunter}},\ }\bibfield  {title} {\bibinfo {title} {Matplotlib: A 2d graphics
  environment},\ }\href {https://doi.org/10.1109/MCSE.2007.55} {\bibfield
  {journal} {\bibinfo  {journal} {Computing In Science \& Engineering}\
  }\textbf {\bibinfo {volume} {9}},\ \bibinfo {pages} {90} (\bibinfo {year}
  {2007})}\BibitemShut {NoStop}%
\end{thebibliography}%

\end{document}